\documentclass[aps, prd, twocolumn, nofootinbib, superscriptaddress, 9pt, floatfix, rmp, amsfonts, amssymb, amsmath]{revtex4-1}

\usepackage{lipsum} 
\usepackage[utf8]{inputenc}
\usepackage{color}
\usepackage{graphicx}
\usepackage[printonlyused]{acronym}
\usepackage{array}
\usepackage{subfigure}
\usepackage{color}
\usepackage{listings}
\usepackage{natbib}
\usepackage{url}
\usepackage{bm}
\usepackage{multirow}
\usepackage{etoolbox}
\usepackage[utf8]{inputenc}
\usepackage[dvipsnames]{xcolor}
\usepackage[breaklinks, plainpages=false, colorlinks=true, anchorcolor=cyan, linkcolor=magenta, citecolor=blue, urlcolor=blue, bookmarks=false]{hyperref}
\usepackage[normalem]{ulem}
\usepackage{tabu}
\usepackage{tikz}
\usepackage{bbding}
\usetikzlibrary{arrows,shapes,trees,decorations.pathreplacing}
\usepackage{import}
\usepackage{bigints}
\usepackage{svn-multi}
\usepackage{lmodern}
\usepackage{lineno}

\DeclareMathAlphabet{\mathpzc}{OT1}{pzc}{m}{it}
\newcommand{\GW}{\ensuremath{\mathrm{GW~}}}
\newcommand{\IMBH}{\ensuremath{\mathrm{IMBH~}}}
\newcommand{\sw}[1]{\texttt{#1}}

\begin{document}

\title{Fishing massive black hole binaries with THAMES}

\author{Kritti Sharma}
    \email{kritti@caltech.edu}
    \affiliation{Department of Mechanical Engineering, Indian Institute of Technology Bombay, Powai, Mumbai 400 076, India}
\author{Koustav Chandra}
    \email{koustav.chandra@iitb.ac.in}
    \affiliation{Department of Physics, Indian Institute of Technology Bombay, Powai, Mumbai 400 076, India}
\author{Archana~Pai}
    \email{archanap@iitb.ac.in}
    \affiliation{Department of Physics, Indian Institute of Technology Bombay, Powai, Mumbai 400 076, India}
\date{\today}

\begin{abstract}
Hierarchical mergers in a dense environment are one of the primary formation channels of intermediate-mass black hole (IMBH) binary system. We expect that the resulting massive binary system will exhibit mass asymmetry. The emitted gravitational-wave (GW) carry significant contribution from higher-order modes and hence complex waveform morphology due to superposition of different modes. Further, IMBH binaries exhibit lower merger frequency and shorter signal duration in the LIGO detector which increases the risk of them being misclassified as short-duration noisy glitches. Deep learning algorithms can be trained to discriminate noisy glitches from short GW transients. We present the $\mathtt{THAMES}$ -- a deep-learning-based end-to-end signal detection algorithm for GW signals from quasi-circular nearly edge-on, mass asymmetric IMBH binaries in advanced GW detectors. Our study shows that it outperforms matched-filter based $\mathtt{PyCBC}$ searches for higher mass asymmetric,  nearly edge-on IMBH binaries. The maximum gain in the sensitive volume-time product for mass ratio $q \in (5, 10)$ is by a factor of 5.24 (2.92) against $\mathtt{PyCBC-IMBH}$ ($\mathtt{PyCBC-HM}$) search at a false alarm rate of 1 in 100 years.
Compared to the broad $\mathtt{PyCBC}$ search this factor is $\sim100$ for the $q \in (10,18)$.  One of the reasons for this leap in volumetric sensitivity is its ability to discriminate between signals with complex waveform morphology and noisy transients,  clearly demonstrating the potential of deep learning algorithms in probing into complex signal morphology in the field of gravitational wave astronomy. With the current training set, $\mathtt{THAMES}$ slightly underperforms with respect to $\mathtt{PyCBC}$-based searches targeting intermediate-mass black hole binaries with mass ratio $q \in (5, 10)$ and detector frame total mass $M_T(1+z) \in (100,200)~M_\odot$.

\end{abstract}

\maketitle

\acrodef{BBH}[BBH]{binary black hole}
\acrodef{BH}[BH]{black hole}
\acrodef{LVC}[LVC]{LIGO Scientific and Virgo Collaborations}
\acrodef{GW}[GW]{gravitational-wave}
\acrodef{CI}[CI]{confidence interval}
\acrodef{IMBH}[IMBH]{intermediate-mass black hole}
\acrodef{SNR}[SNR]{signal-to-noise ratio}
\acrodef{FAR}[FAR]{false alarm rate}
\acrodef{IFAR}[IFAR]{inverse false alarm rate}
\acrodef{PSD}[PSD]{power spectral density}
\acrodefplural{PSD}[PSDs]{power spectral densities}
\acrodef{LVK}[LVK]{LIGO Scientific, Virgo, and KAGRA Collaborations}
\acrodef{GR}[GR]{General Relativity}
\acrodef{O3}[O3]{third observing run}
\acrodef{THAMES}[THAMES]{Transfer learning for Higher-order modes from Asymmetric Massive Edge-on Systems}
\acrodef{BNS}[BNS]{binary neutron star}
\acrodef{NS}[NS]{neutron star}
\acrodef{NSBH}[NSBH]{neutron star--black hole binary}
\acrodefplural{NSBH}[NSBHs]{neutron star--black hole binaries}
\acrodef{CNN}[CNN]{convolutional neural network}
\acrodef{CWT}[CWT]{continuous wavelet transformation}

\section{Introduction}
\label{sec:introduction}
\Acf{GW} observations of compact binary mergers have to date unveiled an emerging population of \aclp{BBH}, \aclp{BNS} and \aclp{NSBH}~\citep{LIGOScientific:2018mvr, Nitz:2018imz, Venumadhav:2019lyq, Zackay:2019btq, Zackay:2019tzo, Nitz:2020oeq, LIGOScientific:2020ibl, LIGOScientific:2021djp, Nitz:2021uxj, Nitz:2021zwj, Olsen:2022pin}. These detections have helped us gain insights into compact binaries' nature, formation mechanism, population distribution, and demography. They have also allowed us to test our current understanding of gravity and have provided us with information that complements conventional astronomy and cosmology.

While most of these events correspond to \textit{canonical} compact binary mergers with their components consistent with first-generation black holes formed from stellar collapse, there have been few exceptional events. Among them, GW190521 is particularly exciting~\citep{LIGOScientific:2020iuh, LIGOScientific:2020ufj}. Despite being $\sim0.1$~s long in the detector bandwidth, the event was detected at high confidence, with a \acf{FAR} of $<1/4900$ years~\citep{Szczepanczyk:2020osv}. Parameter estimation studies reveal that the source of this event is most likely a quasi-spherical merger of two highly spinning black holes with masses $85^{+21}_{-14}~M_\odot$ and $66^{+17}_{-18}~M_\odot$. This indicates that at least the more massive of the two black holes has a high probability of being inside the pair-instability gap where no first-generation black holes can exist~\citep{Woosley:2021xba}. Further, the merger remnant with a mass of $\sim142~M_\odot$ is arguably the first unambiguously detected \acf{IMBH} --- black holes with masses between $10^2-10^5~M_\odot$. These two make GW190521 a promising hierarchical merger candidate~\citep{LIGOScientific:2020ufj, Gerosa:2021hsc}.

Hierarchical mergers usually occur in environments where the black hole number density is high such as stellar clusters, globular clusters, gaseous AGN disks, etc~\citep{Gerosa:2021hsc}. Because of their origin, they contain at least one component black hole that originated in a previous merger. Therefore, these binaries tend to have both massive and high spin components, which is in stark contrast to ``field binaries'' composed purely of stellar remnants with lower masses and spins~\citep{Baibhav:2020xdf}. Further, because of random pairing, these binaries are largely mass and spin asymmetric. All these complexities are directly reflected into the emitted \GW signal, which depending on the binary's orientation, and degree of asymmetry, can have both lower signal amplitude than their symmetric counterparts and a detectable amount of subdominant higher harmonics~\citep{Berti:2007fi, Mills:2020thr, CalderonBustillo:2015lrt, CalderonBustillo:2016rlt}.

Confident detections of these higher-order harmonics can help in many ways. For example, the observation of higher-order modes enables us to measure the properties of the systems with higher accuracies~\citep{Varma:2014jxa, CalderonBustillo:2015lrt, Usman:2018imj, Vitale:2018wlg}. This is because different modes have different dependencies on the binary's orientation and thus help break the degeneracy between certain combinations of the binary's parameters. Secondly, if the source is massive, their frequency at merger and ringdown can be near the optimal sensitivity of the current detectors. Such an occurrence can allow us to test general relativity with higher-order modes of ringdown signals~\citep{Carullo:2019flw, Isi:2019aib, Bustillo:2020buq}. Thirdly, it also allows us to study important phenomena of astrophysical relevance, such as recoil kicks that play a decisive role in the black-hole growth process~\citep{Varma:2020nbm, CalderonBustillo:2018zuq}. These make detecting and parametrizing \GW higher-order modes necessary. 

Current model-based \GW searches use dominant harmonic of quasi-circular binary as templates that have an optimum response for the face-on/off binaries. As a result, these searches are inefficient in detecting nearly edge-on binary systems which carry measurable higher-order modes~\citep{CalderonBustillo:2015lrt, CalderonBustillo:2016rlt, CalderonBustillo:2017skv}. Therefore, recently,~\citet{Chandra:2022ixv} developed and deployed the first-ever matched-filter-based search that uses higher-order mode templates. This helped them improve their detection efficiency, especially towards massive sources that emit a measurable amount of higher harmonics by a factor of almost 4.5 at a \ac{FAR} of 1 in 100 years, compared to the \sw{PyCBC-IMBH} search that uses just the dominant harmonics from quasi-circular binary~\citep{Chandra:2021wbw}. However, like contemporary searches, they are also limited in sensitivity due to data quality issues resulting from the presence of non-Gaussian and/or non-stationary noisy transients or glitches. Despite using a combination of vetoes~\citep{LIGOScientific:2017tza}, signal-noise discriminators~\citep{Allen:2004gu}, and informed/automatic gating of loud glitches~\citep{Usman:2015kfa}, these glitches continue to affect the sensitivity of searches adversely. Therefore, there is a pressing need to model these glitches well to identify and alleviate them better. 

To this end, machine learning techniques, and in particular \ac{CNN}, have been investigated as a method to complement or potentially replace traditional matched-filtering techniques \citep{Cuoco:2020ogp}. This approach takes advantage of CNN's ability to recognize and classify distinct images. It also benefits from its ability to automatically detect and learn important features that might be otherwise difficult to model. We, in this work, develop the \sw{THAMES} algorithm, derived from \sw{T}\textit{ransfer learning for} \sw{H}\textit{igher order modes from} \sw{A}\textit{symmetric} \sw{M}\textit{assive} \sw{E}\textit{dge-on} \sw{S}\textit{ystems}, to perform a coincident analysis of third observing run data from Advanced LIGO detectors~\citep{LIGOScientific:2014pky}. Our method includes generating a time-frequency representation of whitened time series using \acl{CWT} \citep{flandrin_2018}, fine-tuning the image recognition Inception-v3 model~\citep{szegedy2014going, szegedy2015rethinking} to address our classification problem, developing a set of signal-noise discriminators to mitigate the damaging effects of misclassified transient glitches and devising a novel coincident ranking statistic. We restrict ourselves to quasi-circular nearly edge-on mass asymmetric massive \aclp{BBH} with detector frame (redshifted) total masses between $100-500~M_\odot$, the mass ratio between $5-18$ and binary inclination angle between $75^\circ-105^\circ$. In terms of sensitive volume-time product, we find that our search, \sw{THAMES}, outperforms optimized matched-filter-based search for \IMBH binaries, \sw{PyCBC-IMBH}, in overlapping target space at most by a factor of $5.24$ at a \ac{FAR} of 1 in 100 years.  

The remainder of this paper is organized as follows. In Sec.~\ref{sec:background}, we briefly review our target signals and existing modeled search algorithms using either matched-filtering or \acp{CNN}. We elaborate on the morphology of \GW  signals and transient noises in the time-frequency plane in Sec.~\ref{sec:transients}. We then motivate our target binary parameters in Sec.~\ref{sec:binary parameters} and discuss our datasets in Sec.~\ref{sec:datasets}. In Sec.~\ref{sec:analysis_framework}, we present the details of our search algorithm, \texttt{THAMES}. In the following section, we report our search sensitivity when deployed on a representative section of the \acl{O3} data from the Advanced LIGO detectors to which we have added simulated \acl{BBH} waveforms. Next, we investigate the robustness of our model in Sec.~\ref{sec:robustness_investigation}. Finally, we conclude with a summary and outlook in Sec.~\ref{sec:conclusion}.

\section{Background}
\label{sec:background}

\Acl{GW} strain data from interferometric detectors are recorded as a collection of evenly spaced time-series with a spacing of $\Delta t$ seconds. If the sampling frequency is $f_s=1/\Delta t$~Hz and the data analysis time period is $T_a$ seconds, then the detector data consist of $N = f_s \times T_a$ number of samples. We can model this strain data $d(t)$ as a detector's response, $s(t|\boldsymbol{\theta})$, to a \acl{GW} signal that is buried in the wide-sense stationary noise $n(t)$ which is polluted with intermittent non-Gaussian transients or ``glitches'' $g(t)$:
\begin{equation}
    d(t) = s(t|\boldsymbol{\theta}) + n(t) + g(t), \qquad t = 1, \ldots, N.
\end{equation}
The noise term consists of contributions from a myriad of noise sources. It is assumed to obey a zero-mean, wide-sense stationary, Gaussian process with one-sided noise power spectral density $S_n(f)$. The glitch term $g(t)$ is the detector's response to noisy artifacts from either well-known causes or sources yet to be understood. On the other hand, the signal strain is deterministic and depends on parameters $\boldsymbol{\theta}$.  It can be written down in terms of two \GW polarizations $h_{+/\times}$ as
\begin{equation}
    s(t|\boldsymbol{\theta}) = F_+ (\alpha, \delta, \psi)h_+(t|\boldsymbol{\lambda}) + F_\times (\alpha, \delta, \psi) h_\times(t|\boldsymbol{\lambda})~.
\end{equation}
Here, $F_{+/\times}$ denotes the antenna response of the detectors to the two \GW polarizations. They depend on the source's sky-location $(\alpha,\delta)$ and polarization angle $\psi$. In general, $F_{+/\times}$, like the \GW polarizations, depend on the Earth's motion. However, for short-lived \GW transients, we consider them to be constants in time for this work. In the above equation, the vector $\boldsymbol{\lambda}$ parameterizes the polarizations and determines the observed signal morphology, loudness, and evolution. We briefly review our target signals and their dependence on $\boldsymbol{\lambda}$ next. 

\subsection{Gravitational Waves from Massive Quasi-Circular Black Hole Binaries}
\label{sec:Signal}

The primary goal of this work is to develop a deployable modeled search that can efficiently identify signals from massive, mass asymmetric, nearly edge-on quasi-circular binaries with significant higher-order modes contributions that might be present in the \GW detector data. We can conveniently represent signals from such sources using a single complex, dimensionless strain, $\mathpzc{h}=h_+ -ih_\times$ that can be decomposed into a sum of emission modes, $h_{\ell,m}$ with weights ${}^{-2}Y_{\ell,m}$ that depend on the binary's inclination $\iota$ and azimuthal angle $\phi$:
\begin{equation}\label{eq:signal}
    \mathpzc{h}(t|\boldsymbol{\lambda}) = \frac{1}{D_L}\sum_{\ell \leq 2} \sum_{m=-\ell}^{\ell} h_{\ell,m}(t|\boldsymbol{\Xi}){}^{-2}Y_{\ell,m}(\iota,\phi)~.
\end{equation}
The term $D_L$ denotes the luminosity distance to the source. The vector $\boldsymbol{\Xi}$ spans a four-dimensional parameter space characterised by detector frame total mass $M_T(1+z)$ of the binary, mass ratio $q=m_1/m_2$ between the component masses ($m_1$, $m_2$) and the component spins ($\chi_{1z}$, $\chi_{2z}$) aligned with the orbital angular momentum. Together, the component masses and spins determine the waveform characteristics and the signal duration.

During most of the binary's evolution, the quadrupole mode or $\ell = | m | = 2 $ dominates the sum in Eq.~\eqref{eq:signal}~\citep{Blanchet:2013haa}. However, during the late inspiral, merger, and ringdown stages, the contributions of the higher-order radiation multipoles become important. This is especially true if the binary is massive and mass asymmetric. Further, since the emission modes enter the polarization through the spin-2 weighted spherical harmonics, the relative amplitude also depends on the binary's orbital geometry, particularly on the binary's inclination as:
\begin{equation}
    {}^{-2}Y_{\ell,m}(\iota,\phi) = A_{\ell,m}(\iota)e^{im \phi}~.
\end{equation}

\begin{figure} 
    \includegraphics[width=\columnwidth]{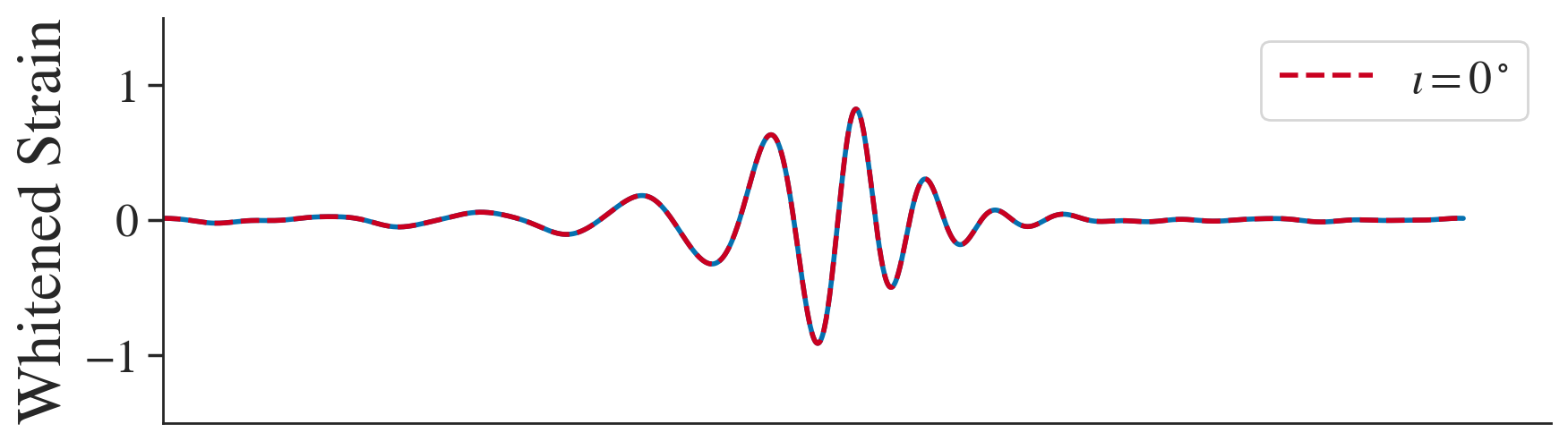}
    \includegraphics[width=\columnwidth]{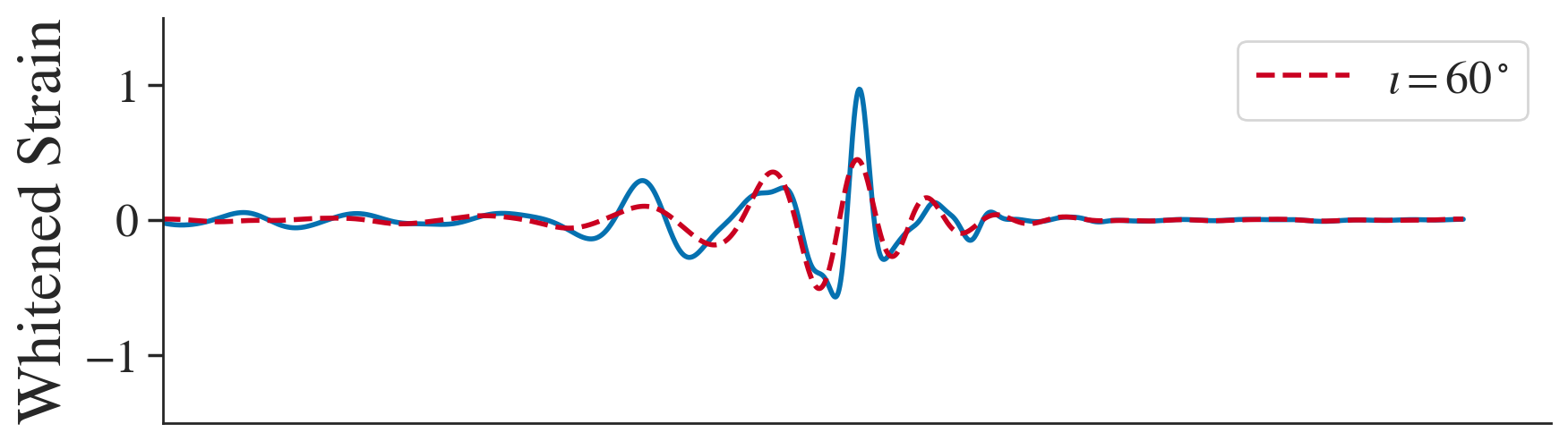}
    \includegraphics[width=\columnwidth]{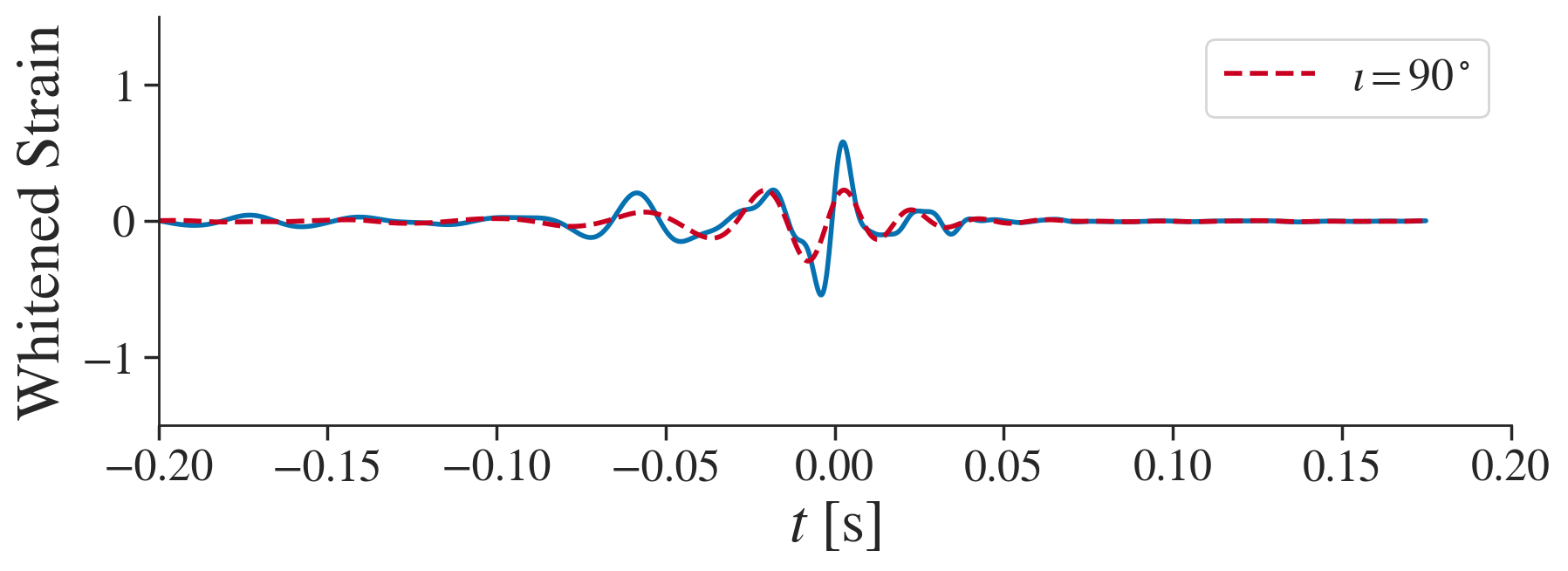}
    \caption{Comparison between the whitened waveform containing higher-order modes (blue) and the whitened dominant harmonic waveform (maroon) from a binary at different orbital inclinations. Except for the inclination and mass ratio (which we have set to $q=10$), the signal parameters are consistent with the maximum likelihood values of GW190521~\citep{LIGOScientific:2020iuh}. The top panel corresponds to the case when the binary is face-on, while the bottom two panels correspond to cases when the binary has a higher inclination. The figure shows that the dominant harmonic appropriately mimics the complete waveform only when the source is face-on.}
    \label{fig:strain}
\end{figure}

To illustrate the latter effect qualitatively, we, in Fig.~\ref{fig:strain}, plot the whitened time-series strain with the Fourier domain representation as $\tilde{d}_w(f) = \tilde{d}(f)/\sqrt{S_n(f)}$ for a simulated quasi-circular binary black hole signal at three different inclinations. The signals are simulated using the multipolar waveform model \texttt{SEOBNRv4HM}~\citep{Cotesta:2018fcv}.

The figure shows that when the binary is optimally oriented, i.e., $\iota=0^\circ$ (face-on), the \GW strain is dominated by the dominant $(2,\pm 2)$ harmonic. However, the relative contributions from the higher-order modes increase for more generic orientations, which is evident in the bottom two panels of Fig.~\ref{fig:strain}. Therefore,  the amplitudes of the whitened strain due to waveforms with the higher-order modes are larger than those omitting it. The plots also show that the dominant harmonic cannot always adequately represent the full gravitational waveform, especially when the source is at a high inclination.

\subsection{Review of Existing Modeled Search Methodology}
\label{sec:modelled searches}

The classic problem of signal detection involves identifying the presence of a weak signal in noisy data. In the context of \GW data analysis, it has traditionally been solved using standard techniques of time series analysis, such as linear filtering and spectral analysis~\citep{Sathyaprakash:1991mt}. Alternative approaches such as those involving coherent analysis of time-frequency representation of the detector data have also been successful, especially when detecting short-duration \GW transients~\citep{Klimenko:2015ypf}. The latter approaches are versatile and are better at detecting transient signals from a wide variety of sources, even those that are not sufficiently modeled. However, the presence of noisy glitches in the detector data limits both approaches' sensitivity. So far, there is no optimal way to address this problem, especially when searching for massive black-hole binaries.

Therefore, more recently, deep-learning-based detection techniques are being explored as alternatives because of their lower computational cost, lower latency, and better promises at dealing with non-stationary instrumental artifacts. Like the linear filtering technique, they are also dependent on signal models. The main advantage of this approach is that a deep-learning-based detection algorithm learns the underlying features. This makes it better at detecting the various glitch morphologies, which are otherwise difficult to model, hence better classifying the data into signals and glitches.

\subsubsection{Matched-Filter-based Searches} 

When the detector noise is Gaussian, and the amplitude and phase evolution of the target signal is well known, matched filtering provides the optimal solution to the signal detection problem. It involves cross-correlating the calibrated detector output with a template waveform, representing our best guess of the expected signal. Assuming that the noise is wide-sense stationary, the cross-correlation operation involves calculating the noise-weighted inner product between whitened detector data $d_w(t)$ and whitened template $s_w(t|\boldsymbol{\theta})$ as follows:
\begin{equation}
    \langle d | s \rangle =  4\mathbb{R}e\int^{f_\mathrm{max}}_{f_\mathrm{min}}df ~\tilde{d}_w^\ast(f) \tilde{s}_w(f|\boldsymbol{\theta})
\end{equation}
where $f_\mathrm{min}$ and $f_\mathrm{max}$ are the lower and upper frequency cut-offs, respectively and $\ast$ denotes the complex conjugation.  By construction, this inner product operation is designed to suppress contributions coming from frequencies where the noise is large. We use the output of this noise-weighted inner product to define the matched-filter \acf{SNR} as:
\begin{equation}
    \rho = \langle d| \hat{s} \rangle
\end{equation}
where $\hat{s}=s/\sqrt{\langle s | s \rangle}$. The maximum possible value of $\rho$ is the optimal \ac{SNR} $\rho_\mathrm{opt}=\sqrt{\langle s | s \rangle}$, which gives a measure of the loudness of the signal. However, the signal parameters, in general, are not known a priori, requiring us to use a \textit{bank} of template waveforms. As a result, we often use a sub-optimal template $h(t|\boldsymbol{\theta})$ that may not perfectly match the signal in the data. Therefore, the detected \ac{SNR}, $\rho=\langle d| \hat{h} \rangle $ is less than $\rho_\mathrm{opt}$. 

The signal waveform's complexity determines the template bank's size and the subsequent computational cost. Therefore, current matched-filter searches, like \sw{PyCBC-Broad}~\citep{Usman:2015kfa} and \sw{PyCBC-IMBH}~\citep{Chandra:2021wbw}, make simplifying assumptions about the morphology of the expected signal. The former tries to find quasi-circular compact binary merger signals over a broad parameter space ranging from binary neutron stars to low-mass range \IMBH binaries. The latter is optimized to specifically target only low mass range quasi-circular \IMBH binaries with $M_T(1+z) \in(100,600)~M_\odot$ and $q\in(1,10)$. Both these searches assume that the incoming signal can be adequately modeled by the dominant quadrupole harmonic from a quasi-circular black hole binary. 
This allows them to express the observed signal in terms of an overall amplitude scaling, a time shift, and an overall phase shift, which we can easily maximize over. Since these overall factors are dependent on the sky-location angles, binary orientation angles, polarization angle, and distance, the searches only need to match-filter with an amplitude-phase maximized \ac{SNR} statistic and waveforms that are parameterized entirely by $(m_1,m_2, \chi_{1z},\chi_{2z})$. However, making such an assumption creates an ``observational bias'' against massive, mass asymmetric, nearly edge-on sources, which tend to emit \aclp{GW} with detectable higher-harmonics and are intrinsically less luminous. So, newly proposed matched-filter searches, like \sw{PyCBC-HM} search~\citep{Chandra:2022ixv}, use a generic detection statistic that allows the usage of template waveforms with higher harmonics that additionally depends on binary's orientation parameters $(\iota, \phi)$. In particular, like \sw{PyCBC-IMBH} search, it targets quasi-circular binaries with $M_T(1+z)\in(100,500)~M_\odot$ and $q\in(1,10)$ that are highly inclined with respect to the observer --- $\iota\in(75^\circ,105^\circ)$.

A search claims a trigger detection if any of the matched-filter outputs from the template bank is above a predetermined threshold. In the presence of wide-sense stationary noise, a coincident observation of high \ac{SNR} trigger across the detector network suffices to claim a confident detection. However, the detector data is plagued with glitches whose morphology resembles short-duration gravitational wave transient such as \aclp{GW} from \IMBH  binary merger. They are thus responsible for adversely affecting the search sensitivity. Therefore, current matched filter searches use a combination of vetoes~\citep{LIGOScientific:2017tza}, glitch robust search statistics~\citep{Allen:2004gu, Babak:2012zx, Nitz:2017svb, Nitz:2017lco}, gating/autogating~\citep{Usman:2015kfa} and glitch subtraction~\citep{Cornish:2020dwh} to mitigate their effects. Additionally, most modeled gravitational wave search routines demand that if a trigger is due to astrophysical origin, it must be seen across the detector network with the same best-matched template within appropriate time delays. This further reduces the number of spurious events due to noise, as coincident observation of glitches is highly unlikely.

The surviving coincident triggers are assigned a rank $\mathcal{R}$ based on their matched-filter \acp{SNR} and their response to multiple signal-glitch discriminator tests. Following the frequentist approach,  a \textit{degree of confidence} is assigned to each coincident trigger. We do so empirically using the ``method of time-slides'', which involves introducing artificial time delays between the detector data and then looking for chance coincidences between a pair of triggers to create an empirical noise statistic. The resulting set of simulated coincident triggers are called as \textit{background triggers}, and we use their rank distribution $\{\mathcal{R}'\}$ to estimate the \ac{FAR} of an event~\citep{Babak:2012zx}:
\begin{equation}\label{eq:far}
    \mathrm{FAR} = \frac{n_b(\mathcal{R}' \geq \mathcal{R})+1}{T_b}~.
\end{equation}
Here $n_b$ is the number of background triggers with rank $\mathcal{R}'$ which is either greater than or equal to the rank $\mathcal{R}$ of the event, and $T_b$ is the amount of background time generated by the time-shift analysis. A \ac{FAR} of one per $T_b$ years implies the frequency of a noise event coincident between two detectors is as high as one in $T_b$ years. It also indicates that the statistical significance of the event is limited by the background statistics generated for the estimation process.

\subsubsection{Existing Deep-Learning-based Approaches} 
\label{sec:cnn-bkg}

As mentioned earlier, even though current \GW searches are fortified against various glitch classes, these continue to adversely affect the search sensitivity, especially when looking for short-duration \acl{GW} transients with complex morphology. Hence, in this work, we use the classification capability of machine learning techniques to better separate glitches from short-duration transients. Below, we briefly review existing deep-learning-based \GW search methodologies.

Deep learning techniques have surpassed the efficiency of statistical techniques in computer vision~\citep{dl_vs_st_in_cv}. This consistently evolving field has helped make impressive contributions to the dynamic field of data-driven astronomy~\citep{Mahabal_2019, 2022scio.confE..17S, Vibho:2022jkn, Barchi:2019yda, GONZALEZ2018103, Duev_2019}. Machine learning has recently featured in many areas of \GW research, including signal detection, signal characterization, glitch mitigation, etc.~(for an overview, see \citet{Cuoco:2020ogp}).

For the first time, \citet{George:2017vlv} illustrated that deep learning could have comparable detection sensitivity to optimal matched-filtering-based techniques in real LIGO detector data, with added benefits of lower computational expense and resilience to glitches. They developed and presented a method called Deep Filtering, which uses a curriculum learning strategy for stellar-mass black hole binaries signal detection and parameter estimation. \citet{Gabbard:2017lja} demonstrated the similarity in the optimality of deep learning algorithms to that of matched-filter-based techniques in simulated Gaussian noise. They constructed a deep \ac{CNN}, where the architecture developed using hyperparameter tuning alone emulated the sensitivity of matched-filtering.

\citet{Alvares:2020bjg} proposed a Residual Network (Resnet) based classification model for detecting black hole binaries and a Cross-Resnet regression architecture for parameter estimation. They use the data from the three detectors as color channels of the spectrograms for non-spinning, quasi-circular black hole binaries. \citet{Chatterjee:2021lit} developed a \ac{CNN} architecture with bidirectional Long Short-Term Memory (LSTM) denoising autoencoder in deep learning algorithms to extract \acl{GW} from real LIGO detector noise. They affirmed the consistency of extracted waveforms with Numerical Relativity templates and robustness to glitches.

\citet{PhysRevD.100.063015} remarked on the possibility of deep learning algorithms rapidly flagging potential candidates. They developed a fully convolutional architecture for detecting spinning stellar-mass black hole binaries in real LIGO noise. They argued that one could not use \acp{CNN} alone to claim statistically significant \acl{GW} events because the prediction probability only represents the network's confidence in a particular signal. However, given their high inference rate, we can utilize these algorithms to generate low-latency triggers.

In~\citep{Schafer:2021fea, PhysRevD.105.043003}, the authors examined the impact of various training strategies on a detection model's sensitivity in a single and multi-detector framework. They inferred that curriculum learning did not significantly enhance the network's performance. They also established that fixed interval training with low \ac{SNR} signals gives an optimal performance, with detection capability generalization to high \ac{SNR} signals. Further, they also compared the detection efficiency when inferring data from multiple detectors independently and then running a coincidence search \textit{vs} using a single neural network for coincidence search in multi-detector data. They illustrated that the fraction of retained matched-filter sensitivity significantly drops in the latter case, thus supporting the usage of a single-detector neural network to do an independent search on data from multiple detectors.

Furthermore, significant literature exists developing machine learning ideas for detecting \aclp{GW} from supernovae and classifying glitches. \citet{Lopez:2021rci} explored the capability of deep learning in detecting \aclp{GW} from core-collapse supernova explosions with a neural network architecture based on Mini-Inception Resnet modules. The authors trained this network on spectrograms of simulated phenomenological signals from core-collapse supernova with high \acp{SNR}. \citet{George:2018awu} explored the capacities of a fairly distinct regime of deep learning algorithms entitled \textit{Transfer Learning}. They used deep transfer learning for glitch classification, where the features were transferred directly from multiple pre-trained models and fine-tuned based on domain-specific requirements.

\citet{Lopez:2021ikt} developed an alternative approach where the time-frequency-based unmodeled search, cWB~\citep{Klimenko:2015ypf}, is used to project the data into a lower dimensional attribute space. The authors used a supervised Gaussian Mixture Model, developed in the likelihood ratio-based framework, to classify the data into short-duration \acl{GW} transients \textit{vs} noisy glitches. They tested the developed search on the first part of the third observing run, and the authors demonstrated that the algorithm could detect most of the massive binary black hole events in the data with improved sensitivity.

\section{Morphology of Transients in Time-Frequency Maps}
\label{sec:transients}

To study the morphology of transients, we obtain the time-frequency maps by calculating the \acf{CWT}~\citep{flandrin_2018}:
\begin{equation}
    X(\tau, a) = \frac{1}{\sqrt{a}}\bigintsss_{-\infty}^{\infty} dt ~d_w(t)~\mathrm{W}_s^\ast(f_0, \eta)
\end{equation}
of the whitened data $d_w(t)$ in the Gabor-Morlet wavelet basis $\mathrm{W}_s(f_0, \eta)$:
\begin{equation}
    \mathrm{W}_s(f_0, \eta) = \frac{1}{\pi^{1/4}} \Big(e^{2i\pi f_0 \eta} - e^{-(2\pi f_0)^2/2}\Big) e^{-\eta^2/2}.
\end{equation}
Here, $f_0$ is the center frequency of the mother wavelet $\mathrm{W}$ and $\eta = (t - \tau)/a$, where the parameter $\tau$ is the time translation parameter, and $a$ is a scaling parameter that controls compression or dilation state of a wavelet.

In what follows, we review the morphological characteristics of our target signals in the time-frequency domain, namely \aclp{GW} from \IMBH binaries with mass asymmetry and a variety of noisy transients that affect the searches of massive black hole binaries.

\begin{figure}
    \includegraphics[width=\columnwidth]{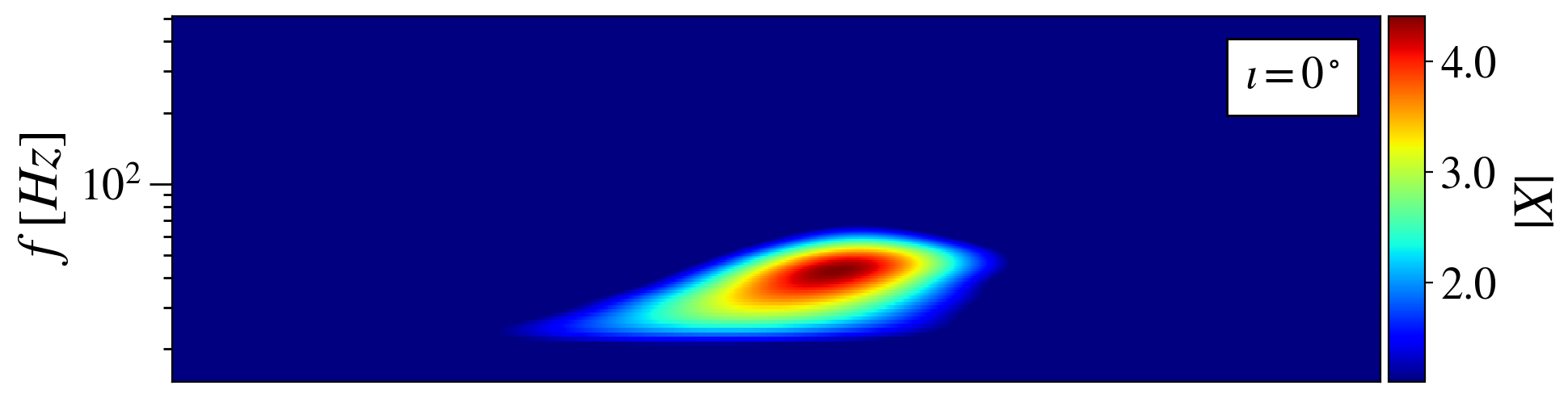}
    \includegraphics[width=\columnwidth]{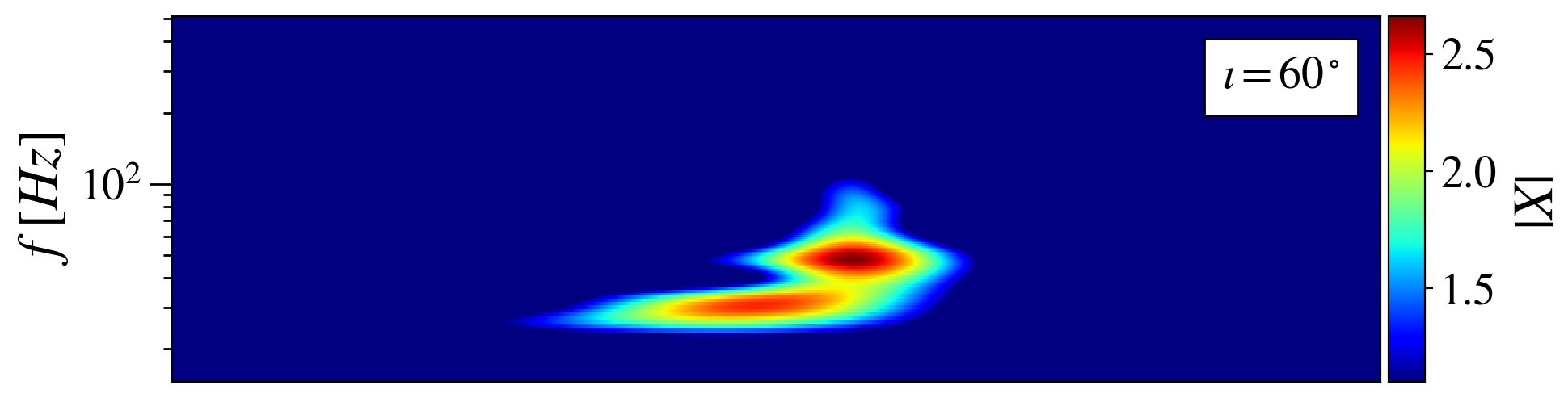}
    \includegraphics[width=\columnwidth]{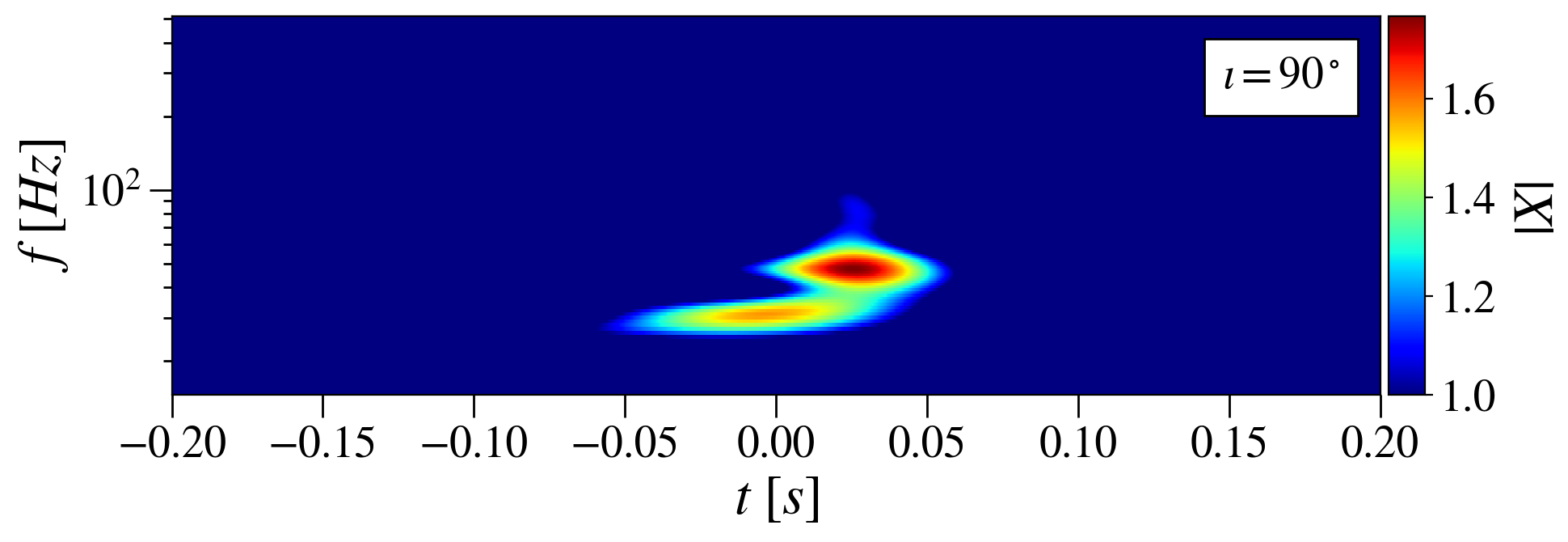}
    \caption{
    The corresponding time-frequency maps of the whitened higher-order mode waveforms in Fig.~\ref{fig:strain}. The plots show a double blob feature when the binary is at higher inclinations. It also shows a decrease in the signal's intrinsic luminosity with an increased orbital inclination, as shown by a decrease in the limits of $|X|$.}
    \label{fig:cwt}
\end{figure}

\subsection{Morphology of Gravitational Waves from Massive Quasi-Circular Binaries}
\label{sec:gw-cwt}

\begin{figure} 
    \includegraphics[width=\columnwidth]{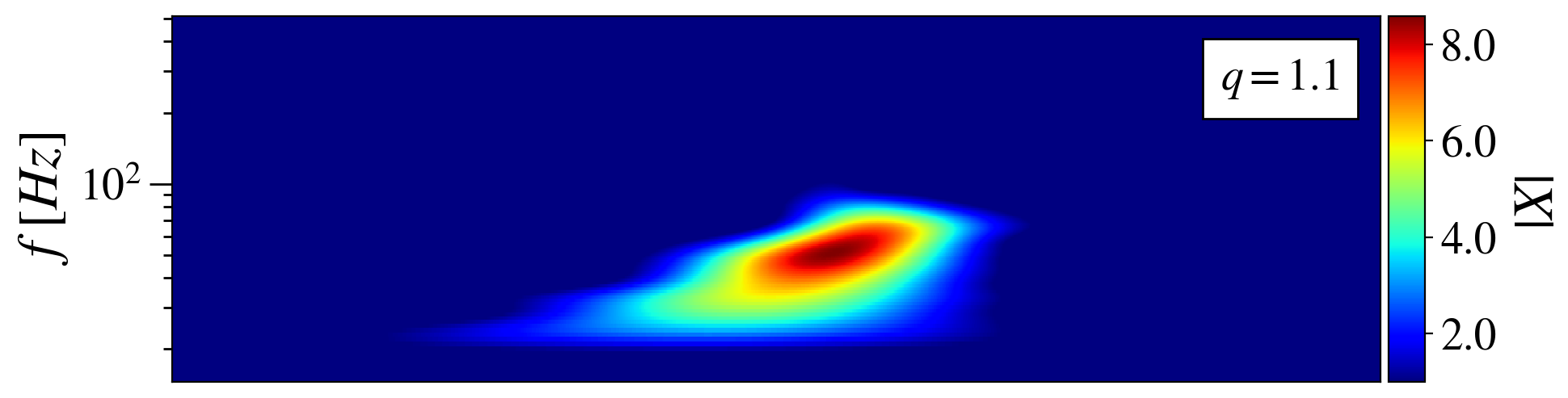}
    \includegraphics[width=\columnwidth]{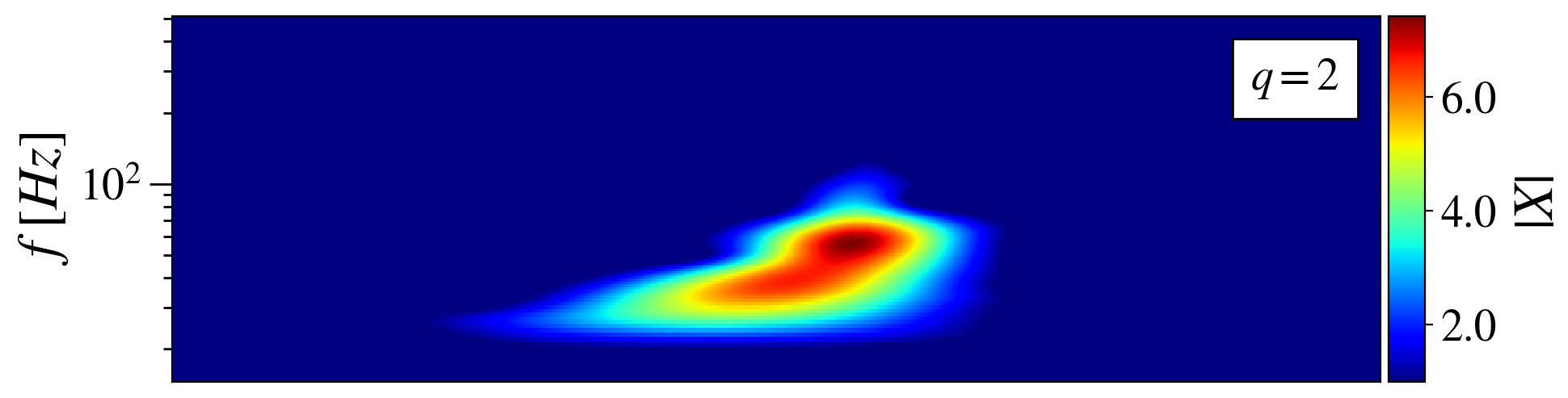}
    \includegraphics[width=\columnwidth]{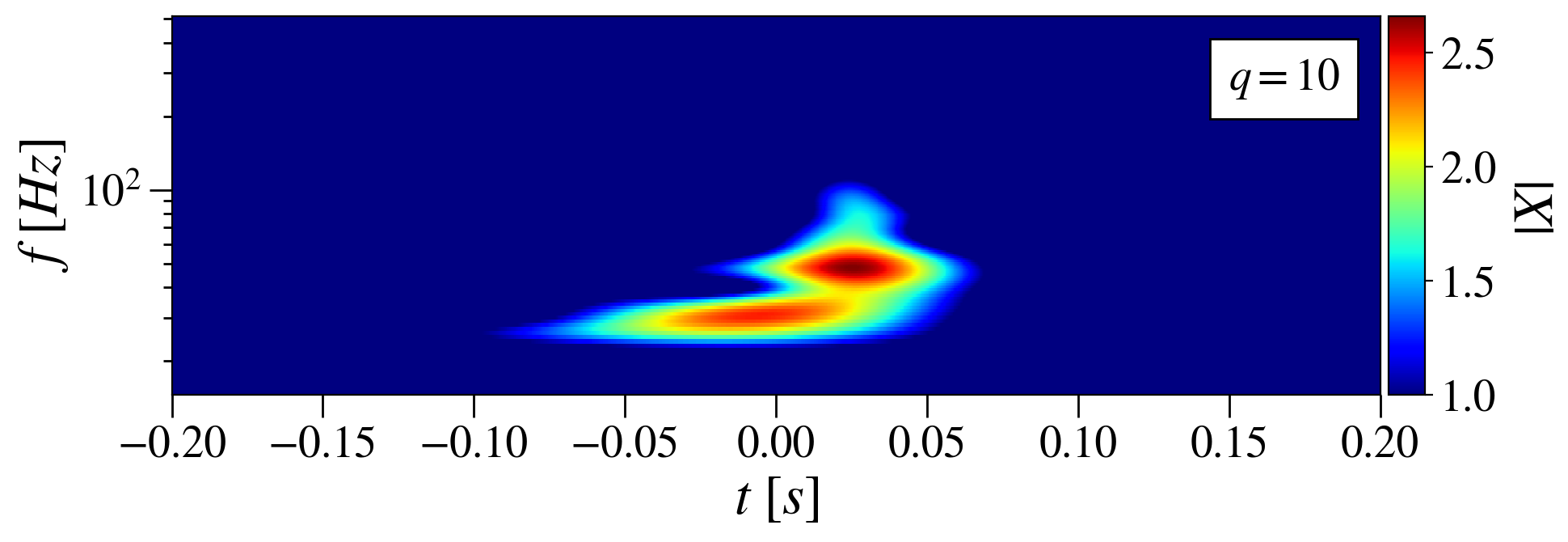}
    \caption{
    The time-frequency representation of the whitened waveforms emitted by binaries with a corresponding mass ratio $q=$ (1.1, 2, 10) at a fixed inclination angle of $\iota=60^\circ$. Like Fig.~\ref{fig:strain}, all the remaining binary parameters are consistent with the maximum likelihood values of GW190521~\citep{LIGOScientific:2020iuh}. The distinctive feature becomes more prominent as the binary's mass ratio increases. Also, the waveform's amplitude decreases with an increase in the mass ratio, as shown by decreasing $|X|$ limits.}
    \label{fig:q-cwt}
\end{figure}

Asymmetric massive black hole binaries emit \aclp{GW} with strong sub-dominant higher-order modes during and after the late-inspiral phase. This leads to a morphology complexity which one can best visualize in the time-frequency plane. Fig.~\ref{fig:cwt} shows the time-frequency representation of the whitened time-domain signals in Fig.~\ref{fig:strain}. The frequencies in the spectrograms are log-scaled between $15-512$~Hz to improve the visibility of low-frequency features. The top panel shows the \ac{CWT} map of the signal when the source is face-on, while the bottom two panels correspond to cases when the orbital inclinations are higher. A clear double blob-like structure is prominent in the last two panels, which is absent when the binary is face-on. As expected, the time-frequency morphology aptly captures multi-modality features that are integrated in either time or frequency domain. Projecting the data in both the time and frequency domain provides the spectral and temporal location of various modes and their superposition quite effectively. Notably, this is also true for varying mass ratio $q=1.1-10$ as shown in Fig.~\ref{fig:q-cwt}. The plots make it evident that the double blob structure becomes more and more pronounced as the binary becomes more and more asymmetric at a fixed orbital inclination of $\iota=60^\circ$. The plots also indicate that the net amplitude of the signal decreases with an increasing mass ratio, as indicated by the decreasing limits of $|X|$ in the color bar.

\subsection{Glitch Morphology}
\label{subsec:glitch-cwt}

\begin{figure}
    \includegraphics[width=\columnwidth]{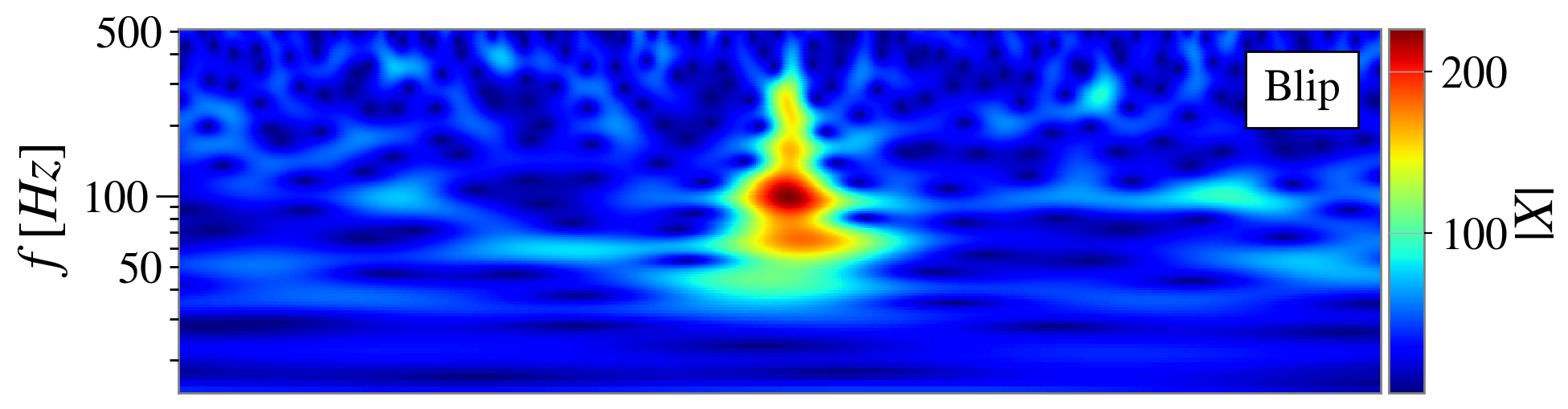}
    \includegraphics[width=\columnwidth]{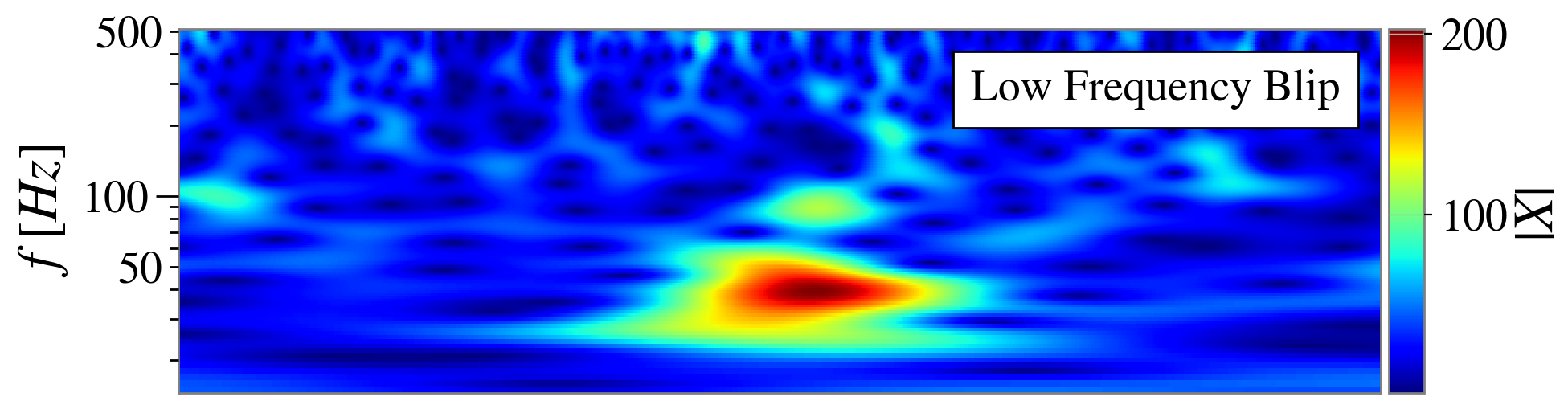}
    \includegraphics[width=\columnwidth]{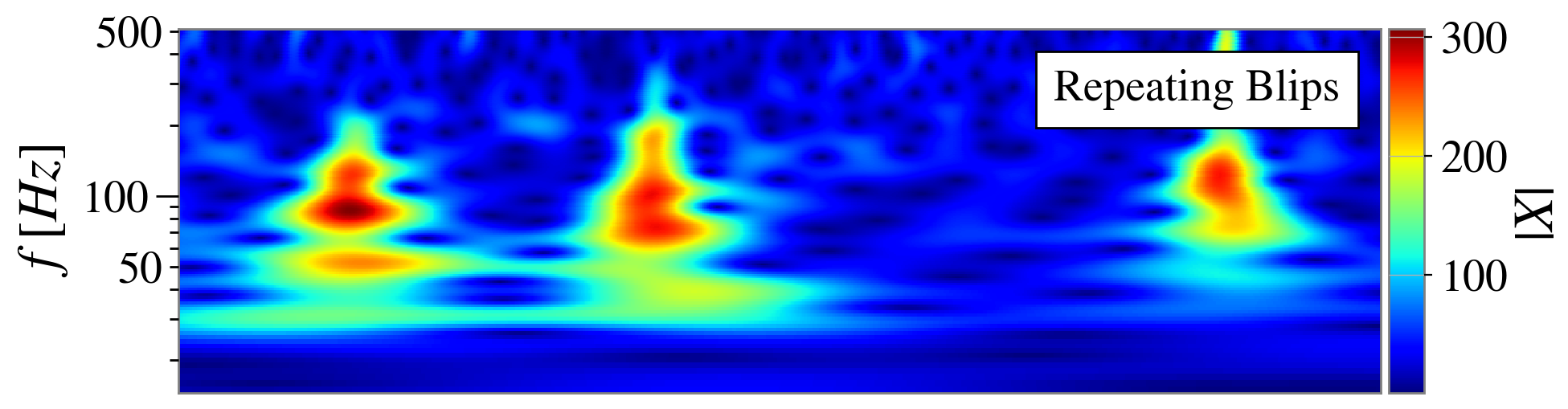}
    \includegraphics[width=\columnwidth]{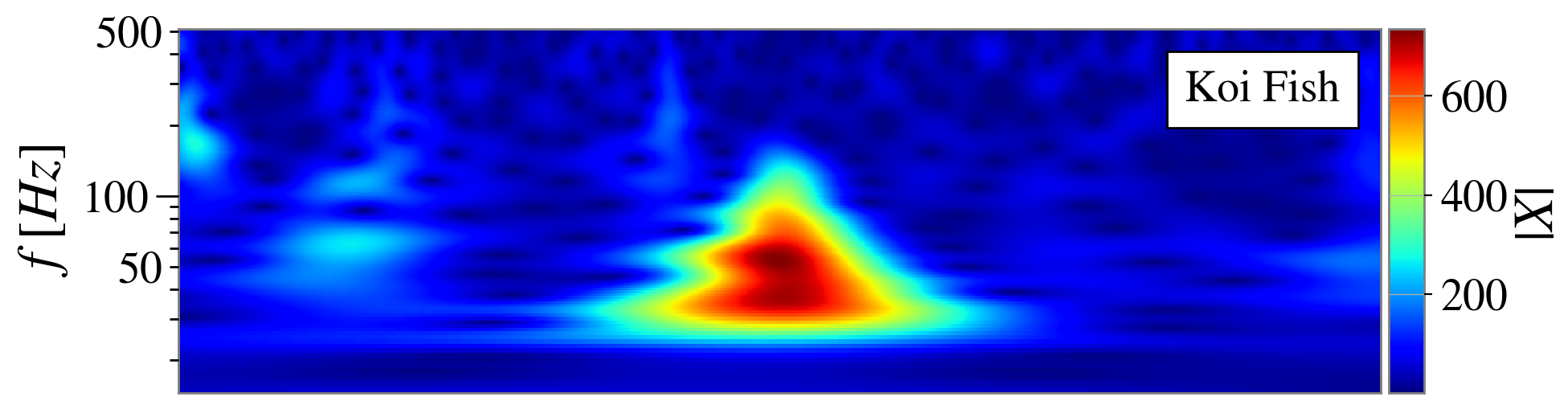}
    \includegraphics[width=\columnwidth]{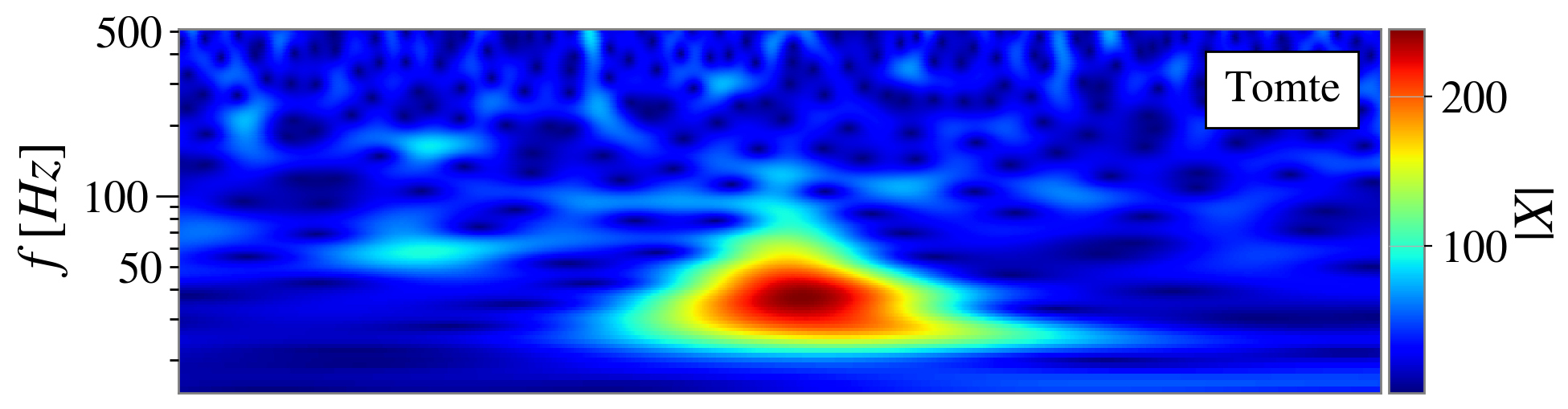}
    \includegraphics[width=\columnwidth]{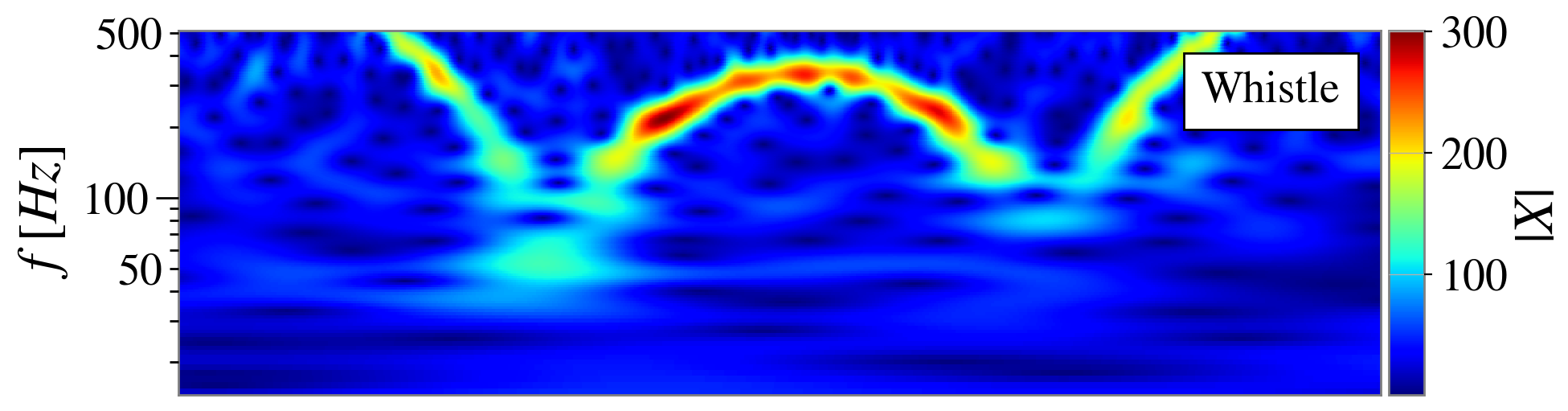}
    \includegraphics[width=\columnwidth]{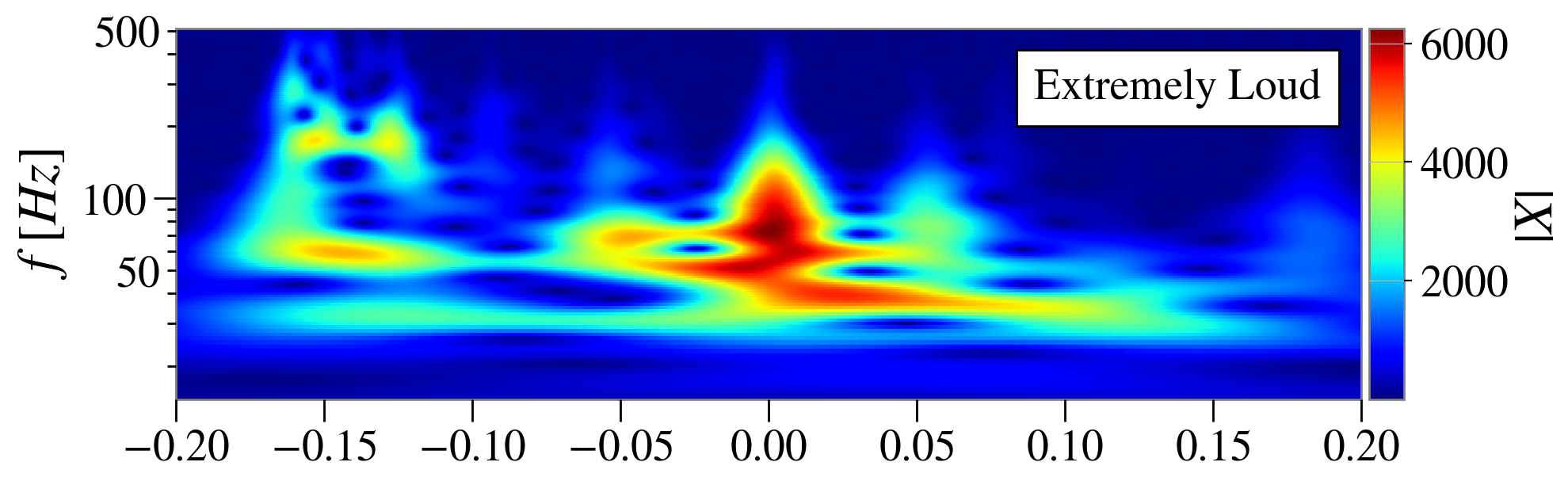}
    \caption{Time-frequency maps of representative samples of frequently occurring short-duration glitches in \acl{GW} detector data.}
\label{training_dataset_appx}
\end{figure}

A variety of noisy transients affect the \acl{GW} transient searches. Some of these glitches are short-lived and reside in the frequency space of massive black hole binary signals, thus affecting both modeled and unmodeled searches. The main consequence is that they raise the number of false alarms and thus subsequently reduce the statistical significance of a putative candidate event. A large number of vetoes are developed in both modeled as well as unmodeled searches encapsulating the glitch characteristics in either frequency domain or time-frequency domain. Here, we focus on a few frequently-occurring glitches in the detector data with a morphology resembling \GW from \IMBH binaries. We aim to develop a deep learning detection technique that can leverage these glitches' morphology and help improve our target signal's detection sensitivity. For this purpose, we have identified seven classes of glitches. We show their representative \ac{CWT} maps in Fig.~\ref{training_dataset_appx}~\footnote{Samples of each glitch classes are a subset of Gravity Spy dataset~\citep{Zevin:2016qwy}.}.

\begin{figure*}
    \includegraphics[width=\columnwidth]{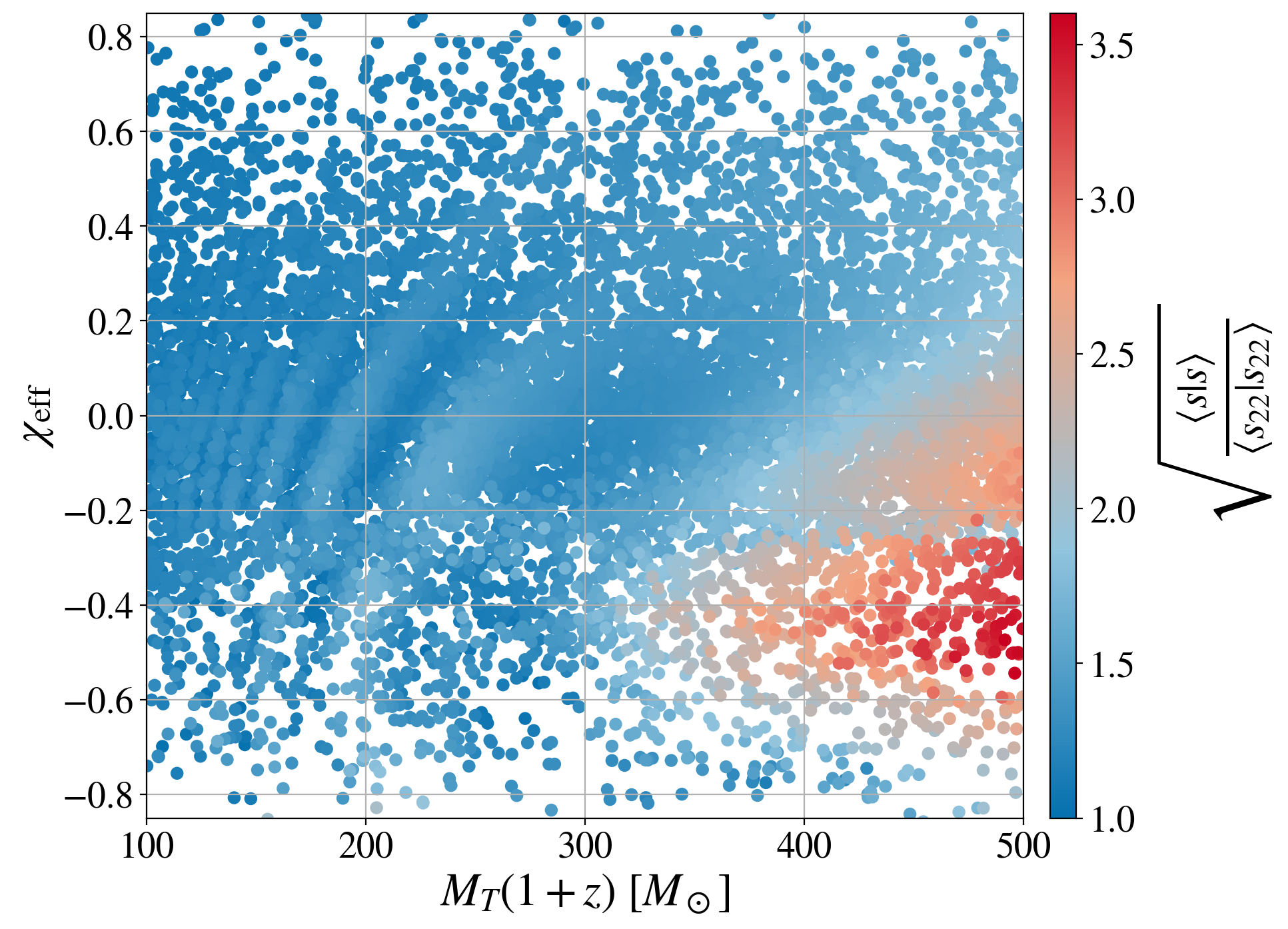}
    \includegraphics[width=\columnwidth]{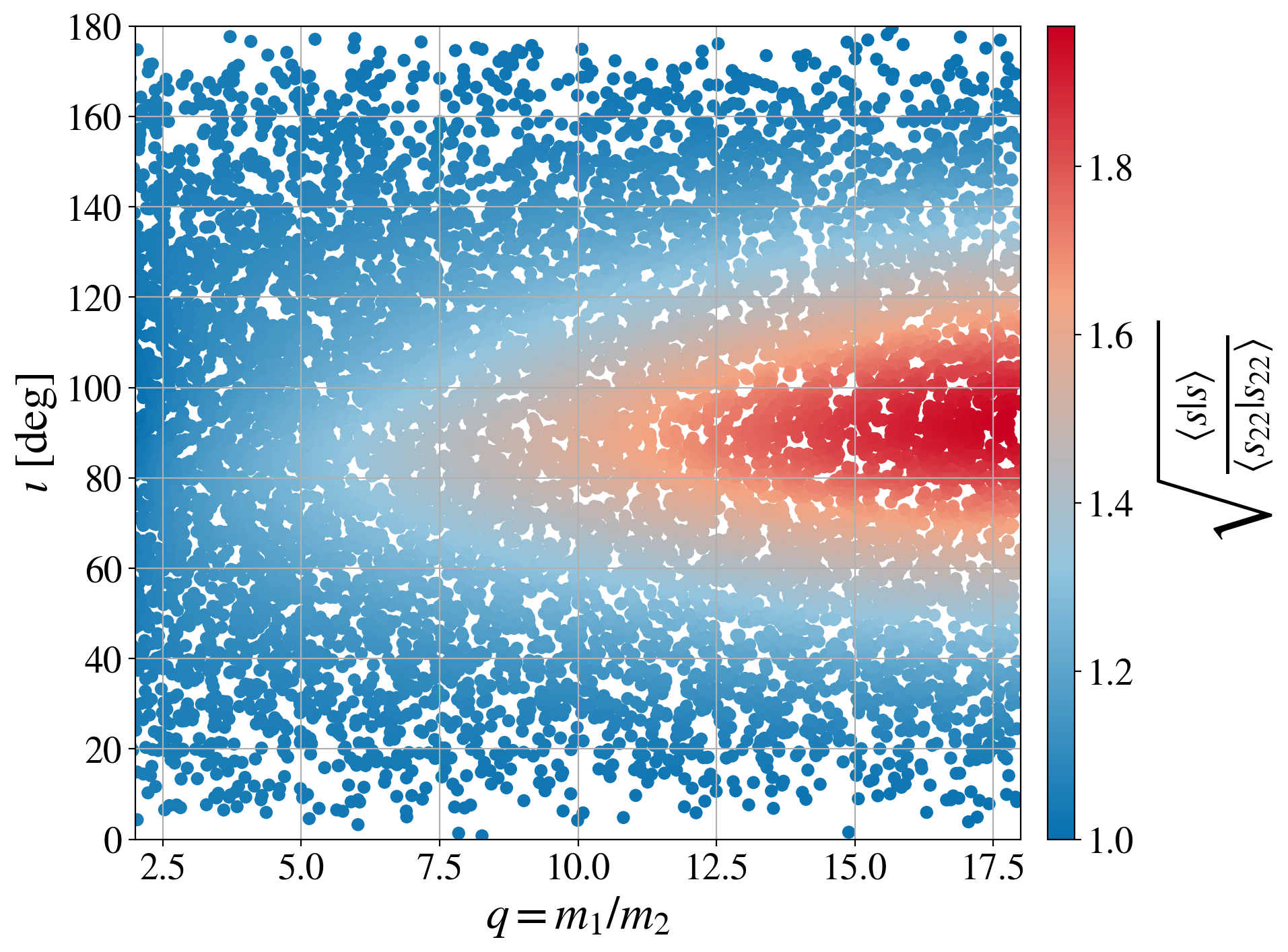}
    \caption{The ratio of the loudness of a signal with higher-order modes to the $(2,\pm 2)$ harmonic as a function of mass ratio and inclination (right panel) and as a function of detector frame total mass and effective aligned spin (left panel) in a detector with sensitivity matching Advanced LIGO-O3 sensitivity. The binaries in the right panel are non-spinning and have $M_T(1+z)=300~M_\odot$ while those in the left panel have mass ratio $q=10$ and have orbital inclination $\iota=60^\circ$. In both cases, the sources are isotropically distributed over the sky and placed at $440$ Mpc, consistent with GW150914. When the system is mass asymmetric and/or highly inclined, contributions from the higher-order modes become important. We also see a similar effect when the mass of the binary increases and/or when the effective aligned spin decreases.}
    \label{fig:optimal-snr}
\end{figure*}

A \textit{blip} glitch is a sine-Gaussian transient with few short-duration loud cycles, tear-drop morphology, and large frequency bandwidth of 60-500~Hz~\citep{Zevin:2016qwy, Cabero:2019orq, Davis:2019mgu}. They are frequently identified by \IMBH binary templates and thus interfere with searches targeting systems with $M_T(1+z) \gtrsim 100~M_\odot$ regime. \textit{Low-frequency blips} have a blob-like appearance and are usually present in the frequency range 30-120~Hz~\citep{Soni:2021cjy}, thus having a near-perfect overlap with peak frequencies of massive binaries. At times, both these blips are repeated with an irregular cadence ($0.25-0.5$~s) and are labeled as \textit{Repeating blips}~\citep{Zevin:2016qwy}. \textit{Koi fish} glitches also closely resemble massive black hole binary waveforms but have relatively higher \acp{SNR} and larger frequency bandwidth of 10-1000~Hz~\citep{Davis:2019mgu}. Another class of frequently occurring glitches with a morphological resemblance to blips are \textit{Tomtes}~\citep{LIGO:2021ppb}. They have comparatively lower \ac{SNR} and relatively different spectral characteristics.

The comparably long-lived but frequently occurring \textit{whistle} glitches look like a ``V'' or ``W''. They are radio frequency signals that beat with the LIGO Voltage Controlled Oscillators~\citep{LIGOScientific:2016gtq, Nuttall:2015dqa}. They occur at frequencies $\gtrsim 100$~Hz~\citep{Davis:2019mgu}. Therefore, they do not regularly hamper \IMBH binary searches. Significant disturbances in the Advanced LIGO detectors cause \textit{extremely loud} glitches, which are often the gated-out~\citep{LIGO:2021ppb}. They often leak out of the gated region due to their loudness, and these leaked features tend to fit \IMBH binary templates, thus harming search sensitivity~\citep{Chandra:2021wbw}. We use the gated-out region surrounding extremely loud glitches.

Despite morphological resemblance to \ac{IMBH} binary signals, each of these glitch classes have distinct time-frequency features. Thus, we develop a \ac{CNN}-based \acl{GW} signal detection algorithm, which can learn these underlying features directly from the time-frequency maps without explicit modeling.

\vspace{-3mm}
\section{Dependence of the loudness of higher harmonics on binary's parameters}
\label{sec:binary parameters}

As noted in the earlier sections, the two important binary parameters that determine the relative power in the higher-order modes are mass ratio and orbital inclination. To investigate these effects across the binary parameter space, we calculate the ratio of the optimal \ac{SNR} between a waveform, including the harmonics $(\ell,|m|)=(2,1),~(2,2),~(3,3),~(4,4),~(5,5)$ and a waveform with just the dominant quadrupolar mode for a simulated population of massive \textit{non-spinning} binary black hole mergers~\citep{Cotesta:2018fcv}. As an example,  we consider binary black hole systems with a fixed detector frame total mass, $M_T(1+z)$ of $268.83~M_\odot$ \footnote{This is consistent with the detector-frame total mass of GW190521.}. We set $f_\mathrm{min}=15~\mathrm{Hz}$, $f_\mathrm{max}=512~\mathrm{Hz}$ and assume that the entire population is observed only at the Advanced Livingston detector at $t_c=1242442967.41$~s and at fixed sky location $(\alpha,\delta)=(0.16,-1.14)$ with phase $\phi=0.004$. We have also assumed that the sources are at a distance $D_L=440~\mathrm{Mpc}$, and the polarization angle $\psi$ is fixed at $2.38$. The mass ratio is chosen to be uniform in $q=(2,~18)$, and the inclination is isotropically sampled in $(0,~\pi)$.

The left panel of Fig.~\ref{fig:optimal-snr} shows the SNR ratio in the detector frame total mass $M_T(1+z)$ and the effective aligned spin parameter~\citep{Santamaria:2010yb}
\begin{equation}
    \chi_\mathrm{eff} = \frac{q\chi_{1z} + \chi_{2z}}{1+q}~
\end{equation}
space. Both these quantities, along with the detector noise power spectrum's shape, govern the duration of the signal in the detector bandwidth. We see that the SNR becomes larger at high detector frame total mass and negative values of the effective spin parameter, implying that the contributions from higher harmonics are significant for such configuration. We can understand this by mapping the two-body dynamics to the motion of a test particle with a reduced mass of the two-body system in an effective Kerr spacetime characterized by $M_T$,  $q$ and $\chi_\mathrm{eff}$. The total energy for a binary with fixed component masses is higher for an aligned system compared to its non-spinning counterpart,  which in turn has higher total energy than its anti-aligned counterpart~\citep{Blanchet:2013haa}. This is directly reflected in the higher amplitudes for aligned-spin systems.

Also, due to the ``orbital hang-up'' effect~\citep{Campanelli:2006uy}, black holes with aligned spin inspiral to much closer separations,  thus resulting in a significantly longer higher \ac{SNR} \GW signal, as compared to a non-spinning counterpart. If we assume that the merger occurs close to the radius at innermost stable circular orbit, then general relativity predicts that the radius at innermost stable circular orbit,  $r_\mathrm{ISCO}$, is $GM_T/c^2$ for a binary with ${\chi_\mathrm{eff} = 1}$ and $9GM_T/c^2$ for ${\chi_\mathrm{eff} = -1}$. The corresponding orbital frequency in the detector frame is given by
\begin{equation}
    \omega_\mathrm{orb}^\mathrm{ISCO} = \frac{c^3}{GM_T(1+z)}\Bigg[ \chi_\mathrm{eff} + \Big(\frac{c^2r_\mathrm{ISCO}}{GM_T}\Big)^{3/2} \Bigg]^{-1}
\end{equation}
which crucially depends on the $\chi_\mathrm{eff}$ and total binary mass. For a fixed total mass,  the \acl{GW} frequency of the dominant $(2,\pm 2)$ mode at innermost-stable circular orbit is $f^\mathrm{ISCO}_{22}=32~\mathrm{kHz}~M_\odot/M_T(1+z)$ for ${\chi_\mathrm{eff} = 1}$ and $f^\mathrm{ISCO}_{22}=2.88~\mathrm{kHz}~M_\odot/M_T(1+z)$ for ${\chi_\mathrm{eff} = -1}$.

During the inspiral, the \acl{GW} frequency of the $(\ell,m)$  emission mode is related to this orbital frequency as $\omega_{\ell,m} \sim m\omega_\mathrm{orb}$, meaning that the frequency of the $(3,\pm 3)$ mode is $3/2$ more than the $(2, \pm 2)$ mode. Therefore, the sub-dominant harmonics can contribute significantly even if the $f^\mathrm{ISCO}_{22}$ mode may lie below the optimal sensitivity of the detector. Thus, the SNR becomes large for signals with higher detector frame total mass and/or lower $\chi_\mathrm{eff}$ (either due to larger $q$ or anti-aligned spins), making the higher-order modes important. Therefore, the observed loudness of the signal depends on its amplitude and duration within the detector bandwidth, along with the detector's noise \acl{PSD} estimate. It also explains how the frequency content of each mode determines the relative strength of a given mode. The ``fringe-shaped'' pattern in the \ac{SNR} ratio distribution is due to superposition between dominant $(2,\pm2)$ mode and sub-dominant $(3,\pm3)$ mode.

The right panel of Fig.~\ref{fig:optimal-snr} shows the variation in the SNR ratio in the $q - \iota$ plane . We find that the contribution from the higher-order harmonics is significant over most of the parameter space. For approximately $\sim37\%$ of the cases,  the full signal's loudness is higher than the dominant harmonic by at least a factor of $\sim1.33$. Furthermore,  these effects increase with an increase in binary's mass ratio and nearly edge-on systems with a maximum SNR ratio value as high as 2.

\begin{figure}
    \includegraphics[width=\columnwidth]{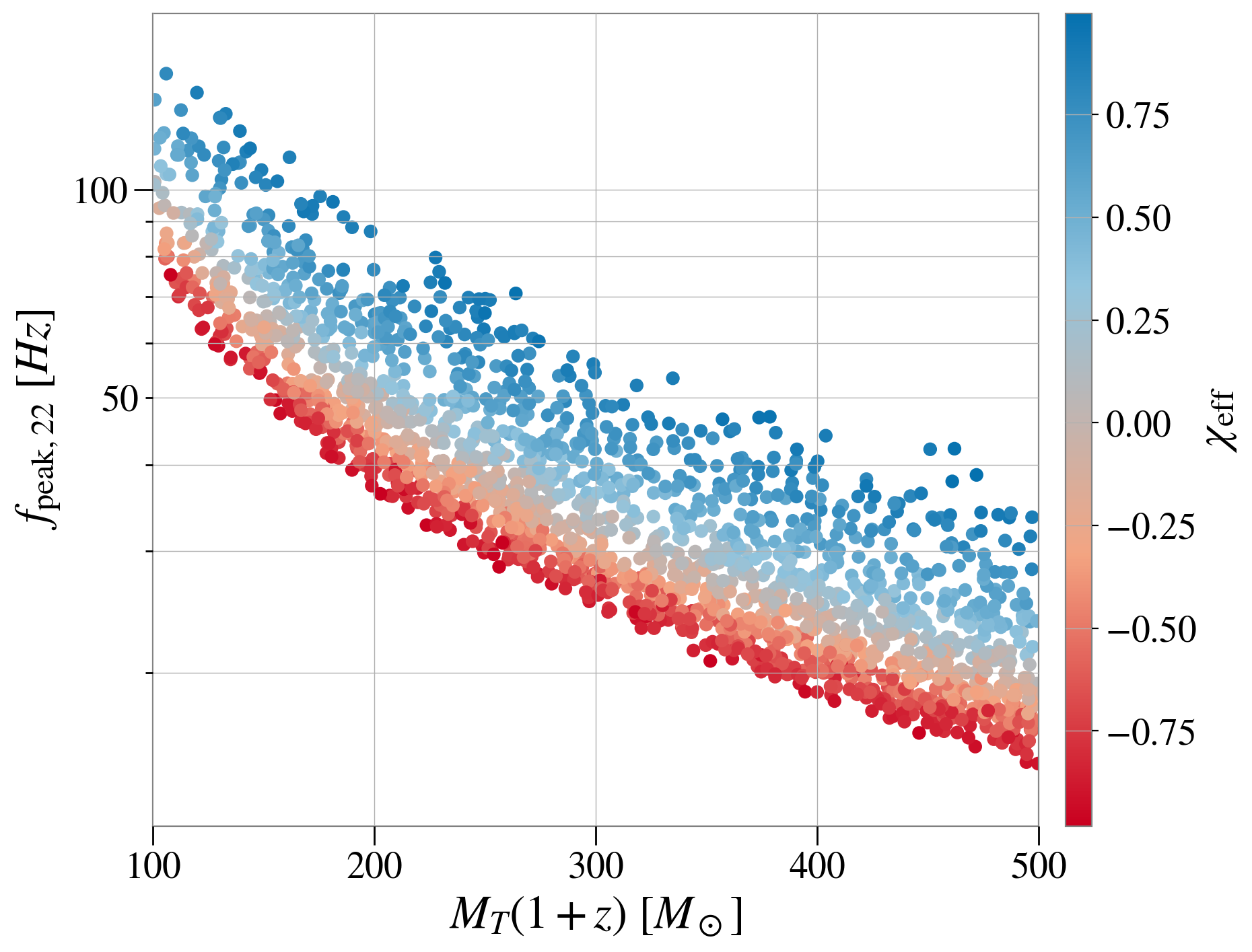}
    \caption{Effect of detector frame total mass and effective aligned spin parameter on maximum merger frequency of the dominant waveform harmonic. We observe a decrease in peak frequency with increasing detector frame total mass and/or a decrease in effective aligned spin value.}
    \label{training_dataset}
\end{figure}

\section{Dataset}
\label{sec:datasets}

Motivated by Fig.~\ref{fig:optimal-snr}, we choose to target \acl{GW} signals from nearly edge-on, mass asymmetric \IMBH binaries. Our training set comprises of \ac{CWT} maps of glitches discussed in Sec~\ref{subsec:glitch-cwt}, Gaussian Noise colored by Advanced LIGO O3 noise power spectrum, and simulated signals that are added to a representative section of O3 Advanced LIGO Livingston data. Table~\ref{table:training-inj-parameters} lists our target parameter space and the waveform used to simulate these signals. The $M_T(1+z)$, $q$, $\chi_{1z,2z}$ and $\rho_\mathrm{opt}$ are uniformly and independently sampled from the intervals listed in the table while the orbital inclination is isotropically sampled between $(75^\circ,105^\circ)$. We obtain a uniform distribution of $\rho_\mathrm{opt}$ by re-scaling the simulated waveform's \ac{SNR} to desired values. This is physically equivalent to moving the source closer or further from the detector. Therefore, our choice in \ac{SNR} distribution is equivalent to a uniform distribution in luminosity distance between $0.1$ Gpc and $2~\mathrm{Gpc}$. Such a uniform distribution over distance, even though ``unphysical'' prevents biasing the trained model against intrinsically weaker mass asymmetric sources located at close distance. Also, we distribute these binaries isotropically over the sky-sphere and uniformly over the polarization angle and azimuth in the interval $(0, 2\pi)$.

\begin{table}[ht]
    \centering
    \begin{tabular}{c c c c}
          Parameter & Symbol & Value  \\
         \hline \hline
         Waveform model& \sw{SEOBNRv4HM} & \\
         Mass ratio &  $q = m_1/m_2$ & (5, 18) \\
         Detector-frame total mass & $M_T(1+z)$ & (100, 500)~$M_\odot $ \\
         Spin z-component  & $\chi_\mathrm{1z}$, $\chi_\mathrm{2z}$ & (-1, 1) \\
         Orbital inclination & $\iota$ & ($75^\circ, 105^\circ$) \\
         Optimal \ac{SNR} & $\rho_{\mathrm{opt}}$ & (5, 40) \\ 
         \hline \hline
         \end{tabular}
         \caption{Target parameter space of \sw{THAMES}.}
    \label{table:training-inj-parameters}
\end{table}

\begin{figure*}[htb]
    \includegraphics[width=18cm]{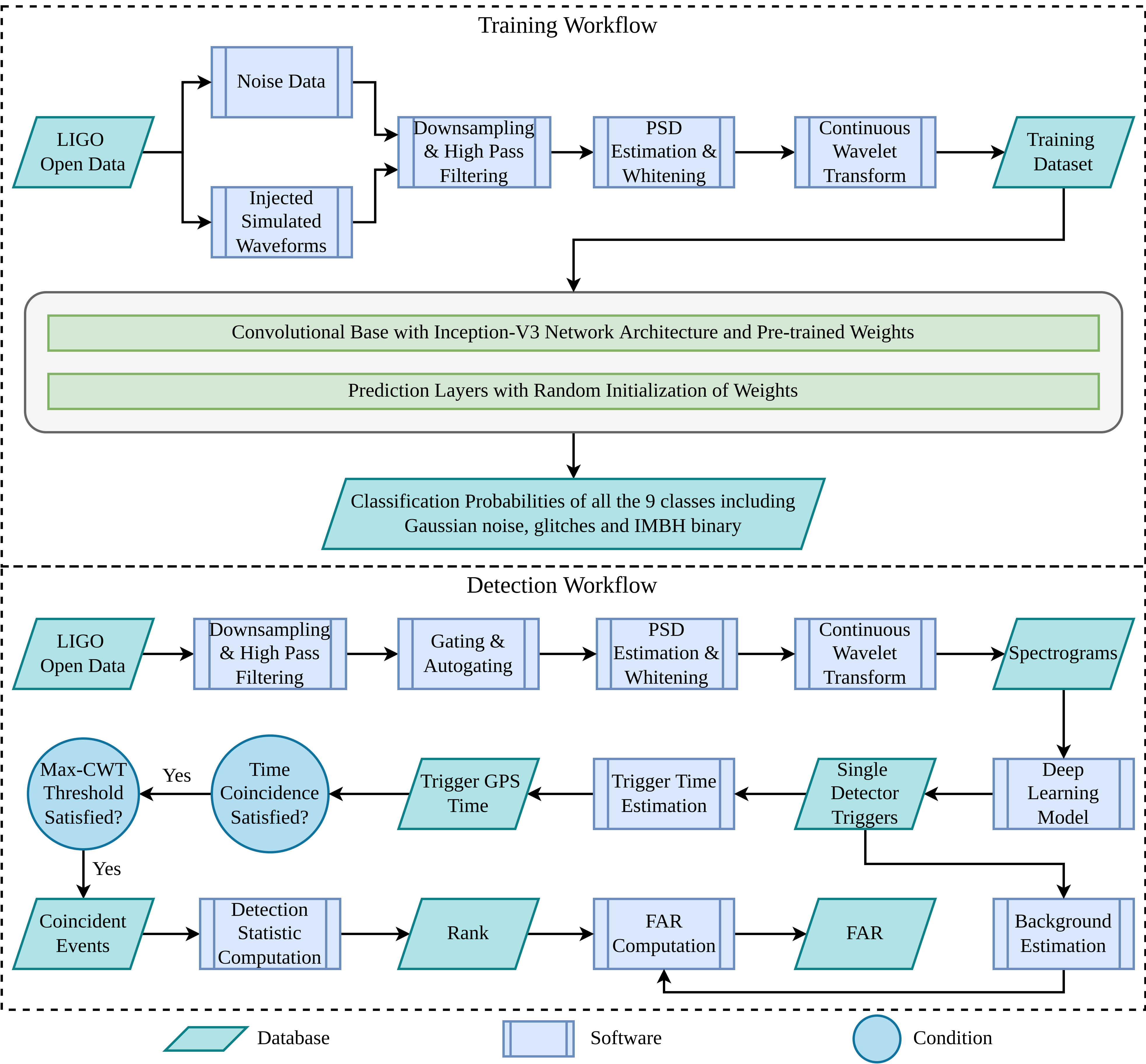}
    \caption{Flowchart of the steps involved in training and deploying \sw{THAMES}.}
    \label{detection_flowchart}
\end{figure*}

To generate the \ac{CWT} maps of our training set, we first downsample 100~s of data around the center time of the transient (time corresponding to the maximum amplitude of the transient) from its original sampling rate of 4096 Hz of the GWOSC data to 2048 Hz. This corrupts both the ends of the data, so we crop off four seconds of data from both edges. We then \textit{gate} out flagged segments of loud non-Gaussianity by applying an inverse Tukey window~\citep{Usman:2015kfa} and then band-pass the gated data between 15 Hz and 1024 Hz to remove low and high-frequency noise. The lower-frequency cut-off is based on the trade-off between a decrease in the merger frequency ($f_{\mathrm{peak}}$) with increasing detector frame total mass (see Fig.~\ref{training_dataset}) and a steep increase in noise power spectral density below 15~Hz for Advanced LIGO detectors~\citep{LIGOScientific:2021tfm}. We whiten the resulting data using its noise power spectrum so that any excess power within the desired frequency range becomes apparent. Subsequently, we create spectrograms with a duration of 2~s around the center time of the transient using \acl{CWT}. We use a balanced training dataset with $\sim1000$ spectrograms in each training class to reduce false alarms during classification due to class imbalance.

\section{Analysis Framework}
\label{sec:analysis_framework}

\begin{figure}
    \includegraphics[width=\columnwidth]{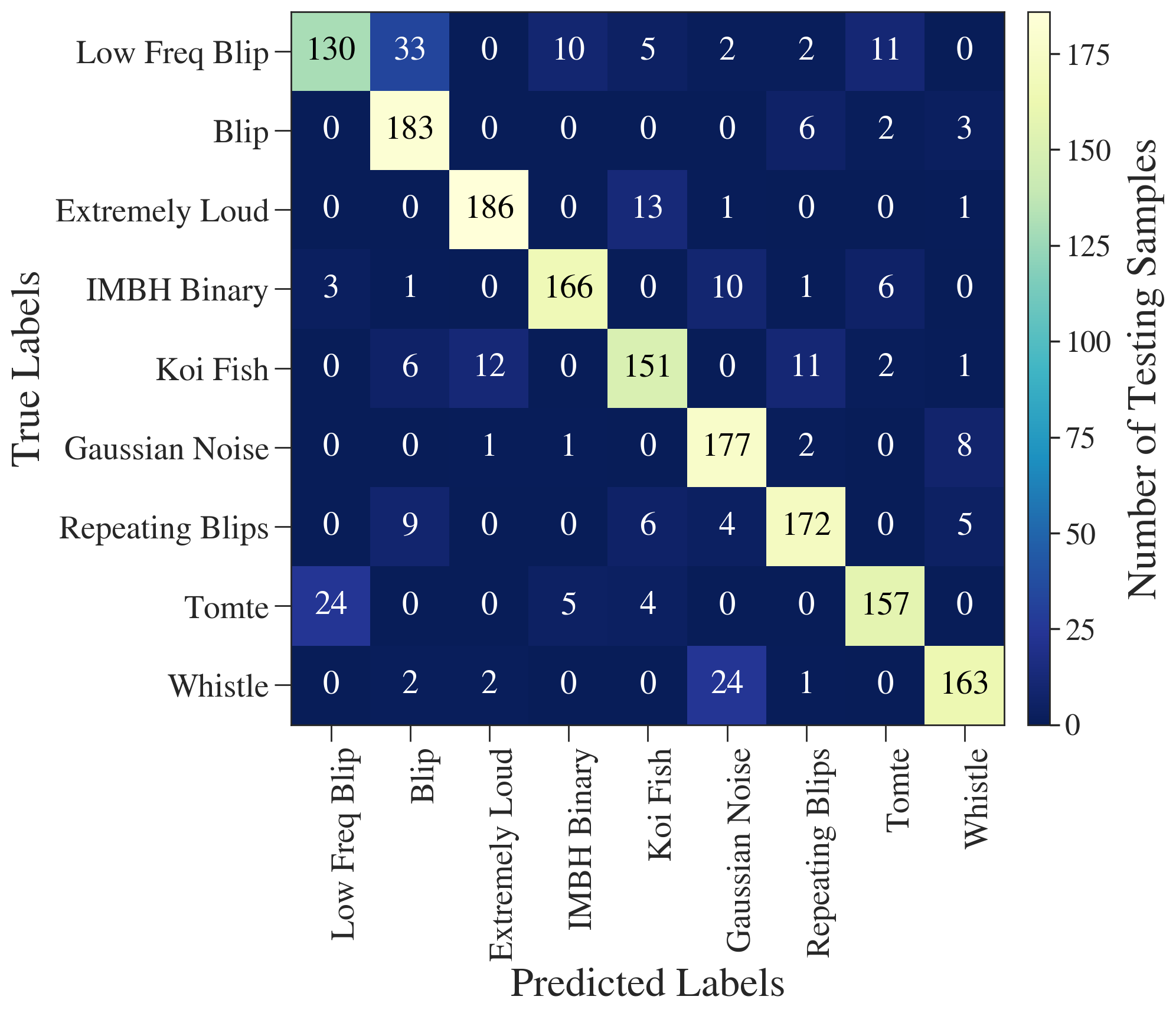}
    \caption{Confusion matrix of \sw{THAMES}'s deep transfer learning model for detection when tested on a held-out dataset of mass asymmetric, nearly edge-on intermediate-mass black hole binaries.}
    \label{confusion_matrix_high_mass_ratio}
\end{figure}

This section elaborates on the deep learning analysis framework we use to classify the data into glitches, Gaussian noise, or an \IMBH binary. Particularly, we use the deep transfer learning approach~\citep{NIPS1992_67e103b0}, where the convolutional base of the network has been optimally trained on a different dataset to extract generic features, such as edges, blobs, and corners. We optimize this network to extract abstract features specific to our dataset for prediction in the final output layers. This approach has several advantages --- it saves training time, gives better baseline performance, and circumvents the need for large training datasets~\citep{https://doi.org/10.48550/arxiv.1911.02685, 8920338}. In Sec.~\ref{subsec:detection_model}, we describe the deep learning model used. Next, we discuss our choice of detection statistic in Sec.~\ref{subsec:ranking_statistic}. We further discuss the event time estimation technique and the time coincidence threshold used in our detection search in Sec.~\ref{subsec:event_time_estimation}. We elaborate on our Max-CWT threshold, which we use to prevent misclassified glitches from hampering search sensitivity in Sec.~\ref{subsec:glitch_veto}. Lastly, we examine the improvement in the noise background distribution after applying these two thresholds in Sec.~\ref{subsec:efficacy_of_vetoes}. The entire workflow of our search, \sw{THAMES} is summarized in Fig.~\ref{detection_flowchart}.

\subsection{Deep Learning Model}
\label{subsec:detection_model}

For the transfer learning model, we use the Inception-v3 network~\citep{szegedy2015rethinking} from the Inception family~\citep{szegedy2014going}. The authors originally trained this network for classification on the \href{https://www.image-net.org/}{ImageNet dataset}, which has 1000 classes of natural images. We initialize the convolutional base weights with the pre-trained weights and randomly initialize the dense prediction layers. We present the data to the network in batches of 16. The data is scaled by a factor of 1/255 and augmented with shear, zoom, and width shift~\citep{https://doi.org/10.48550/arxiv.2204.08610}. We undergo 100 steps in each epoch and aim to maximize the F1-Score~\footnote{F1-Score is the harmonic mean of precision and recall. The precision represents the fraction of correct identifications (TP/(TP+FP)), and recall is the fraction of actual positives that the model correctly identified (TP/(TP+FN)), where TP = true positives, FP = false positives, and FN = false negatives \citep{https://doi.org/10.48550/arxiv.2001.09636}.}. We choose F1-Score as it optimizes both precision and recall simultaneously. We use callback functions during training for early stopping with the patience of 10 \citep{Goodfellow-et-al-2016} and train our model with a 4:1 train-test split.

Our model achieves a training (testing) accuracy of 86.87\% (86.50\%) and an F1-Score of 86.76\% (85.96\%), each of which is robust to train-test split ratios and training datasets with 1\% uncertainty. We ensure that our model does not over-fit the training dataset by using 30\% drop-out in fully convolutional layers for prediction and an early stopping regularization~\citep{Goodfellow-et-al-2016}. Further, the Inception-v3 network architecture is designed to prevent overfitting. We refer the readers to Appendix~\ref{appx:network_architecture} for a concise discussion on Inception-v3 network architecture.

Fig.~\ref{confusion_matrix_high_mass_ratio} shows our model's $9 \times 9$ confusion matrix, which is largely diagonal, indicating that most images are correctly classified. The fraction of correctly classified \IMBH binaries is $\sim88.8$\%. Most of the misclassified \IMBH transients correspond to signals with $\rho_\mathrm{opt} \sim 5$. We put a threshold on the maximum absolute CWT value to mitigate these false alarms, as discussed in Sec.~\ref{subsec:glitch_veto}.

We use \sw{tensorflow}~\citep{https://doi.org/10.48550/arxiv.1603.04467} backend for model development. The training time of this model is $72$ min which we can reduce by using Nvidia GPU, provided by \href{https://colab.research.google.com/}{Google Colab}~\citep{Bisong2019}. The average inference time for a spectrogram is $\sim44$~ms without GPU and parallelization.

\vspace{-1mm}
\subsection{Detection Statistic}
\label{subsec:ranking_statistic}

\begin{figure}
    \includegraphics[width=\columnwidth]{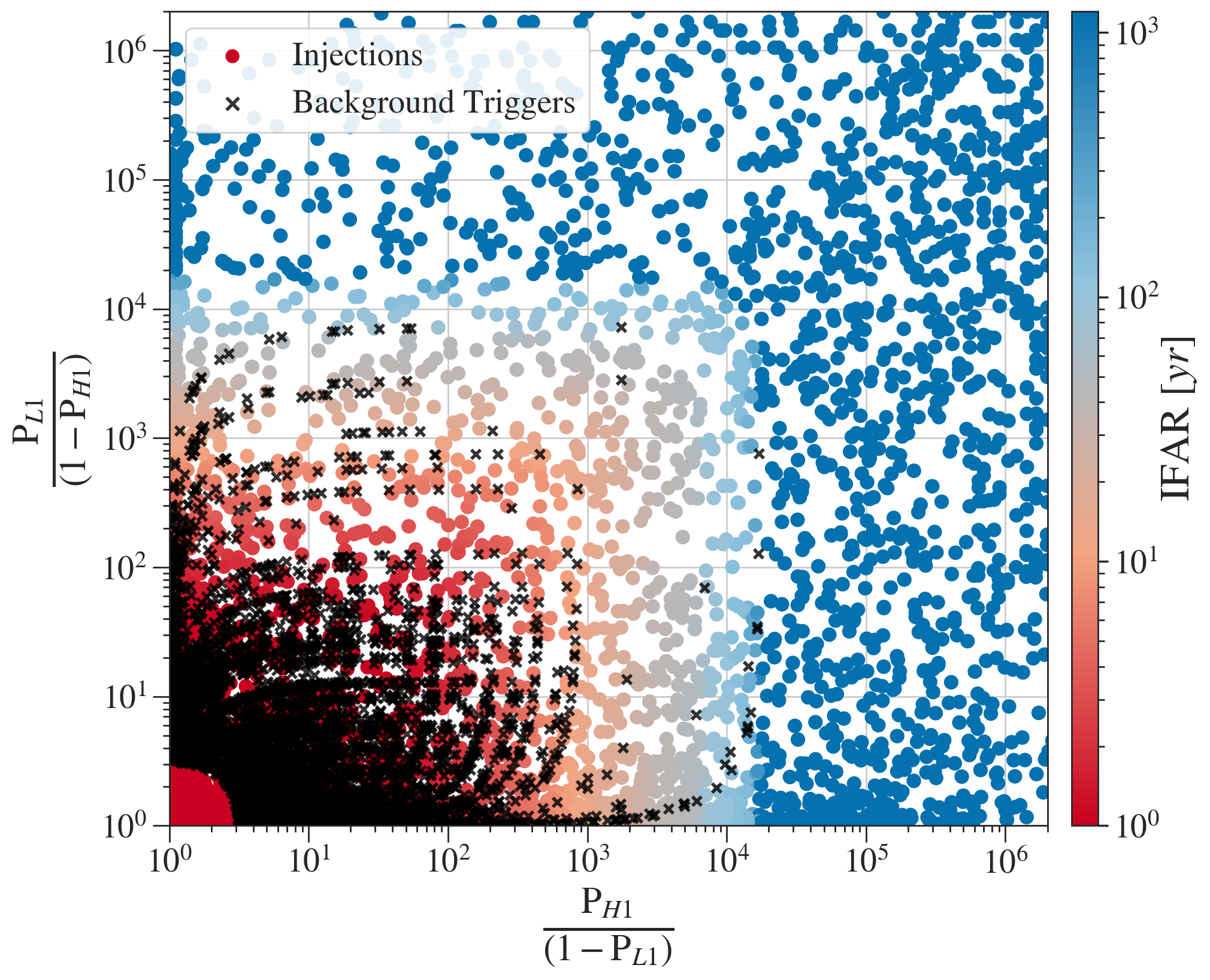}     \caption{IFAR in the space of
    $P_\mathrm{H1}/(1-P_\mathrm{L1})$ and $ P_\mathrm{L1}/(1-P_\mathrm{H1})$ for simulated signals.  A separation between high-rank signals (circles) and loud background triggers (crosses) is apparent.}
\label{ranking_statistic_justification}
\end{figure}

Our deep learning model allots a classification probability to each single detector trigger, $\mathrm{P}(\mathrm{IMBH} \big| \mathrm{H1}) = P_\mathrm{H1}$ in LIGO-Hanford (H1) and $\mathrm{P}(\mathrm{IMBH} \big| \mathrm{L1}) = P_\mathrm{L1}$ in LIGO-Livingston (L1). To rank all the possible candidates, we construct a detection statistic as
\begin{equation}\label{eq:stat-1}
    \hat{\mathcal{R}} = \sqrt{\left(\frac{P_\mathrm{H1}}{1-P_\mathrm{L1}}\right)^2 + \left(\frac{P_\mathrm{L1}}{1-P_\mathrm{H1}}\right)^2}
\end{equation}
By construction, this statistic maximizes the rank when a trigger has high $P_\mathrm{H1}$ and $P_\mathrm{L1}$, and heavily penalizes triggers that are inconsistent with trained signals in either of the two detectors. Fig.~\ref{ranking_statistic_justification} displays the estimated \acf{IFAR} as a function of the detection statistic's constituents, demonstrating the separation between the high-rank signals and loud background triggers. 

The detection statistic efficiently utilizes this feature which is evident in the sensitivity shown in Sec.~\ref{Sec:eta}. The patterns formed by the background triggers in this figure are due to the same single-detector trigger coinciding with multiple triggers in the other detector. For the sake of completion, we compare this construct against the routinely used joint probability statistic in Appendix~\ref{appx:detection_statistic}.

\subsection{Time Coincidence Threshold} 
\label{subsec:event_time_estimation}

\begin{figure}
    \centering
    \includegraphics[width=\columnwidth]{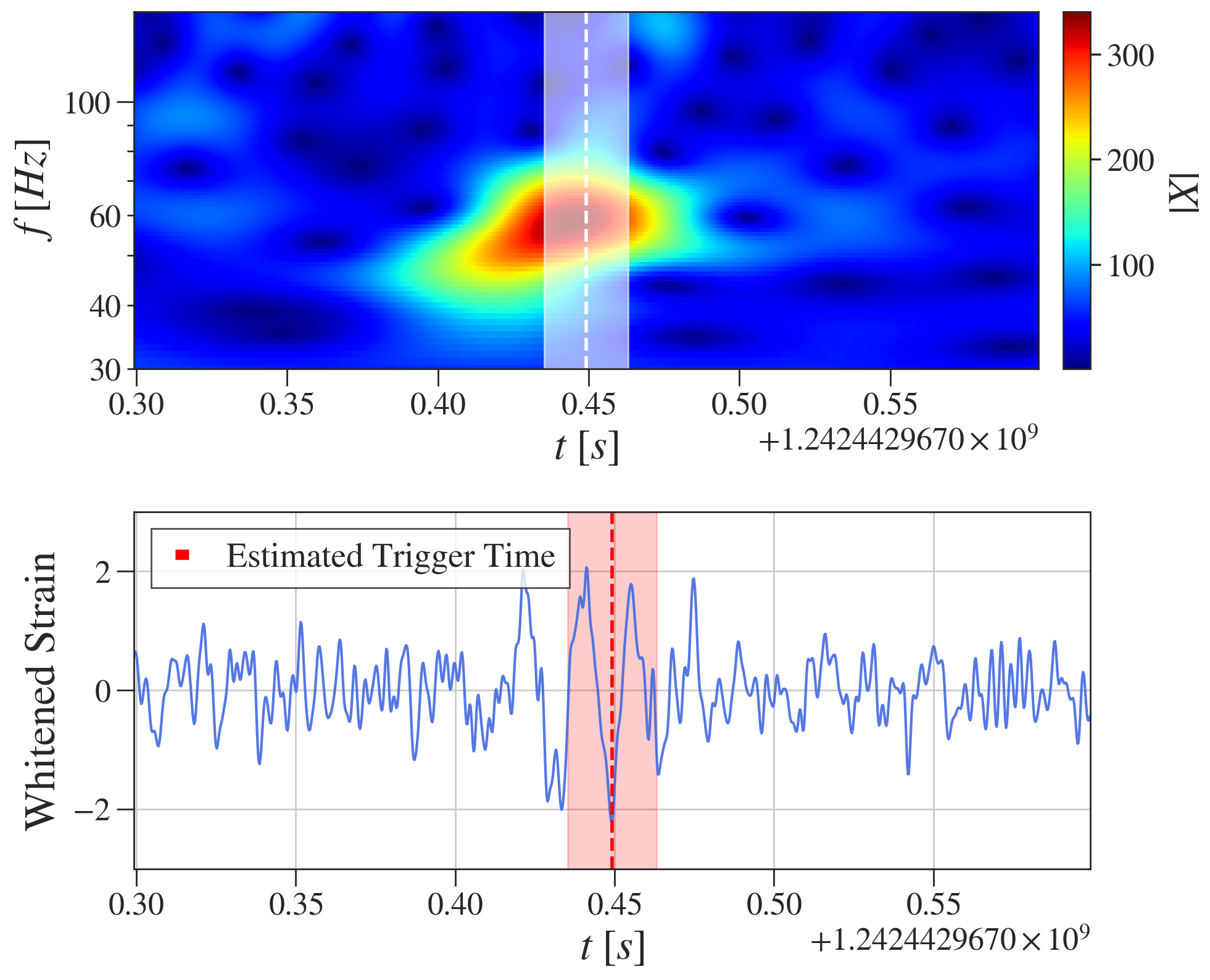}
    \caption{The \ac{CWT} map (top) and the whitened time-series (bottom) around GW190521 in Advanced LIGO Livingston data. The dotted line in both the plots shows the estimated merger time. The shaded region's upper and lower bounds show $\pm13.9$ ms around the estimated merger time,  which is the threshold used by our search for time coincidence.}
    \label{tdiff_veto_explanation}
\end{figure}

To claim a signal detection, we impose the signal coincidence criterion; namely, astrophysical triggers must be seen across the detector network within physically allowed time delays. We assume that the time at the observed peak amplitude of the trigger corresponds to the spectrogram's maximum energy. We identify the same by considering the mean temporal location of the top $\sim20\%$ energetic pixels in the spectrogram~\footnote{We locate these pixels using color image segmentation technique~\citep{color_img_seg} and implement in \sw{THAMES} using the \sw{OpenCV} library.}. Fig.~\ref{tdiff_veto_explanation} illustrates our trigger time estimation technique on GW190521~\citep{LIGOScientific:2020iuh} data in Advanced LIGO Livingston detector. The estimated time delay between the two detectors, $\big| T_\mathrm{H1} - T_\mathrm{L1} \big|$, for this event is 4.2~ms, which is appreciable, considering each pixel has a width of $\sim2.7$~ms.

\begin{figure}
    \centering
    \includegraphics[width=8cm]{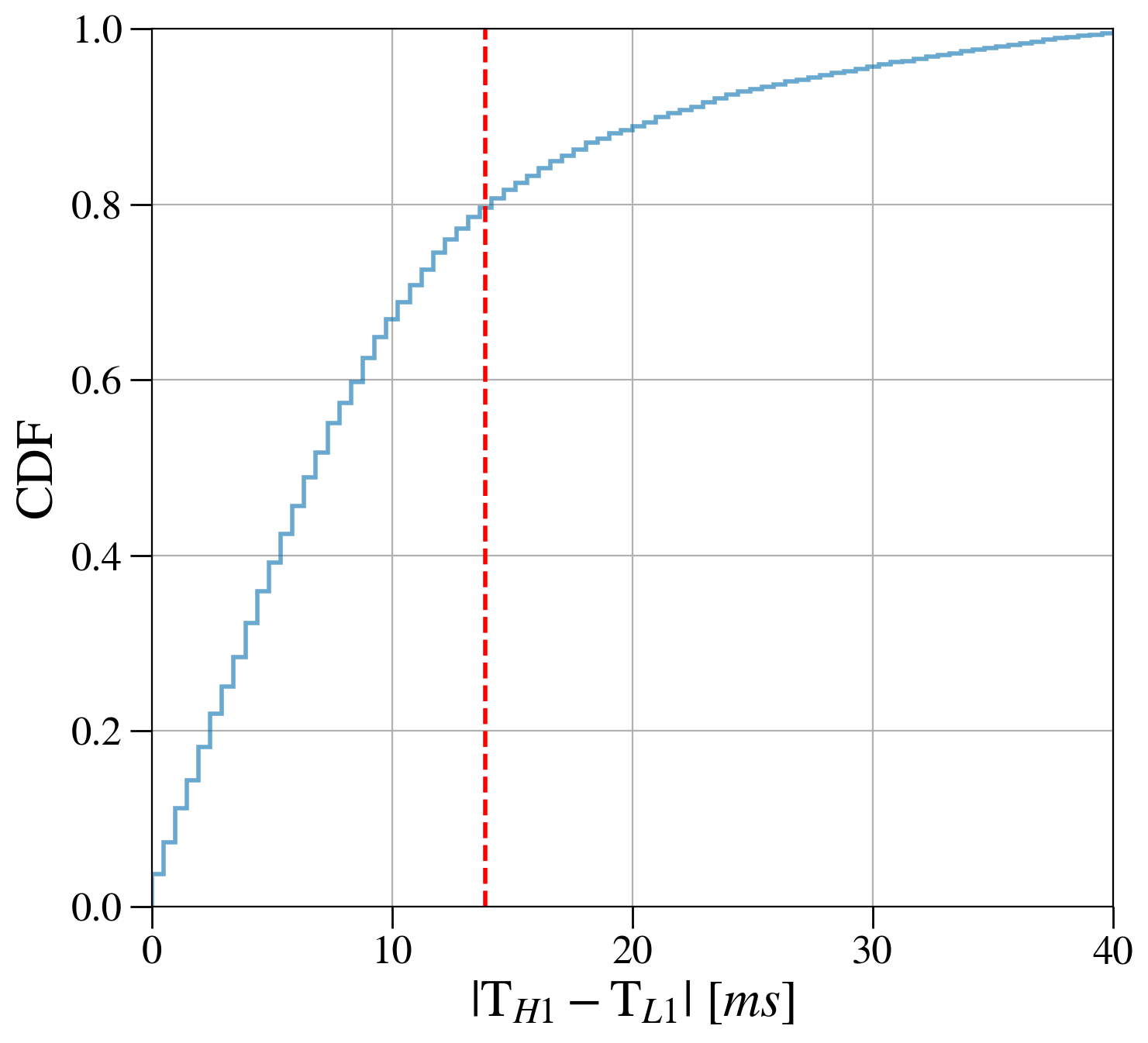}
    \caption{Distribution of estimated time delay, computed for simulated signals added to a representative section of Advanced LIGO data.  We observe that 80\% of the detected simulated signals have $\big|\mathrm{T}_{H1}-\mathrm{T}_{L1}\big| \leq 13.9$~ms. We use this as the threshold for time coincidence.}
    \label{tdiff_veto_justification}
\end{figure}

We compute the time delays by applying the trigger location algorithm to the simulated signals in the testing dataset. Fig.~\ref{tdiff_veto_justification} shows the cumulative distribution of the measured time delays with a binning of $\sim0.5$~ms. While physically, we expect the time delays to fall within $10$~ms duration owing to the maximum light travel time between the two LIGO detectors, it extends to $40$~ms for our dataset. Out of this, $\sim80\%$ of the signals have a measured time delay $\leq 13.9$~ms. Those beyond this bound are largely due to incorrect trigger time estimation of low network \ac{SNR} signals. Their estimated trigger time is compromised owing to poor contrast between the simulated signal and the Gaussian noise in the CWT maps. Thus, for time coincidence, we impose the criterion of $\big| T_\mathrm{H1} - T_\mathrm{L1} \big| \leq 13.9$~ms for detection.

\begin{figure}
    \includegraphics[width=\columnwidth]{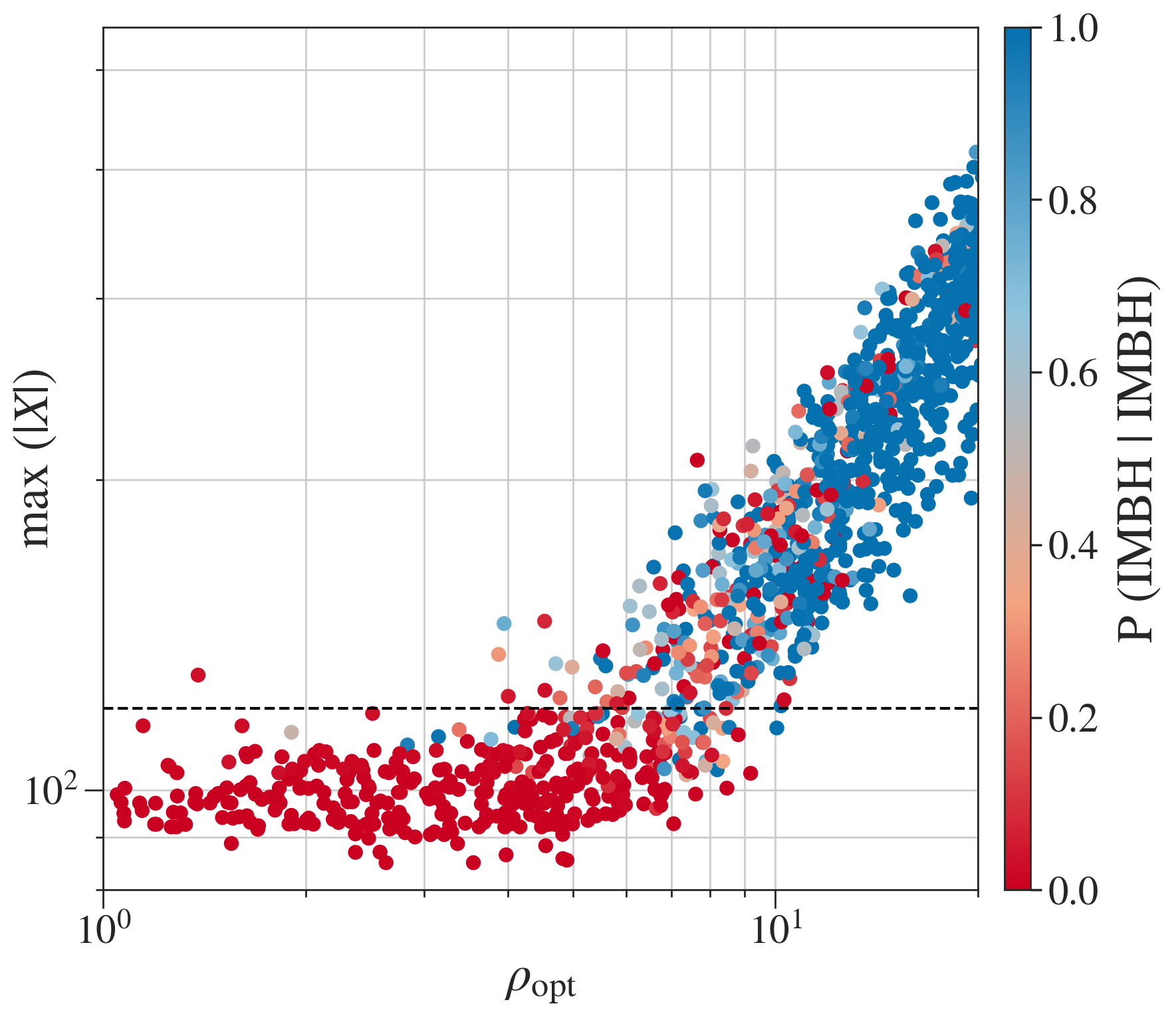}
    \caption{Distribution of the prediction probabilities for a simulated population of a highly inclined massive mass asymmetric binaries in the optimal \ac{SNR} and maximum absolute \ac{CWT} space (See Table~\ref{table:training-inj-parameters}). The dotted line shows the Max-CWT threshold below which we reject triggers.}
    \label{cwt_veto_explanation}
\end{figure}

\subsection{Max-CWT Threshold}
\label{subsec:glitch_veto}

\begin{figure}
    \includegraphics[width=\columnwidth]{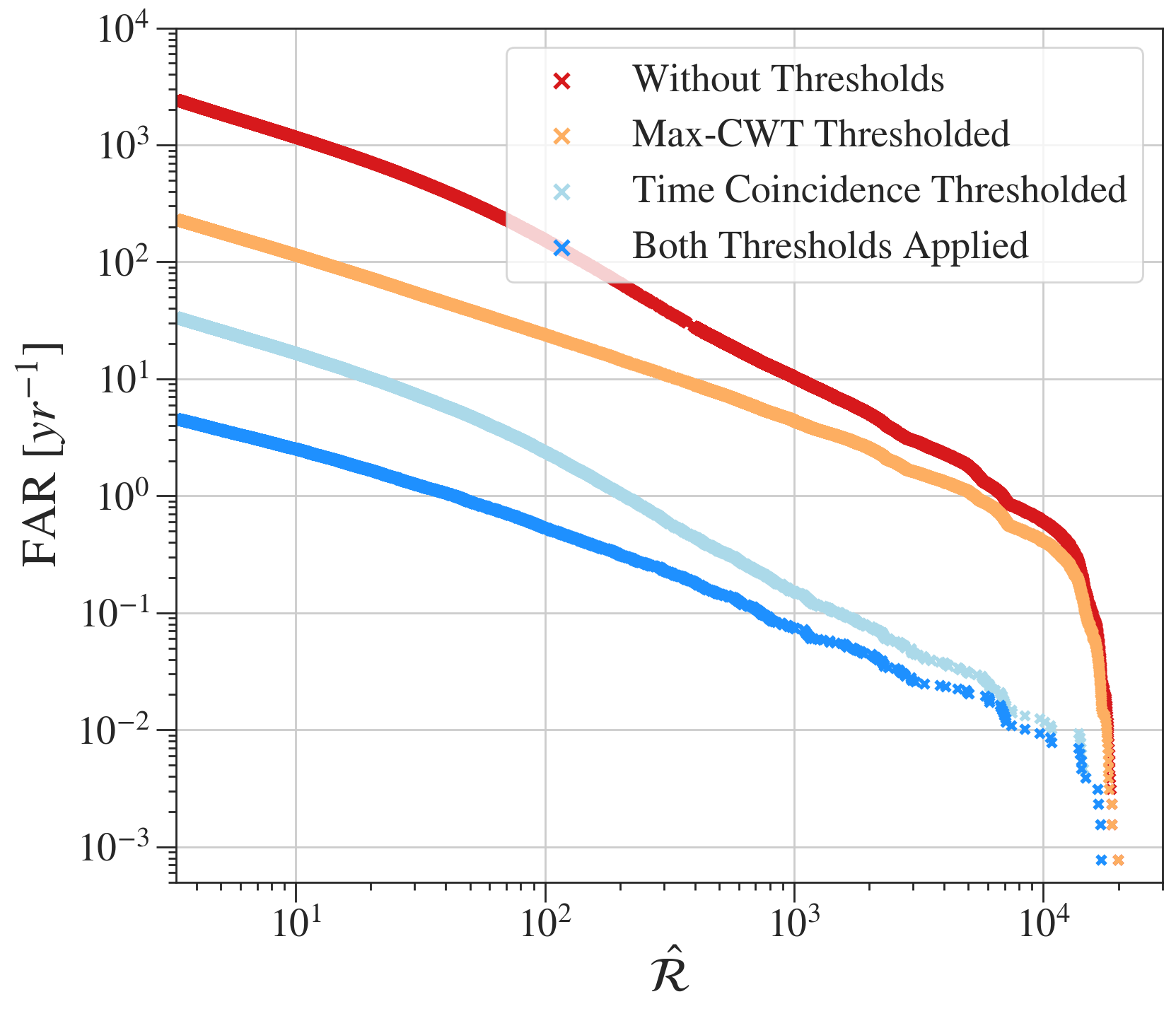}
    \caption{Ablation study of the time coincidence and max-CWT thresholds using the background noise curves. We apply both these thresholds in the search.}
    \label{rank_vs_far_bkg_triggers_veto_justification}
\end{figure}

We observe that the SNR of the simulated signals and the corresponding maximum absolute CWT value max($|X|$) follows a definite trend (see Fig.~\ref{cwt_veto_explanation}).  Capitalizing on this information, we implement a \textit{Max-CWT Threshold} to reject triggers with max($|X|$) $\leq$ 120.
The Max-CWT threshold is similar to the matched-filter SNR threshold often used in matched-filter searches. This constraint rejects approximately $\sim1.5$\% of the total injections used, most with low optimal \ac{SNR}. This also removes some of the noisy transients, thus improving our overall sensitivity.

\subsection{Efficacy of Thresholds}
\label{subsec:efficacy_of_vetoes}

We compare the effectiveness of the time coincidence and Max-CWT thresholds by computing the distribution of the noise background in FAR {\it vs} rank space. For our purpose, we generate $T_b=1284.8$ years of background data free of all coincident simulated signals and use it to estimate the \ac{FAR} (see Eq.~\eqref{eq:far}). The results from our ablation study are shown in Fig.~ \ref{rank_vs_far_bkg_triggers_veto_justification}. A significant drop in \ac{FAR} at low rank comes from the Max-CWT threshold. The time coincidence threshold reduces loud false alarms which are not coincident across the detectors. In \sw{THAMES} we use both the time coincidence and Max-CWT thresholds.


\section{Sensitivity Comparison with PyCBC-based searches}
\label{subsec:model_sensitivity} 

To assess the benefits of our newly developed search, we compare its performance against existing \sw{PyCBC}-based modeled searches in overlapping target parameter space with the simulated signals. Later in this section, we also test the robustness of our search algorithm \sw{THAMES}
by testing it outside the target space.

\subsection{Detection Efficiency}
\label{Sec:eta}

We characterize the sensitivity of \sw{THAMES} with the Receiver Operating Characteristics (ROC) curves. We compute the detection efficiency as the fraction of recovered simulated signals at a given \ac{FAR}. We compare the sensitivity of \sw{THAMES} with that of the \sw{PyCBC-IMBH} and \sw{PyCBC-HM} searches. The simulations used for this comparison are in the overlapping parameter space of mass ratio $q \in (5, 10)$ between these searches. The rest of the binary parameters of the simulated waveforms are similar to the ones in Table \ref{table:training-inj-parameters}. These signals are isotropically distributed in sky location and uniformly in comoving volume with network optimal \ac{SNR} in the range of 10-56.

\begin{figure}
    \includegraphics[width=\columnwidth]{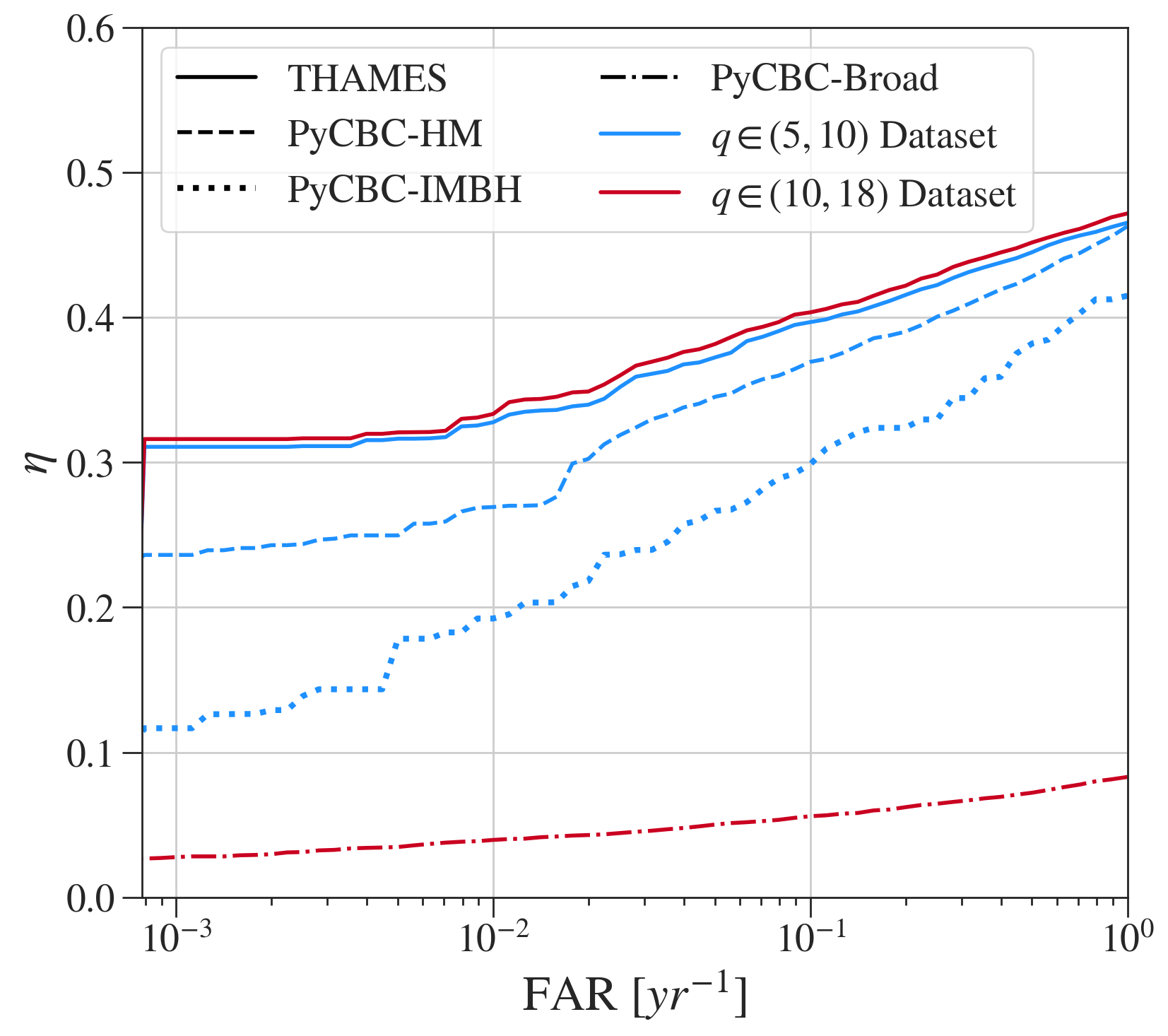}
        \caption{Comparison of \sw{THAMES}'s detection efficiency with \sw{PyCBC-IMBH}, \sw{PyCBC-HM} and \sw{PyCBC-Broad} search. We find that at an \ac{IFAR} threshold of 100 years, \sw{THAMES} outperforms \sw{PyCBC-IMBH}, \sw{PyCBC-HM} and \sw{PyCBC-Broad} by 70\%, 22\% and 730\% respectively.}
\label{compare_with_pycbc_all_in_one}
\end{figure}

Fig.~\ref{compare_with_pycbc_all_in_one} shows the result of our study. We find that \sw{THAMES} outperforms the \sw{PyCBC-IMBH} as well as the \sw{PyCBC-HM} searches.  At $\mathrm{IFAR} = 100$~years and $\mathrm{IFAR} = 1000$~years,  \sw{THAMES} is 1.7 (1.22) and 2.66 (1.32) times more sensitive than \sw{PyCBC-IMBH} (\sw{PyCBC-HM}) respectively. This gain in detection efficiency is because \sw{THAMES} is better equipped to detect signals and glitches than the other two searches, which allows it to discriminate glitches better. The improvement compared to \sw{PyCBC-IMBH} is also because \sw{THAMES} is trained on signals with higher harmonics as opposed to the latter, which uses the dominant harmonic of the \IMBH binary as templates.

Since the only \sw{PyCBC}-based search that overlaps with the training space of \sw{THAMES} is \sw{PyCBC-Broad}, we compare our search's sensitivity against it. To do that, we use signals with $q\in(10,18)$ but keep all the other parameters similar to the other comparison study. Fig.~\ref{compare_with_pycbc_all_in_one} summarizes the detection efficiency comparison results. We find that \sw{THAMES} appreciably outperforms \sw{PyCBC-Broad} search in this parameter space with an average sensitivity gain factor of 8.43 and 11.42 at IFAR = 100~years and IFAR = 1000~years, respectively, thus illustrating the capability of deep learning models to probe the complex signal morphology. The efficiency of \sw{PyCBC-Broad} is low due to both \textit{look elsewhere effect} and the omission of higher-order modes in the search templates.

\begin{figure}
    \includegraphics[width=\columnwidth]{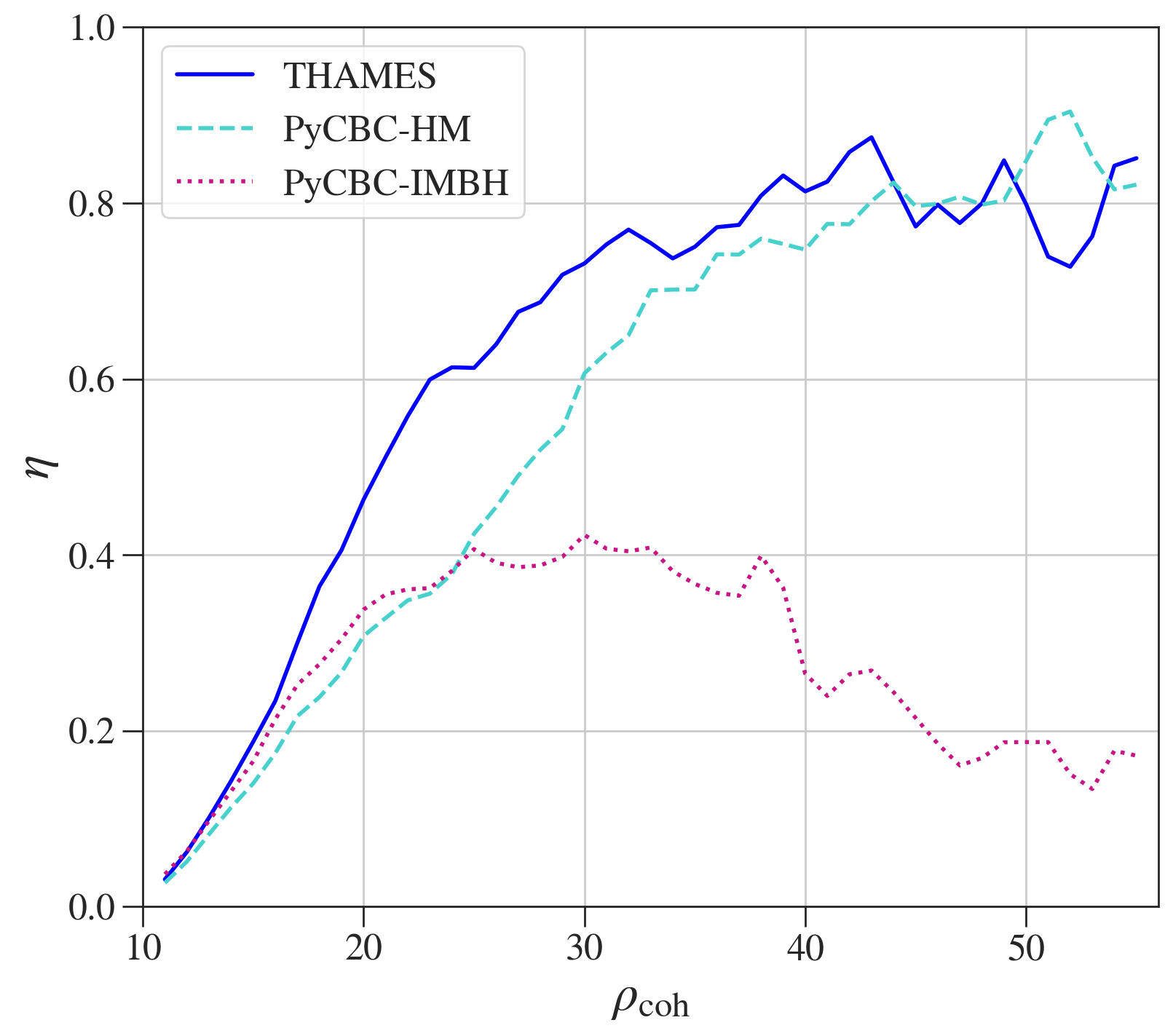}
    \caption{Detection efficiency as a function of network optimal \ac{SNR} at IFAR = 100~years for the three algorithms. \sw{THAMES} outperforms \sw{PyCBC-IMBH} search at all network optimal \ac{SNR}s and \sw{PyCBC-HM} search at network optimal \ac{SNR} below $\sim45$.}
    \label{snr_sensitive_regime}
\end{figure}

In Fig.~\ref{snr_sensitive_regime}, we compare the detection efficiency as a function of the optimal coherent network \ac{SNR} \citep{Pai:2000zt},  $\rho_{coh} = \sqrt{\rho_{\mathrm{opt, H1}}^2 + \rho_{\mathrm{opt, L1}}^2}$ at IFAR of 100~years. We observe that \sw{THAMES} outperforms \sw{PyCBC-IMBH} search at all $\rho_{coh}$ values. However, as compared to the \sw{PyCBC-HM} search, it outperforms only in mid-range $\rho_{coh}$ values and performs comparably at high values of $\rho_{coh}$.

We also observe a progressive drop in the sensitivity of \sw{PyCBC-IMBH} after achieving the peak sensitivity at $\rho_{coh}$ of $\sim30$. This reduction in efficiency is due to an increased contribution from higher harmonics with increasing \ac{SNR}, which \sw{PyCBC-IMBH} search is not tuned to detect. However, the efficiency of \sw{PyCBC-HM} search continues to increase with increasing \acp{SNR}. 

We observe a small decline in sensitivity of \sw{THAMES} at high SNRs. A representative of these transients unveiled that model's capacity to detect large \ac{SNR} signals deteriorates slightly due to confusion with Koi Fishes and Extremely Loud glitch classes, especially at even higher \ac{SNR}s. Here, the \sw{PyCBC-HM} performs better than \sw{THAMES}. Hence, a model trained on optimal \ac{SNR} $\rho_{\mathrm{opt}} \in (5, 40)$ does not generalize to signals with higher optimal \ac{SNR}, especially if the signal brings complex features with an increase in \ac{SNR}.  Further,  the contrast between the signal transient and the background noise in a CWT map depends on the signal's loudness, which is an inherent input to our machinery. We will investigate the latter in a future study.


\subsection{Sensitive Volume-Time Product}

\begin{figure}
    \includegraphics[width=\columnwidth]{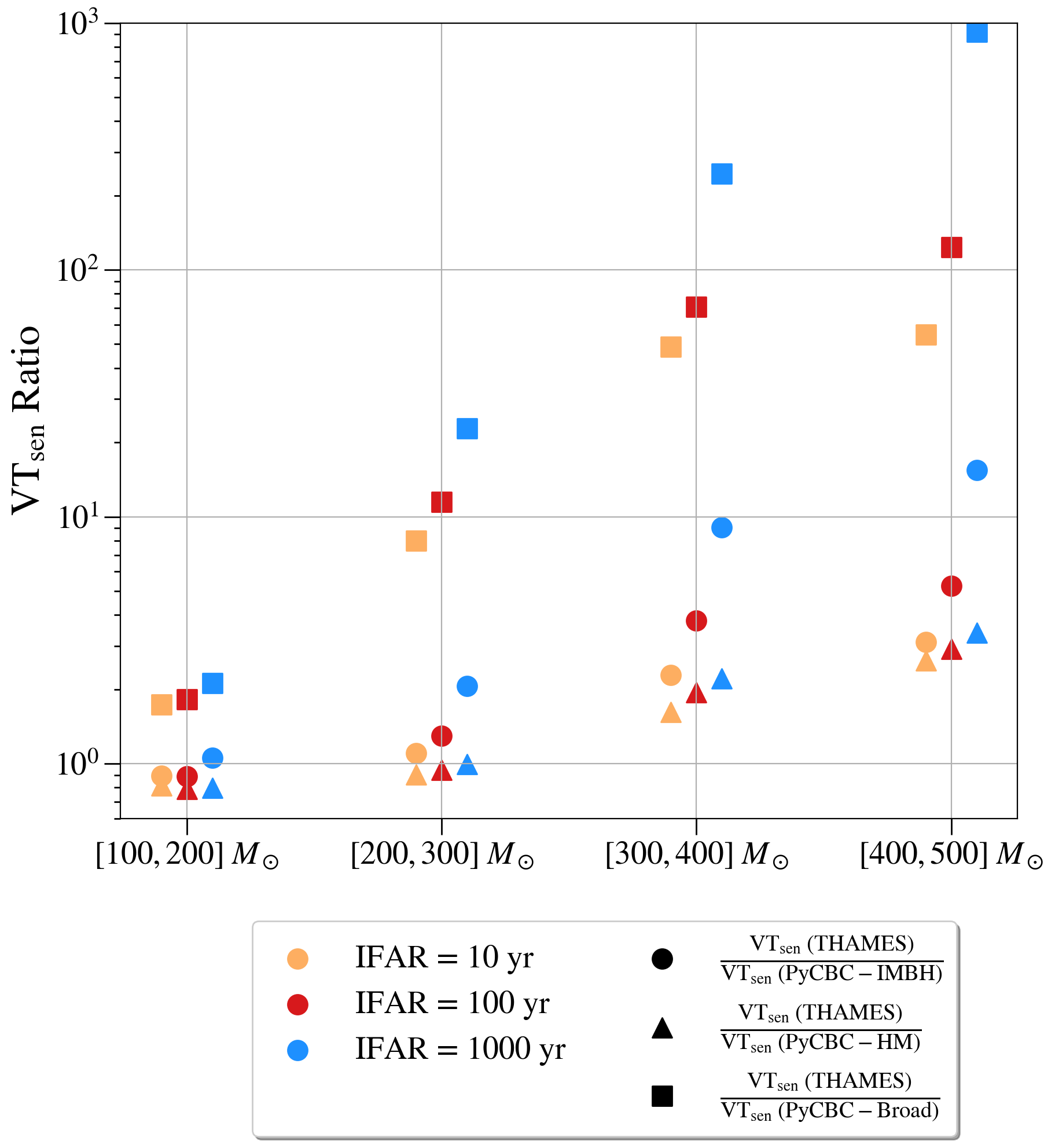}
    \caption{The ratio of the sensitive volume-time for \sw{THAMES} to \sw{PyCBC-IMBH}, \sw{PyCBC-HM} and \sw{PyCBC-Broad} searches at different \ac{IFAR} thresholds. A typical error in the sensitive volume-time ratio estimate ranges between 0.10 and 0.28.}
    \label{mass_ratio-total_mass-sensitive_volume}
\end{figure}

To compare the sensitivity of the four searches in the $M_T(1+z)-q$ plane, we estimate the sensitive volume-time product (VT$_{\mathrm{sen}}$) using Monte-Carlo method~\citep{Usman:2015kfa}. VT$_{\mathrm{sen}}$ is the time and population-averaged spacetime volume to which the detector is sensitive at the chosen \ac{FAR} threshold. The results from this analysis are summarized in Fig.~\ref{mass_ratio-total_mass-sensitive_volume}. All the comparisons shown are carried out in the overlapping parameter set and using the same simulated signals discussed in Sec.~\ref{Sec:eta}.

The relative sensitivity of \sw{THAMES} improves with an increase in detector frame total mass. This is due to \sw{THAMES}'s better ability to discriminate glitches from \IMBH binary signals, especially when the signal duration gets shorter with increasing detector frame total mass. The relatively poor VT$_{\mathrm{sen}}$ at low detector frame total mass and low mass ratio ($M_T(1+z) \lesssim 300~M_\odot$,  $q \in (5,10) $) is due to considerable confusion between loud (louder than the training set) IMBH binaries and blips. We observe that both matched-filter-based searches perform better than \sw{THAMES} for $M_T(1+z) \in (100, 200)~M_\odot$ and $q \in (5,10) $. \sw{THAMES} improves over \sw{PyCBC-IMBH} and \sw{PyCBC-HM} by a factor of 5.24 and 2.92 respectively in $M_T(1+z) \in (400, 500)~M_\odot$ range at an IFAR of 100 years. Further, \sw{THAMES} outperforms \sw{PyCBC-Broad} search at all detector frame total mass, with maximum gain in VT$_{\mathrm{sen}}$ by a factor of $\sim100$ for high detector frame total mass $M_T(1+z) \in (400, 500)~M_\odot$ range. 

Since a full analysis of the \ac{O3} run is beyond the scope of the current work, we \textit{approximate} the statistical significance of a chosen set of O3a events. We summarize this in Appendix.~\ref{appx:open-box-results}.

\section{Robustness Investigation}
\label{sec:robustness_investigation}

Typically the \GW search algorithms are designed to target a subspace of the multi-dimensional signal parameter space. However, they are also tested outside their target space. Likewise, we perform two exercises to test the robustness of \sw{THAMES} against various signals from asymmetric and \IMBH binary systems. {\it First,} we consider signals from precessing binaries in the same $M_T(1+z) - q$ parameter space. {\it Second,} we consider \GW signals from \IMBH binaries from all inclinations rather than only restricted to nearly edge-on binary orientations.

\subsection{Susceptibility to Complex Morphology}

Due to random pairing, hierarchically formed \IMBH binaries are expected to have misalignment between the spins and the orbital angular momentum. Such a configuration induces relativistic precession effects, which causes the orbital plane to change its spatial orientation as the binary evolves. As a result, the emitted \GW signal, depending on the binary's orientation, can have a strong amplitude and phase modulation, which, in general, may not be sufficiently described by quasi-circular binary waveforms. 

To check the detection efficiency of our search algorithm toward such generic sources, we simulated a population of quasi-spherical black hole binaries with mass, distance, binary orientation parameters, and optimal network SNRs similar to the ones used in Sec.~\ref{Sec:eta}. As for the spins, we sampled from a uniform distribution of spin magnitudes $|\chi_{1,2}|\in(0,0.98)$ and isotropic distribution of spin orientations. Like before, we added the resulting set of simulated signals to the same representative section of Advanced LIGO detector data. We find that, at an \ac{IFAR} of 100 years, our sensitive volume-time product drops from 0.52~Gpc$^3$yr for non-precessing binaries to 0.30~Gpc$^3$yr for precessing binaries, thus resulting in a 41.6\% drop in sensitivity towards precessing sources as compared against a non-precessing binary population counterpart. This drop represents the inability of \sw{THAMES} to capture the spin-precession signal morphology, which is distinct from the quasi-circular binary signals carrying only the higher-order modes. A detailed investigation on this is the topic for future work.

\subsection{Generalization to All Inclination Angles}

Most of the observed \acl{GW} events are nearly optimally oriented since such sources are intrinsically loud and easier to detect. However, by construction, \sw{THAMES} is designed to target almost edge-on sources with measurable higher-order mode content. We test our search's capability to detect low-mass range \IMBH binaries with binary parameters described in Table \ref{table:training-inj-parameters} but at an arbitrary inclination. The simulated sources have an isotropic distribution of inclinations between $0^\circ$ and $180^\circ$, isotropic distribution over the sky sphere, and uniform distribution in comoving volume but with $\rho_\mathrm{coh}\in(10, 56)$.

\begin{figure}
    \includegraphics[width=\columnwidth]{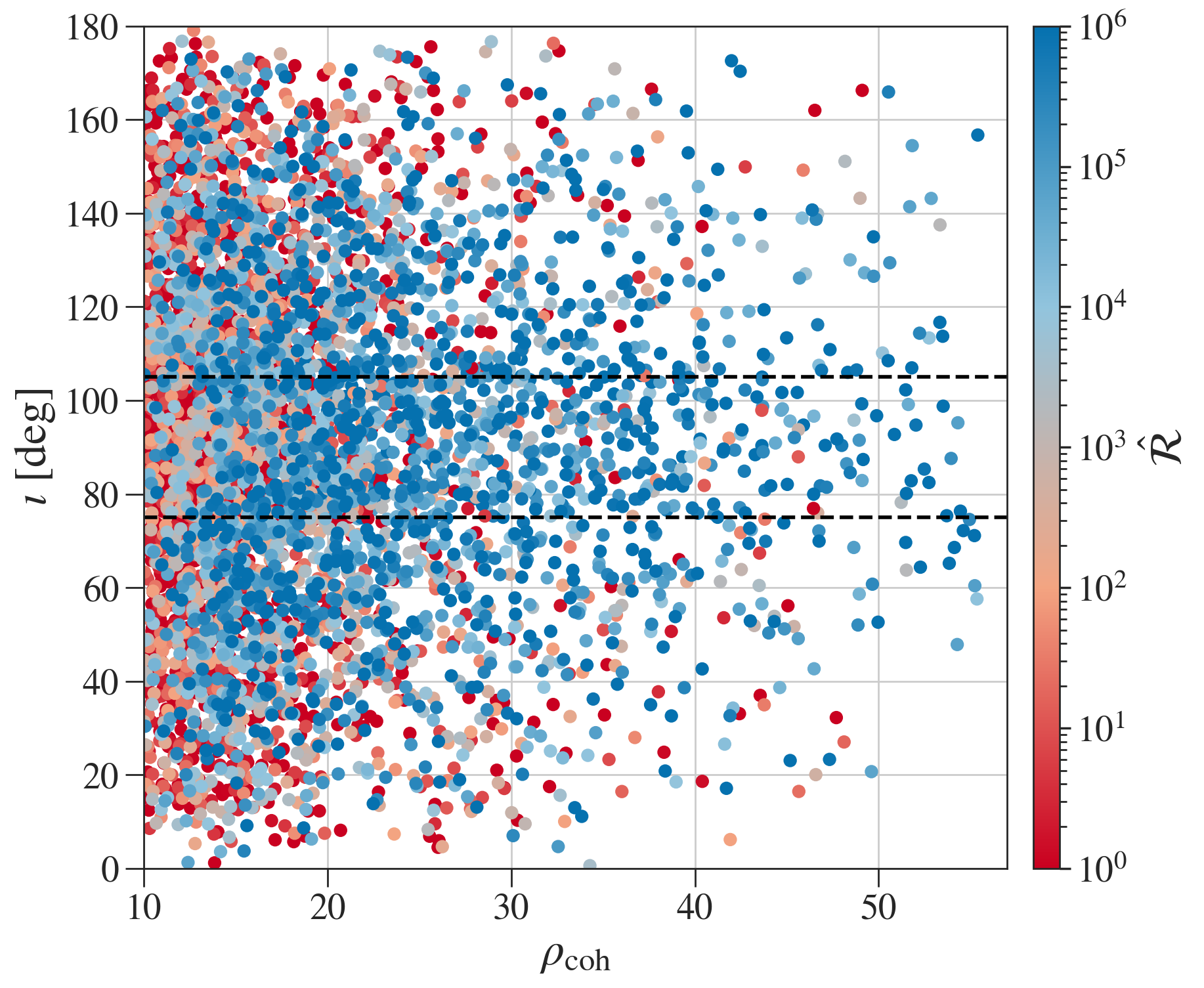}
    \caption{The distribution of the rank of quasi-circular black hole binaries in the inclination and network optimal \ac{SNR} space. The dotted lines enclose the inclination angle upon which \sw{THAMES} is trained.}
    \label{robustness_to_all_inclinations}
\end{figure}

In Fig.~\ref{robustness_to_all_inclinations}, we observe an improvement in the model's sensitivity to detect signals with inclinations beyond the training regime (black dashed lines). The recovered signals with IFAR above 100 years (and a rank greater than $\sim10^4$) are distributed over all inclination angles (blue points). \sw{THAMES}'s recovery outside the trained window is due to higher \ac{SNR} and no additional morphological complexity in the signal.

However, we also observe that a small fraction of signals with high network optimal \ac{SNR} are missed (red points), even for inclination angles in the training regime due to the similar reason explained in Sec. \ref{Sec:eta}.

\section{Conclusion}
\label{sec:conclusion}

We describe the first end-to-end deep-learning-based modeled search, \sw{THAMES}, to detect \GW signals from mass asymmetric, nearly edge-on \IMBH binaries. We use a deep learning model with Inception-v3 network architecture to classify the CWT maps of data from the \GW interferometers into signals from non-optimally oriented mass asymmetric \IMBH binaries and a variety of noisy glitches which resemble \IMBH binary signals. We devise a novel quadrature-sum detection statistic based on the \IMBH evidence provided by the deep learning model to rank all the simulated signals and background triggers. The primary rationale behind the construction is that the transients with inconsistent morphology with the trained \IMBH signals are heavily penalized, and coincident signals with nearly equiprobability in both the LIGO detectors are rewarded. We utilize the time-coincidence condition and Max-CWT threshold to suppress the false alarms in terms of the noisy glitches. 

The detection efficiency of our search is significantly greater than those of other modeled searches with overlapping parameter space. In fact, we find that our search is a factor of 5.24 ($\sim100$) more sensitive than that of the \sw{PyCBC-IMBH} (\sw{PyCBC-Broad}) search at an IFAR = 100 years for systems with total mass in the range 400-500~$M_\odot$. These matched-filter searches are primarily penalized due to the usage of templates without higher-order harmonics. Therefore, this demonstrates the capability of deep learning techniques to effectively probe complex signal morphologies inherent in \GW signals from asymmetric, edge-on \IMBH binary that carry higher-order modes. We also compared our search performance against that of the recently developed \sw{PyCBC-HM} search and found that depending on the system's total mass, the gain in sensitive volume-time is at most by a factor of 2.92. This clearly illustrates deep learning techniques' better signal-noise discrimination power, especially for short-duration signals. However, we report that our search slightly underperforms for \IMBH binaries with a low detector frame total mass of 100-200~$M_\odot$. This is due to significant confusion with blips in this parameter space. Furthermore, \sw{THAMES} is limited by its susceptibility to complex and loud signal morphology, which it is not trained on. Despite such limitations, deep learning techniques are up-and-coming due to their ability to leverage the morphology of glitches to better separate transient noises from signals and the ease of expanding to any search space.

We need to undergo a couple of upgrades to take advantage of the benefits offered by our deep-learning-based detection algorithm. There is a scope for improvement in our time estimation technique which currently misestimates the time for low SNR signals. At present,  the least count of the image is five times the sampling interval of our data ($\Delta t \sim0.5$~ms), which affects our trigger time estimation accuracy. Since the time-coincidence threshold is the root cause for the principle improvement in background trigger distribution,  possibly an improvement in the time estimation technique will further improve the detection efficiency. Secondly, the current search uses continuous wavelet transform to generate spectrograms, which is time-consuming. More time-efficient approaches need to be explored.

Encouraged by the remarkable sensitivity of \sw{THAMES} in detecting \IMBH binary signals with higher-order modes, we look forward to extending this search to precessing and eccentric waveforms that carry additional signal complexity. We also plan on extending the search to a multi-detector network.

\begin{figure*}[ht]
    \includegraphics[width=\textwidth]{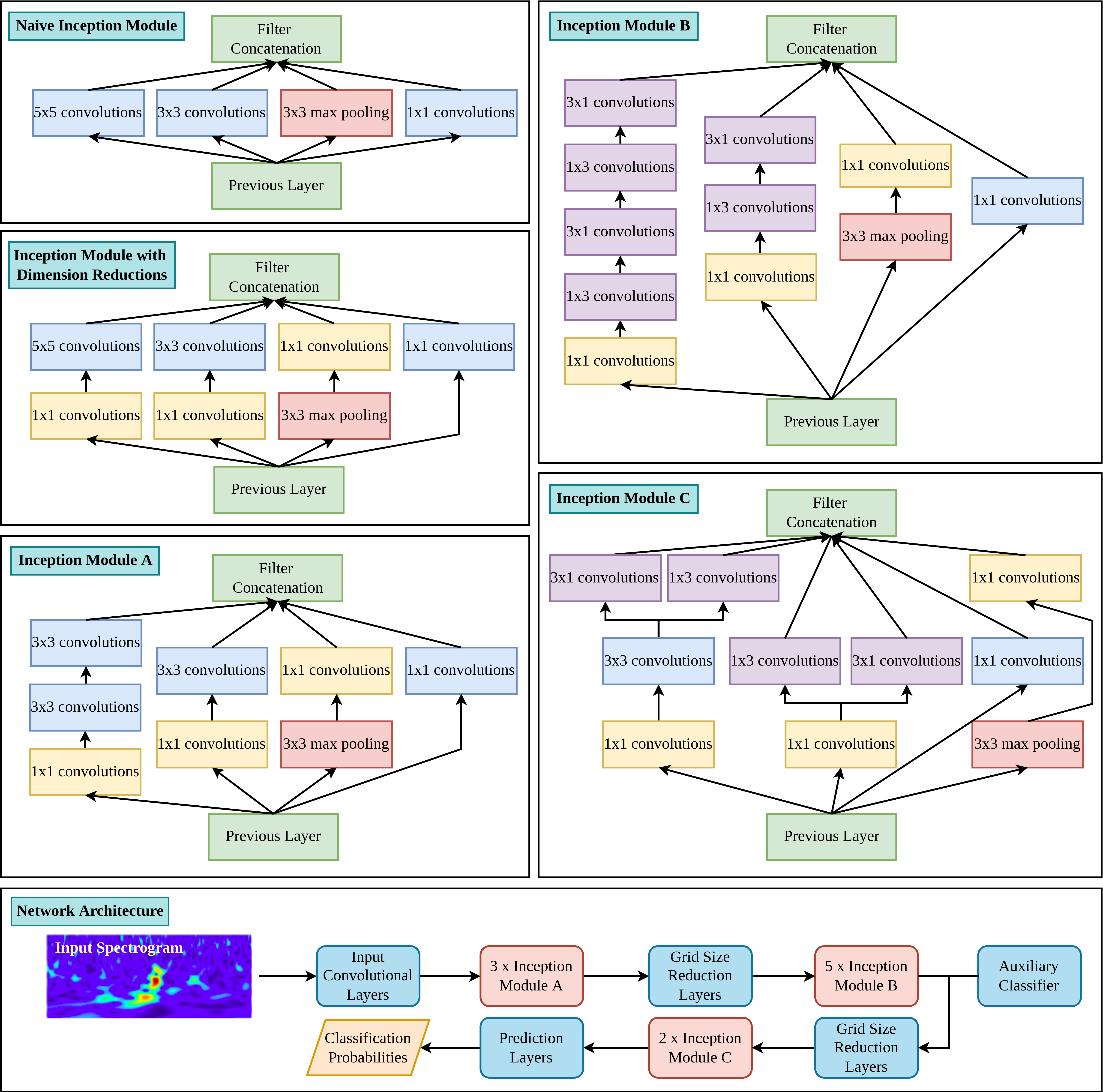}
    \caption{Network architecture used in the classification model of \sw{THAMES}. The network comprises 3 Inception Module A, 5 Inception Module B, and 2 Inception Module C, along with input convolutional layers, grid size reduction convolutional layers, auxiliary classifier, and prediction layers. The size of the input is 135 pixels $\times$ 735 pixels $\times$ 3 channels, and the output size is 9, composing the detection probabilities of all classes.}
\label{network_architecture}
\end{figure*}

\begin{acknowledgments}

We thank Ik Siong Heng and Juan Calder\'{o}n Bustillo for comments on the manuscript. This research has made use of data, software, and/or web tools obtained from the Gravitational Wave Open Science Center (\href{https://www.gw-openscience.org}{\sw{https://www.gw-openscience.org}}), a service of LIGO Laboratory, the LIGO Scientific Collaboration and the Virgo Collaboration. Virgo is funded by the French Centre National de Recherche Scientifique (CNRS), the Italian Istituto Nazionale della Fisica Nucleare (INFN), and the Dutch Nikhef, with contributions by Polish and Hungarian institutes. This material is based upon work supported by NSF’s LIGO Laboratory, which is a major facility fully funded by the National Science Foundation. The authors are grateful for computational resources provided by the LIGO Laboratory and supported by National Science Foundation Grants PHY0757058 and PHY-0823459. The authors thank the G-Suite products and Google Colaboratory for providing the powerful hardware option of GPU (\href{https://colab.research.google.com/}{\sw{https://colab.research.google.com/}}). KC acknowledges the MHRD, the Government of India, for the fellowship support. AP acknowledges the support from SERB-Power-fellowship grant SPF/2021/000036, DST, India. This document has LIGO DCC No LIGO-P2200227.

\end{acknowledgments}

\appendix

\section{Network Architecture}
\label{appx:network_architecture}

The network proposed in \sw{THAMES} uses transfer learning by reusing the Inception-v3 pre-trained model~\citep{szegedy2015rethinking, szegedy2014going}. The key idea behind inception modules is to formulate convolution operations optimally to capture salient parts of an image. Smaller convolution kernels are optimal for capturing local features (with low size variations). Similarly, larger kernels are preferred to capture global features (with high size variations). To cater to this need, the use of multiple stacked convolution operations will make the network too deep, thus causing over-fitting and a tremendous increase in training time~\citep{Ying_2019, https://doi.org/10.48550/arxiv.1901.06566}. Inception modules tackle these problems by having wider networks with various convolution sizes and max-pooling operations to downscale the image appropriately~\citep{2020arXiv200907485G}. These max-pooling layers also have the added advantage of adding translation, rotation, and scale invariance to the deep learning model. Therefore, the construct of a \textit{Naive Inception Module} will have various convolutional kernels ($1\times1$, $3\times3$, $5\times5$) and a max-pooling filter ($3\times3$), the output from which will be concatenated and fed into the next layer (see top-left panel of Fig. \ref{network_architecture}).
Since the time complexity of an $m \times n$ convolution operation is $O(m \cdot n)$, this construct of the naive inception module increases the computational cost. The developers of inception modules tackle this problem by adding $1\times1$ convolutions before the $3\times3$ and $5\times5$ convolutions and after the $3\times3$ max-pooling layer to reduce the number of input channels and hence, the computational cost. This construct is called \textit{Inception Module with Dimension Reductions}. Despite this, the overall network is very deep, leading to a problem of vanishing gradients. This limitation is lifted by adding \textit{Auxiliary Classifiers} so that the network takes into account an ``auxiliary loss", in addition to the training loss when learning~\citep{7298594, pmlr-v38-lee15a}.

The computational cost of larger convolutions can be further reduced by factorizing them into two smaller convolution operations, such that the receptive fields of the two operations remains the same~\citep{szegedy2015rethinking}. This way, a $5\times5$ convolution can be factorized into two $3\times3$ convolutions, thus reducing the computational complexity by $O(2 \cdot 3^2/5^2)$, while retaining the same receptive field. We refer to the module with these amendments as \textit{Inception Module A}. Similarly, we further modify this module by factorizing $3\times3$ convolutions into $3\times1$ and $1\times3$ convolutions, thus reducing the computational complexity by $O((3\cdot1 + 1\cdot3)/3^2)$ which we refer as \textit{Inception Module B}. The loss of information due to a stack of convolutions which significantly reduces the image's dimensions can be fixed by expanding the filters. This we refer to as \textit{Inception Module C}.

\begin{table*}[t]
    \centering
    \begin{tabular}{llcccccccc}
        \hline
        \multirow{2}{*}{Event Name} &\multirow{2}{*}{GPS Time~[s]}~& \multirow{2}{*}{$M_T(1+z)$ [$M_\odot$]} & \multirow{2}{*}{$q$} &  \multicolumn{5}{c}{IFAR~[yr]} \\ 
        & & & & \sw{THAMES} & \sw{PyCBC-IMBH} & \sw{PyCBC-HM} & \sw{GWTC-2} & \sw{IAS}  \\
         \hline
          $\GW190408\_181802$ & $1238782700.3$ & 55.5 & 1.34 & 47.52 & 62.66 & 60.60 & $>1000$ & $>1000$  \\
          $\GW190412$ & $1239082262.2$ & 44.2 & 3.63 & 38.88 & - & 10.50 & $>1000$ & $>1000$ \\
          $\GW190426\_190642$ & $1240340820.6$ & 315.4 & 1.39 & 2.12 & - & - & $0.24$ & $0.19$ \\
          $\GW190513\_205428$ & $1241816086.8$ & 73.8 & 1.98 & 47.52 & - & 1.49 & $>1000$ & $>1000$ \\
         $\GW190519\_153544$ & $1242315362.4$ & 154.2 & 1.65 & 4.95 & $> 1000$ & 31.35 & $>1000$ & $>1000$ \\
         $\GW190521$ & $1242442967.4$ & 278.2 & 1.28 & 11.46 & 726.65 & $1.13$ & $>1000$ & - \\
         $\GW190521\_074359$ & $1242459857.4$ & 92.6 & 1.27 & 4.94 & $>1000$ & $>1000$ & $>1000$ & $>1000$ \\
         $\GW190602\_175927$ & $1243533585.1$ & 172.2 & 1.62 & $>1000$ &  915.02 & 2.07 & $>1000$ & $>1000$ \\
         $\GW190929\_012149$ & $1253755327.5$ & 144.0 & 3.35 & 17.11 & - & - & $6.2$ & $>1000$ \\
    \hline
    \end{tabular}
    \caption{\Acl{GW} events recovered by our search with \ac{IFAR} $>1$ year, inferred by assuming similar detector sensitivity across the first part of the \ac{O3} data from Hanford and Livingston detector. These events have all been reported in~\citet{LIGOScientific:2020ibl, Nitz:2021uxj, Chandra:2021wbw, Chandra:2022ixv, Olsen:2022pin} with a comparatively higher/lower statistical significance. The $>$ sign indicates that the estimated \ac{IFAR} value is larger than the one stated for the respective search. We attribute their comparatively lower significance in \sw{THAMES} to the absence of detectable higher-order modes and/or being outside the bounds of our target search space.}
    \label{tab:events}
\end{table*}
    
Our detailed network architecture is represented in Fig.~\ref{network_architecture}. Our deep learning network comprises three inception modules A, five inception modules B, and two inception modules C. The size of our \ac{CWT} maps are 135~pixels $\times$ 735~pixels $\times$ 3~channels, which are fed into input convolutional layers, followed by inception modules with interleaved grid size reduction layers. As discussed above, an auxiliary classifier is used to avoid the vanishing gradients problem. All these layers are initialized with pre-trained weights and are only fine-tuned during training to extract generic features from our dataset. The features extracted from the input by our convolutional base have a size of $2 \times 21 \times 2048$.

\begin{figure}
    \includegraphics[width=\columnwidth]{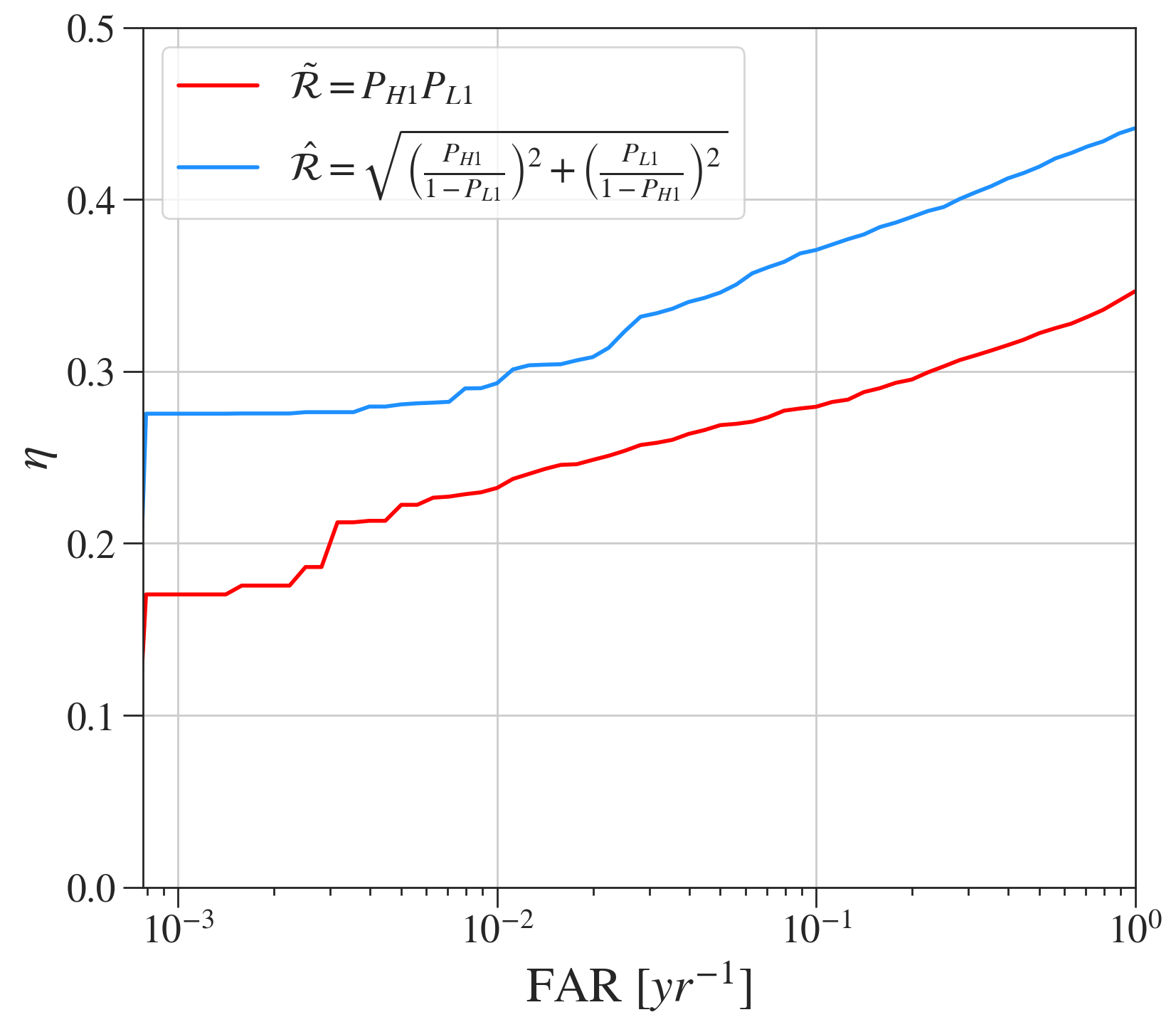}
    \caption{Comparison between the two detection statistics. The quadrature-sum detection statistic ($\hat{\mathcal{R}}$) gives the maximum detection efficiency at all FARs.}
    \label{metric_comparison}
\end{figure}

These feature maps are then fed into a set of prediction layers, which are randomly initialized for training. Our first filter in this set of prediction layers is an average-pooling layer~\citep{https://doi.org/10.48550/arxiv.1312.4400}. Since the entire Inception-v3 network only uses max-pooling layers, which only retain the feature map's most prominent (bright) attributes, we choose to use an average pooling layer in the end to get smooth features. Next, we use a densely connected layer since it passes information from each neuron in the previous layer to each neuron in the following layer, thus increasing each neuron's receptive field to take into account features from the entire image before making any prediction~\citep{https://doi.org/10.48550/arxiv.1608.06993}. We also use a drop-out layer to avoid overfitting when using densely connected layers~\citep{JMLR:v15:srivastava14a}. Finally, we use a dense connected layer with softmax~\citep{Goodfellow-et-al-2016} activation function to output the relative probability of a \ac{CWT} map to belong to one of the nine trained classes.


\section{Detection Statistic Comparison}
\label{appx:detection_statistic}
Here,  we compare the benefits of the quadrature sum statistic defined in Eq.~\eqref{eq:stat-1} with the conventionally used rank defined as the joint probability statistic under the assumption of uncorrelated noise between  H1 and L1 as:
\begin{equation}
    \tilde{\mathcal{R}} = P_\mathrm{H1} \cdot P_\mathrm{L1}~
\end{equation}
with $P_\mathrm{H1}$,  $P_\mathrm{L1}$ being the probability of the data containing an \IMBH binary signal in H1 and L1, respectively.
To understand their impact on search sensitivity, we, in Fig.~\ref{metric_comparison}, compare the detection efficiency of \sw{THAMES} using these two different ranking statistics. We see that the quadrature sum statistic ($\hat{\mathcal{R}}$) performs better than the joint probability statistic ($\tilde{\mathcal{R}}$), reconfirming the ability of $\hat{\mathcal{R}}$ to separate the signal and noise transients, also shown in Fig.~\ref{ranking_statistic_justification}.

\section{\textit{Surrogate} Open-Box Results}
\label{appx:open-box-results}
Though doing a full analysis of the \ac{O3} data of Advanced LIGO detectors is beyond the scope of the current work, we wish to get an estimate of what we should expect from the extended analysis. We, therefore, assume that the estimated background trigger distribution remains more or less the same during the duration of the first part of the \ac{O3} run. We use this to estimate the \ac{IFAR} of a chosen set of statistically significant candidates already reported elsewhere~\citep{LIGOScientific:2020ibl, Olsen:2022pin}. Table~\ref{tab:events} lists the events which are recovered by \sw{THAMES} with \ac{IFAR} $> 1$ year. We note that the (expected) reduced statistical significance of the reported events for those obtained by existing searches highlights that these candidates are outside our target parameter space.

\bibliography{refs}

\begin{thebibliography}{93}%
\makeatletter
\providecommand \@ifxundefined [1]{%
 \@ifx{#1\undefined}
}%
\providecommand \@ifnum [1]{%
 \ifnum #1\expandafter \@firstoftwo
 \else \expandafter \@secondoftwo
 \fi
}%
\providecommand \@ifx [1]{%
 \ifx #1\expandafter \@firstoftwo
 \else \expandafter \@secondoftwo
 \fi
}%
\providecommand \natexlab [1]{#1}%
\providecommand \enquote  [1]{``#1''}%
\providecommand \bibnamefont  [1]{#1}%
\providecommand \bibfnamefont [1]{#1}%
\providecommand \citenamefont [1]{#1}%
\providecommand \href@noop [0]{\@secondoftwo}%
\providecommand \href [0]{\begingroup \@sanitize@url \@href}%
\providecommand \@href[1]{\@@startlink{#1}\@@href}%
\providecommand \@@href[1]{\endgroup#1\@@endlink}%
\providecommand \@sanitize@url [0]{\catcode `\\12\catcode `\$12\catcode
  `\&12\catcode `\#12\catcode `\^12\catcode `\_12\catcode `\%12\relax}%
\providecommand \@@startlink[1]{}%
\providecommand \@@endlink[0]{}%
\providecommand \url  [0]{\begingroup\@sanitize@url \@url }%
\providecommand \@url [1]{\endgroup\@href {#1}{\urlprefix }}%
\providecommand \urlprefix  [0]{URL }%
\providecommand \Eprint [0]{\href }%
\providecommand \doibase [0]{http://dx.doi.org/}%
\providecommand \selectlanguage [0]{\@gobble}%
\providecommand \bibinfo  [0]{\@secondoftwo}%
\providecommand \bibfield  [0]{\@secondoftwo}%
\providecommand \translation [1]{[#1]}%
\providecommand \BibitemOpen [0]{}%
\providecommand \bibitemStop [0]{}%
\providecommand \bibitemNoStop [0]{.\EOS\space}%
\providecommand \EOS [0]{\spacefactor3000\relax}%
\providecommand \BibitemShut  [1]{\csname bibitem#1\endcsname}%
\let\auto@bib@innerbib\@empty
\bibitem [{\citenamefont {Abbott}\ \emph {et~al.}(2019)\citenamefont {Abbott}
  \emph {et~al.}}]{LIGOScientific:2018mvr}%
  \BibitemOpen
  \bibfield  {author} {\bibinfo {author} {\bibfnamefont {B.~P.}\ \bibnamefont
  {Abbott}} \emph {et~al.} (\bibinfo {collaboration} {LIGO Scientific,
  Virgo}),\ }\href {\doibase 10.1103/PhysRevX.9.031040} {\bibfield  {journal}
  {\bibinfo  {journal} {Phys. Rev. X}\ }\textbf {\bibinfo {volume} {9}},\
  \bibinfo {pages} {031040} (\bibinfo {year} {2019})},\ \Eprint
  {http://arxiv.org/abs/1811.12907} {arXiv:1811.12907 [astro-ph.HE]}
  \BibitemShut {NoStop}%
\bibitem [{\citenamefont {Nitz}\ \emph {et~al.}(2019)\citenamefont {Nitz},
  \citenamefont {Capano}, \citenamefont {Nielsen}, \citenamefont {Reyes},
  \citenamefont {White}, \citenamefont {Brown},\ and\ \citenamefont
  {Krishnan}}]{Nitz:2018imz}%
  \BibitemOpen
  \bibfield  {author} {\bibinfo {author} {\bibfnamefont {A.~H.}\ \bibnamefont
  {Nitz}}, \bibinfo {author} {\bibfnamefont {C.}~\bibnamefont {Capano}},
  \bibinfo {author} {\bibfnamefont {A.~B.}\ \bibnamefont {Nielsen}}, \bibinfo
  {author} {\bibfnamefont {S.}~\bibnamefont {Reyes}}, \bibinfo {author}
  {\bibfnamefont {R.}~\bibnamefont {White}}, \bibinfo {author} {\bibfnamefont
  {D.~A.}\ \bibnamefont {Brown}}, \ and\ \bibinfo {author} {\bibfnamefont
  {B.}~\bibnamefont {Krishnan}},\ }\href {\doibase 10.3847/1538-4357/ab0108}
  {\bibfield  {journal} {\bibinfo  {journal} {Astrophys. J.}\ }\textbf
  {\bibinfo {volume} {872}},\ \bibinfo {pages} {195} (\bibinfo {year}
  {2019})},\ \Eprint {http://arxiv.org/abs/1811.01921} {arXiv:1811.01921
  [gr-qc]} \BibitemShut {NoStop}%
\bibitem [{\citenamefont {Venumadhav}\ \emph {et~al.}(2020)\citenamefont
  {Venumadhav}, \citenamefont {Zackay}, \citenamefont {Roulet}, \citenamefont
  {Dai},\ and\ \citenamefont {Zaldarriaga}}]{Venumadhav:2019lyq}%
  \BibitemOpen
  \bibfield  {author} {\bibinfo {author} {\bibfnamefont {T.}~\bibnamefont
  {Venumadhav}}, \bibinfo {author} {\bibfnamefont {B.}~\bibnamefont {Zackay}},
  \bibinfo {author} {\bibfnamefont {J.}~\bibnamefont {Roulet}}, \bibinfo
  {author} {\bibfnamefont {L.}~\bibnamefont {Dai}}, \ and\ \bibinfo {author}
  {\bibfnamefont {M.}~\bibnamefont {Zaldarriaga}},\ }\href {\doibase
  10.1103/PhysRevD.101.083030} {\bibfield  {journal} {\bibinfo  {journal}
  {Phys. Rev. D}\ }\textbf {\bibinfo {volume} {101}},\ \bibinfo {pages}
  {083030} (\bibinfo {year} {2020})},\ \Eprint
  {http://arxiv.org/abs/1904.07214} {arXiv:1904.07214 [astro-ph.HE]}
  \BibitemShut {NoStop}%
\bibitem [{\citenamefont {Zackay}\ \emph {et~al.}(2021)\citenamefont {Zackay},
  \citenamefont {Dai}, \citenamefont {Venumadhav}, \citenamefont {Roulet},\
  and\ \citenamefont {Zaldarriaga}}]{Zackay:2019btq}%
  \BibitemOpen
  \bibfield  {author} {\bibinfo {author} {\bibfnamefont {B.}~\bibnamefont
  {Zackay}}, \bibinfo {author} {\bibfnamefont {L.}~\bibnamefont {Dai}},
  \bibinfo {author} {\bibfnamefont {T.}~\bibnamefont {Venumadhav}}, \bibinfo
  {author} {\bibfnamefont {J.}~\bibnamefont {Roulet}}, \ and\ \bibinfo {author}
  {\bibfnamefont {M.}~\bibnamefont {Zaldarriaga}},\ }\href {\doibase
  10.1103/PhysRevD.104.063030} {\bibfield  {journal} {\bibinfo  {journal}
  {Phys. Rev. D}\ }\textbf {\bibinfo {volume} {104}},\ \bibinfo {pages}
  {063030} (\bibinfo {year} {2021})},\ \Eprint
  {http://arxiv.org/abs/1910.09528} {arXiv:1910.09528 [astro-ph.HE]}
  \BibitemShut {NoStop}%
\bibitem [{\citenamefont {Zackay}\ \emph {et~al.}(2019)\citenamefont {Zackay},
  \citenamefont {Venumadhav}, \citenamefont {Dai}, \citenamefont {Roulet},\
  and\ \citenamefont {Zaldarriaga}}]{Zackay:2019tzo}%
  \BibitemOpen
  \bibfield  {author} {\bibinfo {author} {\bibfnamefont {B.}~\bibnamefont
  {Zackay}}, \bibinfo {author} {\bibfnamefont {T.}~\bibnamefont {Venumadhav}},
  \bibinfo {author} {\bibfnamefont {L.}~\bibnamefont {Dai}}, \bibinfo {author}
  {\bibfnamefont {J.}~\bibnamefont {Roulet}}, \ and\ \bibinfo {author}
  {\bibfnamefont {M.}~\bibnamefont {Zaldarriaga}},\ }\href {\doibase
  10.1103/PhysRevD.100.023007} {\bibfield  {journal} {\bibinfo  {journal}
  {Phys. Rev. D}\ }\textbf {\bibinfo {volume} {100}},\ \bibinfo {pages}
  {023007} (\bibinfo {year} {2019})},\ \Eprint
  {http://arxiv.org/abs/1902.10331} {arXiv:1902.10331 [astro-ph.HE]}
  \BibitemShut {NoStop}%
\bibitem [{\citenamefont {Nitz}\ \emph {et~al.}(2020)\citenamefont {Nitz},
  \citenamefont {Dent}, \citenamefont {Davies}, \citenamefont {Kumar},
  \citenamefont {Capano}, \citenamefont {Harry}, \citenamefont {Mozzon},
  \citenamefont {Nuttall}, \citenamefont {Lundgren},\ and\ \citenamefont
  {T\'apai}}]{Nitz:2020oeq}%
  \BibitemOpen
  \bibfield  {author} {\bibinfo {author} {\bibfnamefont {A.~H.}\ \bibnamefont
  {Nitz}}, \bibinfo {author} {\bibfnamefont {T.}~\bibnamefont {Dent}}, \bibinfo
  {author} {\bibfnamefont {G.~S.}\ \bibnamefont {Davies}}, \bibinfo {author}
  {\bibfnamefont {S.}~\bibnamefont {Kumar}}, \bibinfo {author} {\bibfnamefont
  {C.~D.}\ \bibnamefont {Capano}}, \bibinfo {author} {\bibfnamefont
  {I.}~\bibnamefont {Harry}}, \bibinfo {author} {\bibfnamefont
  {S.}~\bibnamefont {Mozzon}}, \bibinfo {author} {\bibfnamefont
  {L.}~\bibnamefont {Nuttall}}, \bibinfo {author} {\bibfnamefont
  {A.}~\bibnamefont {Lundgren}}, \ and\ \bibinfo {author} {\bibfnamefont
  {M.}~\bibnamefont {T\'apai}},\ }\href {\doibase 10.3847/1538-4357/ab733f}
  {\bibfield  {journal} {\bibinfo  {journal} {Astrophys. J.}\ }\textbf
  {\bibinfo {volume} {891}},\ \bibinfo {pages} {123} (\bibinfo {year}
  {2020})},\ \Eprint {http://arxiv.org/abs/1910.05331} {arXiv:1910.05331
  [astro-ph.HE]} \BibitemShut {NoStop}%
\bibitem [{\citenamefont {Abbott}\ \emph
  {et~al.}(2021{\natexlab{a}})\citenamefont {Abbott} \emph
  {et~al.}}]{LIGOScientific:2020ibl}%
  \BibitemOpen
  \bibfield  {author} {\bibinfo {author} {\bibfnamefont {R.}~\bibnamefont
  {Abbott}} \emph {et~al.} (\bibinfo {collaboration} {LIGO Scientific,
  Virgo}),\ }\href {\doibase 10.1103/PhysRevX.11.021053} {\bibfield  {journal}
  {\bibinfo  {journal} {Phys. Rev. X}\ }\textbf {\bibinfo {volume} {11}},\
  \bibinfo {pages} {021053} (\bibinfo {year} {2021}{\natexlab{a}})},\ \Eprint
  {http://arxiv.org/abs/2010.14527} {arXiv:2010.14527 [gr-qc]} \BibitemShut
  {NoStop}%
\bibitem [{\citenamefont {Abbott}\ \emph
  {et~al.}(2021{\natexlab{b}})\citenamefont {Abbott} \emph
  {et~al.}}]{LIGOScientific:2021djp}%
  \BibitemOpen
  \bibfield  {author} {\bibinfo {author} {\bibfnamefont {R.}~\bibnamefont
  {Abbott}} \emph {et~al.} (\bibinfo {collaboration} {LIGO Scientific, VIRGO,
  KAGRA}),\ }\href@noop {} {\  (\bibinfo {year} {2021}{\natexlab{b}})},\
  \Eprint {http://arxiv.org/abs/2111.03606} {arXiv:2111.03606 [gr-qc]}
  \BibitemShut {NoStop}%
\bibitem [{\citenamefont {Nitz}\ \emph
  {et~al.}(2021{\natexlab{a}})\citenamefont {Nitz}, \citenamefont {Capano},
  \citenamefont {Kumar}, \citenamefont {Wang}, \citenamefont {Kastha},
  \citenamefont {Sch\"afer}, \citenamefont {Dhurkunde},\ and\ \citenamefont
  {Cabero}}]{Nitz:2021uxj}%
  \BibitemOpen
  \bibfield  {author} {\bibinfo {author} {\bibfnamefont {A.~H.}\ \bibnamefont
  {Nitz}}, \bibinfo {author} {\bibfnamefont {C.~D.}\ \bibnamefont {Capano}},
  \bibinfo {author} {\bibfnamefont {S.}~\bibnamefont {Kumar}}, \bibinfo
  {author} {\bibfnamefont {Y.-F.}\ \bibnamefont {Wang}}, \bibinfo {author}
  {\bibfnamefont {S.}~\bibnamefont {Kastha}}, \bibinfo {author} {\bibfnamefont
  {M.}~\bibnamefont {Sch\"afer}}, \bibinfo {author} {\bibfnamefont
  {R.}~\bibnamefont {Dhurkunde}}, \ and\ \bibinfo {author} {\bibfnamefont
  {M.}~\bibnamefont {Cabero}},\ }\href {\doibase 10.3847/1538-4357/ac1c03}
  {\bibfield  {journal} {\bibinfo  {journal} {Astrophys. J.}\ }\textbf
  {\bibinfo {volume} {922}},\ \bibinfo {pages} {76} (\bibinfo {year}
  {2021}{\natexlab{a}})},\ \Eprint {http://arxiv.org/abs/2105.09151}
  {arXiv:2105.09151 [astro-ph.HE]} \BibitemShut {NoStop}%
\bibitem [{\citenamefont {Nitz}\ \emph
  {et~al.}(2021{\natexlab{b}})\citenamefont {Nitz}, \citenamefont {Kumar},
  \citenamefont {Wang}, \citenamefont {Kastha}, \citenamefont {Wu},
  \citenamefont {Sch\"afer}, \citenamefont {Dhurkunde},\ and\ \citenamefont
  {Capano}}]{Nitz:2021zwj}%
  \BibitemOpen
  \bibfield  {author} {\bibinfo {author} {\bibfnamefont {A.~H.}\ \bibnamefont
  {Nitz}}, \bibinfo {author} {\bibfnamefont {S.}~\bibnamefont {Kumar}},
  \bibinfo {author} {\bibfnamefont {Y.-F.}\ \bibnamefont {Wang}}, \bibinfo
  {author} {\bibfnamefont {S.}~\bibnamefont {Kastha}}, \bibinfo {author}
  {\bibfnamefont {S.}~\bibnamefont {Wu}}, \bibinfo {author} {\bibfnamefont
  {M.}~\bibnamefont {Sch\"afer}}, \bibinfo {author} {\bibfnamefont
  {R.}~\bibnamefont {Dhurkunde}}, \ and\ \bibinfo {author} {\bibfnamefont
  {C.~D.}\ \bibnamefont {Capano}},\ }\href@noop {} {\  (\bibinfo {year}
  {2021}{\natexlab{b}})},\ \Eprint {http://arxiv.org/abs/2112.06878}
  {arXiv:2112.06878 [astro-ph.HE]} \BibitemShut {NoStop}%
\bibitem [{\citenamefont {Olsen}\ \emph {et~al.}(2022)\citenamefont {Olsen},
  \citenamefont {Venumadhav}, \citenamefont {Mushkin}, \citenamefont {Roulet},
  \citenamefont {Zackay},\ and\ \citenamefont {Zaldarriaga}}]{Olsen:2022pin}%
  \BibitemOpen
  \bibfield  {author} {\bibinfo {author} {\bibfnamefont {S.}~\bibnamefont
  {Olsen}}, \bibinfo {author} {\bibfnamefont {T.}~\bibnamefont {Venumadhav}},
  \bibinfo {author} {\bibfnamefont {J.}~\bibnamefont {Mushkin}}, \bibinfo
  {author} {\bibfnamefont {J.}~\bibnamefont {Roulet}}, \bibinfo {author}
  {\bibfnamefont {B.}~\bibnamefont {Zackay}}, \ and\ \bibinfo {author}
  {\bibfnamefont {M.}~\bibnamefont {Zaldarriaga}},\ }\href@noop {} {\
  (\bibinfo {year} {2022})},\ \Eprint {http://arxiv.org/abs/2201.02252}
  {arXiv:2201.02252 [astro-ph.HE]} \BibitemShut {NoStop}%
\bibitem [{\citenamefont {Abbott}\ \emph
  {et~al.}(2020{\natexlab{a}})\citenamefont {Abbott} \emph
  {et~al.}}]{LIGOScientific:2020iuh}%
  \BibitemOpen
  \bibfield  {author} {\bibinfo {author} {\bibfnamefont {R.}~\bibnamefont
  {Abbott}} \emph {et~al.} (\bibinfo {collaboration} {LIGO Scientific,
  Virgo}),\ }\href {\doibase 10.1103/PhysRevLett.125.101102} {\bibfield
  {journal} {\bibinfo  {journal} {Phys. Rev. Lett.}\ }\textbf {\bibinfo
  {volume} {125}},\ \bibinfo {pages} {101102} (\bibinfo {year}
  {2020}{\natexlab{a}})},\ \Eprint {http://arxiv.org/abs/2009.01075}
  {arXiv:2009.01075 [gr-qc]} \BibitemShut {NoStop}%
\bibitem [{\citenamefont {Abbott}\ \emph
  {et~al.}(2020{\natexlab{b}})\citenamefont {Abbott} \emph
  {et~al.}}]{LIGOScientific:2020ufj}%
  \BibitemOpen
  \bibfield  {author} {\bibinfo {author} {\bibfnamefont {R.}~\bibnamefont
  {Abbott}} \emph {et~al.} (\bibinfo {collaboration} {LIGO Scientific,
  Virgo}),\ }\href {\doibase 10.3847/2041-8213/aba493} {\bibfield  {journal}
  {\bibinfo  {journal} {Astrophys. J. Lett.}\ }\textbf {\bibinfo {volume}
  {900}},\ \bibinfo {pages} {L13} (\bibinfo {year} {2020}{\natexlab{b}})},\
  \Eprint {http://arxiv.org/abs/2009.01190} {arXiv:2009.01190 [astro-ph.HE]}
  \BibitemShut {NoStop}%
\bibitem [{\citenamefont {Szczepa\'nczyk}\ \emph {et~al.}(2021)\citenamefont
  {Szczepa\'nczyk} \emph {et~al.}}]{Szczepanczyk:2020osv}%
  \BibitemOpen
  \bibfield  {author} {\bibinfo {author} {\bibfnamefont {M.}~\bibnamefont
  {Szczepa\'nczyk}} \emph {et~al.},\ }\href {\doibase
  10.1103/PhysRevD.103.082002} {\bibfield  {journal} {\bibinfo  {journal}
  {Phys. Rev. D}\ }\textbf {\bibinfo {volume} {103}},\ \bibinfo {pages}
  {082002} (\bibinfo {year} {2021})},\ \Eprint
  {http://arxiv.org/abs/2009.11336} {arXiv:2009.11336 [astro-ph.HE]}
  \BibitemShut {NoStop}%
\bibitem [{\citenamefont {Woosley}\ and\ \citenamefont
  {Heger}(2021)}]{Woosley:2021xba}%
  \BibitemOpen
  \bibfield  {author} {\bibinfo {author} {\bibfnamefont {S.~E.}\ \bibnamefont
  {Woosley}}\ and\ \bibinfo {author} {\bibfnamefont {A.}~\bibnamefont
  {Heger}},\ }\href {\doibase 10.3847/2041-8213/abf2c4} {\bibfield  {journal}
  {\bibinfo  {journal} {Astrophys. J. Lett.}\ }\textbf {\bibinfo {volume}
  {912}},\ \bibinfo {pages} {L31} (\bibinfo {year} {2021})},\ \Eprint
  {http://arxiv.org/abs/2103.07933} {arXiv:2103.07933 [astro-ph.SR]}
  \BibitemShut {NoStop}%
\bibitem [{\citenamefont {Gerosa}\ \emph {et~al.}(2021)\citenamefont {Gerosa},
  \citenamefont {Giacobbo},\ and\ \citenamefont {Vecchio}}]{Gerosa:2021hsc}%
  \BibitemOpen
  \bibfield  {author} {\bibinfo {author} {\bibfnamefont {D.}~\bibnamefont
  {Gerosa}}, \bibinfo {author} {\bibfnamefont {N.}~\bibnamefont {Giacobbo}}, \
  and\ \bibinfo {author} {\bibfnamefont {A.}~\bibnamefont {Vecchio}},\ }\href
  {\doibase 10.3847/1538-4357/ac00bb} {\bibfield  {journal} {\bibinfo
  {journal} {Astrophys. J.}\ }\textbf {\bibinfo {volume} {915}},\ \bibinfo
  {pages} {56} (\bibinfo {year} {2021})},\ \Eprint
  {http://arxiv.org/abs/2104.11247} {arXiv:2104.11247 [astro-ph.HE]}
  \BibitemShut {NoStop}%
\bibitem [{\citenamefont {Baibhav}\ \emph {et~al.}(2020)\citenamefont
  {Baibhav}, \citenamefont {Gerosa}, \citenamefont {Berti}, \citenamefont
  {Wong}, \citenamefont {Helfer},\ and\ \citenamefont
  {Mould}}]{Baibhav:2020xdf}%
  \BibitemOpen
  \bibfield  {author} {\bibinfo {author} {\bibfnamefont {V.}~\bibnamefont
  {Baibhav}}, \bibinfo {author} {\bibfnamefont {D.}~\bibnamefont {Gerosa}},
  \bibinfo {author} {\bibfnamefont {E.}~\bibnamefont {Berti}}, \bibinfo
  {author} {\bibfnamefont {K.~W.~K.}\ \bibnamefont {Wong}}, \bibinfo {author}
  {\bibfnamefont {T.}~\bibnamefont {Helfer}}, \ and\ \bibinfo {author}
  {\bibfnamefont {M.}~\bibnamefont {Mould}},\ }\href {\doibase
  10.1103/PhysRevD.102.043002} {\bibfield  {journal} {\bibinfo  {journal}
  {Phys. Rev. D}\ }\textbf {\bibinfo {volume} {102}},\ \bibinfo {pages}
  {043002} (\bibinfo {year} {2020})},\ \Eprint
  {http://arxiv.org/abs/2004.00650} {arXiv:2004.00650 [astro-ph.HE]}
  \BibitemShut {NoStop}%
\bibitem [{\citenamefont {Berti}\ \emph {et~al.}(2007)\citenamefont {Berti},
  \citenamefont {Cardoso}, \citenamefont {Gonzalez}, \citenamefont {Sperhake},
  \citenamefont {Hannam}, \citenamefont {Husa},\ and\ \citenamefont
  {Bruegmann}}]{Berti:2007fi}%
  \BibitemOpen
  \bibfield  {author} {\bibinfo {author} {\bibfnamefont {E.}~\bibnamefont
  {Berti}}, \bibinfo {author} {\bibfnamefont {V.}~\bibnamefont {Cardoso}},
  \bibinfo {author} {\bibfnamefont {J.~A.}\ \bibnamefont {Gonzalez}}, \bibinfo
  {author} {\bibfnamefont {U.}~\bibnamefont {Sperhake}}, \bibinfo {author}
  {\bibfnamefont {M.}~\bibnamefont {Hannam}}, \bibinfo {author} {\bibfnamefont
  {S.}~\bibnamefont {Husa}}, \ and\ \bibinfo {author} {\bibfnamefont
  {B.}~\bibnamefont {Bruegmann}},\ }\href {\doibase 10.1103/PhysRevD.76.064034}
  {\bibfield  {journal} {\bibinfo  {journal} {Phys. Rev. D}\ }\textbf {\bibinfo
  {volume} {76}},\ \bibinfo {pages} {064034} (\bibinfo {year} {2007})},\
  \Eprint {http://arxiv.org/abs/gr-qc/0703053} {arXiv:gr-qc/0703053}
  \BibitemShut {NoStop}%
\bibitem [{\citenamefont {Mills}\ and\ \citenamefont
  {Fairhurst}(2021)}]{Mills:2020thr}%
  \BibitemOpen
  \bibfield  {author} {\bibinfo {author} {\bibfnamefont {C.}~\bibnamefont
  {Mills}}\ and\ \bibinfo {author} {\bibfnamefont {S.}~\bibnamefont
  {Fairhurst}},\ }\href {\doibase 10.1103/PhysRevD.103.024042} {\bibfield
  {journal} {\bibinfo  {journal} {Phys. Rev. D}\ }\textbf {\bibinfo {volume}
  {103}},\ \bibinfo {pages} {024042} (\bibinfo {year} {2021})},\ \Eprint
  {http://arxiv.org/abs/2007.04313} {arXiv:2007.04313 [gr-qc]} \BibitemShut
  {NoStop}%
\bibitem [{\citenamefont {Calder\'on~Bustillo}\ \emph
  {et~al.}(2016)\citenamefont {Calder\'on~Bustillo}, \citenamefont {Husa},
  \citenamefont {Sintes},\ and\ \citenamefont
  {P\"urrer}}]{CalderonBustillo:2015lrt}%
  \BibitemOpen
  \bibfield  {author} {\bibinfo {author} {\bibfnamefont {J.}~\bibnamefont
  {Calder\'on~Bustillo}}, \bibinfo {author} {\bibfnamefont {S.}~\bibnamefont
  {Husa}}, \bibinfo {author} {\bibfnamefont {A.~M.}\ \bibnamefont {Sintes}}, \
  and\ \bibinfo {author} {\bibfnamefont {M.}~\bibnamefont {P\"urrer}},\ }\href
  {\doibase 10.1103/PhysRevD.93.084019} {\bibfield  {journal} {\bibinfo
  {journal} {Phys. Rev. D}\ }\textbf {\bibinfo {volume} {93}},\ \bibinfo
  {pages} {084019} (\bibinfo {year} {2016})},\ \Eprint
  {http://arxiv.org/abs/1511.02060} {arXiv:1511.02060 [gr-qc]} \BibitemShut
  {NoStop}%
\bibitem [{\citenamefont {Calder\'on~Bustillo}\ \emph
  {et~al.}(2017)\citenamefont {Calder\'on~Bustillo}, \citenamefont {Laguna},\
  and\ \citenamefont {Shoemaker}}]{CalderonBustillo:2016rlt}%
  \BibitemOpen
  \bibfield  {author} {\bibinfo {author} {\bibfnamefont {J.}~\bibnamefont
  {Calder\'on~Bustillo}}, \bibinfo {author} {\bibfnamefont {P.}~\bibnamefont
  {Laguna}}, \ and\ \bibinfo {author} {\bibfnamefont {D.}~\bibnamefont
  {Shoemaker}},\ }\href {\doibase 10.1103/PhysRevD.95.104038} {\bibfield
  {journal} {\bibinfo  {journal} {Phys. Rev. D}\ }\textbf {\bibinfo {volume}
  {95}},\ \bibinfo {pages} {104038} (\bibinfo {year} {2017})},\ \Eprint
  {http://arxiv.org/abs/1612.02340} {arXiv:1612.02340 [gr-qc]} \BibitemShut
  {NoStop}%
\bibitem [{\citenamefont {Varma}\ \emph {et~al.}(2014)\citenamefont {Varma},
  \citenamefont {Ajith}, \citenamefont {Husa}, \citenamefont {Bustillo},
  \citenamefont {Hannam},\ and\ \citenamefont {P\"urrer}}]{Varma:2014jxa}%
  \BibitemOpen
  \bibfield  {author} {\bibinfo {author} {\bibfnamefont {V.}~\bibnamefont
  {Varma}}, \bibinfo {author} {\bibfnamefont {P.}~\bibnamefont {Ajith}},
  \bibinfo {author} {\bibfnamefont {S.}~\bibnamefont {Husa}}, \bibinfo {author}
  {\bibfnamefont {J.~C.}\ \bibnamefont {Bustillo}}, \bibinfo {author}
  {\bibfnamefont {M.}~\bibnamefont {Hannam}}, \ and\ \bibinfo {author}
  {\bibfnamefont {M.}~\bibnamefont {P\"urrer}},\ }\href {\doibase
  10.1103/PhysRevD.90.124004} {\bibfield  {journal} {\bibinfo  {journal} {Phys.
  Rev. D}\ }\textbf {\bibinfo {volume} {90}},\ \bibinfo {pages} {124004}
  (\bibinfo {year} {2014})},\ \Eprint {http://arxiv.org/abs/1409.2349}
  {arXiv:1409.2349 [gr-qc]} \BibitemShut {NoStop}%
\bibitem [{\citenamefont {Usman}\ \emph {et~al.}(2019)\citenamefont {Usman},
  \citenamefont {Mills},\ and\ \citenamefont {Fairhurst}}]{Usman:2018imj}%
  \BibitemOpen
  \bibfield  {author} {\bibinfo {author} {\bibfnamefont {S.~A.}\ \bibnamefont
  {Usman}}, \bibinfo {author} {\bibfnamefont {J.~C.}\ \bibnamefont {Mills}}, \
  and\ \bibinfo {author} {\bibfnamefont {S.}~\bibnamefont {Fairhurst}},\ }\href
  {\doibase 10.3847/1538-4357/ab0b3e} {\bibfield  {journal} {\bibinfo
  {journal} {Astrophys. J.}\ }\textbf {\bibinfo {volume} {877}},\ \bibinfo
  {pages} {82} (\bibinfo {year} {2019})},\ \Eprint
  {http://arxiv.org/abs/1809.10727} {arXiv:1809.10727 [gr-qc]} \BibitemShut
  {NoStop}%
\bibitem [{\citenamefont {Vitale}\ and\ \citenamefont
  {Chen}(2018)}]{Vitale:2018wlg}%
  \BibitemOpen
  \bibfield  {author} {\bibinfo {author} {\bibfnamefont {S.}~\bibnamefont
  {Vitale}}\ and\ \bibinfo {author} {\bibfnamefont {H.-Y.}\ \bibnamefont
  {Chen}},\ }\href {\doibase 10.1103/PhysRevLett.121.021303} {\bibfield
  {journal} {\bibinfo  {journal} {Phys. Rev. Lett.}\ }\textbf {\bibinfo
  {volume} {121}},\ \bibinfo {pages} {021303} (\bibinfo {year} {2018})},\
  \Eprint {http://arxiv.org/abs/1804.07337} {arXiv:1804.07337 [astro-ph.CO]}
  \BibitemShut {NoStop}%
\bibitem [{\citenamefont {Carullo}\ \emph {et~al.}(2019)\citenamefont
  {Carullo}, \citenamefont {Del~Pozzo},\ and\ \citenamefont
  {Veitch}}]{Carullo:2019flw}%
  \BibitemOpen
  \bibfield  {author} {\bibinfo {author} {\bibfnamefont {G.}~\bibnamefont
  {Carullo}}, \bibinfo {author} {\bibfnamefont {W.}~\bibnamefont {Del~Pozzo}},
  \ and\ \bibinfo {author} {\bibfnamefont {J.}~\bibnamefont {Veitch}},\ }\href
  {\doibase 10.1103/PhysRevD.99.123029} {\bibfield  {journal} {\bibinfo
  {journal} {Phys. Rev. D}\ }\textbf {\bibinfo {volume} {99}},\ \bibinfo
  {pages} {123029} (\bibinfo {year} {2019})},\ \bibinfo {note} {[Erratum:
  Phys.Rev.D 100, 089903 (2019)]},\ \Eprint {http://arxiv.org/abs/1902.07527}
  {arXiv:1902.07527 [gr-qc]} \BibitemShut {NoStop}%
\bibitem [{\citenamefont {Isi}\ \emph {et~al.}(2019)\citenamefont {Isi},
  \citenamefont {Giesler}, \citenamefont {Farr}, \citenamefont {Scheel},\ and\
  \citenamefont {Teukolsky}}]{Isi:2019aib}%
  \BibitemOpen
  \bibfield  {author} {\bibinfo {author} {\bibfnamefont {M.}~\bibnamefont
  {Isi}}, \bibinfo {author} {\bibfnamefont {M.}~\bibnamefont {Giesler}},
  \bibinfo {author} {\bibfnamefont {W.~M.}\ \bibnamefont {Farr}}, \bibinfo
  {author} {\bibfnamefont {M.~A.}\ \bibnamefont {Scheel}}, \ and\ \bibinfo
  {author} {\bibfnamefont {S.~A.}\ \bibnamefont {Teukolsky}},\ }\href {\doibase
  10.1103/PhysRevLett.123.111102} {\bibfield  {journal} {\bibinfo  {journal}
  {Phys. Rev. Lett.}\ }\textbf {\bibinfo {volume} {123}},\ \bibinfo {pages}
  {111102} (\bibinfo {year} {2019})},\ \Eprint
  {http://arxiv.org/abs/1905.00869} {arXiv:1905.00869 [gr-qc]} \BibitemShut
  {NoStop}%
\bibitem [{\citenamefont {Bustillo}\ \emph {et~al.}(2021)\citenamefont
  {Bustillo}, \citenamefont {Lasky},\ and\ \citenamefont
  {Thrane}}]{Bustillo:2020buq}%
  \BibitemOpen
  \bibfield  {author} {\bibinfo {author} {\bibfnamefont {J.~C.}\ \bibnamefont
  {Bustillo}}, \bibinfo {author} {\bibfnamefont {P.~D.}\ \bibnamefont {Lasky}},
  \ and\ \bibinfo {author} {\bibfnamefont {E.}~\bibnamefont {Thrane}},\ }\href
  {\doibase 10.1103/PhysRevD.103.024041} {\bibfield  {journal} {\bibinfo
  {journal} {Phys. Rev. D}\ }\textbf {\bibinfo {volume} {103}},\ \bibinfo
  {pages} {024041} (\bibinfo {year} {2021})},\ \Eprint
  {http://arxiv.org/abs/2010.01857} {arXiv:2010.01857 [gr-qc]} \BibitemShut
  {NoStop}%
\bibitem [{\citenamefont {Varma}\ \emph {et~al.}(2020)\citenamefont {Varma},
  \citenamefont {Isi},\ and\ \citenamefont {Biscoveanu}}]{Varma:2020nbm}%
  \BibitemOpen
  \bibfield  {author} {\bibinfo {author} {\bibfnamefont {V.}~\bibnamefont
  {Varma}}, \bibinfo {author} {\bibfnamefont {M.}~\bibnamefont {Isi}}, \ and\
  \bibinfo {author} {\bibfnamefont {S.}~\bibnamefont {Biscoveanu}},\ }\href
  {\doibase 10.1103/PhysRevLett.124.101104} {\bibfield  {journal} {\bibinfo
  {journal} {Phys. Rev. Lett.}\ }\textbf {\bibinfo {volume} {124}},\ \bibinfo
  {pages} {101104} (\bibinfo {year} {2020})},\ \Eprint
  {http://arxiv.org/abs/2002.00296} {arXiv:2002.00296 [gr-qc]} \BibitemShut
  {NoStop}%
\bibitem [{\citenamefont {Calder\'on~Bustillo}\ \emph
  {et~al.}(2018{\natexlab{a}})\citenamefont {Calder\'on~Bustillo},
  \citenamefont {Clark}, \citenamefont {Laguna},\ and\ \citenamefont
  {Shoemaker}}]{CalderonBustillo:2018zuq}%
  \BibitemOpen
  \bibfield  {author} {\bibinfo {author} {\bibfnamefont {J.}~\bibnamefont
  {Calder\'on~Bustillo}}, \bibinfo {author} {\bibfnamefont {J.~A.}\
  \bibnamefont {Clark}}, \bibinfo {author} {\bibfnamefont {P.}~\bibnamefont
  {Laguna}}, \ and\ \bibinfo {author} {\bibfnamefont {D.}~\bibnamefont
  {Shoemaker}},\ }\href {\doibase 10.1103/PhysRevLett.121.191102} {\bibfield
  {journal} {\bibinfo  {journal} {Phys. Rev. Lett.}\ }\textbf {\bibinfo
  {volume} {121}},\ \bibinfo {pages} {191102} (\bibinfo {year}
  {2018}{\natexlab{a}})},\ \Eprint {http://arxiv.org/abs/1806.11160}
  {arXiv:1806.11160 [gr-qc]} \BibitemShut {NoStop}%
\bibitem [{\citenamefont {Calder\'on~Bustillo}\ \emph
  {et~al.}(2018{\natexlab{b}})\citenamefont {Calder\'on~Bustillo},
  \citenamefont {Salemi}, \citenamefont {Dal~Canton},\ and\ \citenamefont
  {Jani}}]{CalderonBustillo:2017skv}%
  \BibitemOpen
  \bibfield  {author} {\bibinfo {author} {\bibfnamefont {J.}~\bibnamefont
  {Calder\'on~Bustillo}}, \bibinfo {author} {\bibfnamefont {F.}~\bibnamefont
  {Salemi}}, \bibinfo {author} {\bibfnamefont {T.}~\bibnamefont {Dal~Canton}},
  \ and\ \bibinfo {author} {\bibfnamefont {K.~P.}\ \bibnamefont {Jani}},\
  }\href {\doibase 10.1103/PhysRevD.97.024016} {\bibfield  {journal} {\bibinfo
  {journal} {Phys. Rev. D}\ }\textbf {\bibinfo {volume} {97}},\ \bibinfo
  {pages} {024016} (\bibinfo {year} {2018}{\natexlab{b}})},\ \Eprint
  {http://arxiv.org/abs/1711.02009} {arXiv:1711.02009 [gr-qc]} \BibitemShut
  {NoStop}%
\bibitem [{\citenamefont {Chandra}\ \emph {et~al.}(2022)\citenamefont
  {Chandra}, \citenamefont {Calder\'on~Bustillo}, \citenamefont {Pai},\ and\
  \citenamefont {Harry}}]{Chandra:2022ixv}%
  \BibitemOpen
  \bibfield  {author} {\bibinfo {author} {\bibfnamefont {K.}~\bibnamefont
  {Chandra}}, \bibinfo {author} {\bibfnamefont {J.}~\bibnamefont
  {Calder\'on~Bustillo}}, \bibinfo {author} {\bibfnamefont {A.}~\bibnamefont
  {Pai}}, \ and\ \bibinfo {author} {\bibfnamefont {I.}~\bibnamefont {Harry}},\
  }\href@noop {} {\  (\bibinfo {year} {2022})},\ \Eprint
  {http://arxiv.org/abs/2207.01654} {arXiv:2207.01654 [gr-qc]} \BibitemShut
  {NoStop}%
\bibitem [{\citenamefont {Chandra}\ \emph {et~al.}(2021)\citenamefont
  {Chandra}, \citenamefont {Villa-Ortega}, \citenamefont {Dent}, \citenamefont
  {McIsaac}, \citenamefont {Pai}, \citenamefont {Harry}, \citenamefont
  {Davies},\ and\ \citenamefont {Soni}}]{Chandra:2021wbw}%
  \BibitemOpen
  \bibfield  {author} {\bibinfo {author} {\bibfnamefont {K.}~\bibnamefont
  {Chandra}}, \bibinfo {author} {\bibfnamefont {V.}~\bibnamefont
  {Villa-Ortega}}, \bibinfo {author} {\bibfnamefont {T.}~\bibnamefont {Dent}},
  \bibinfo {author} {\bibfnamefont {C.}~\bibnamefont {McIsaac}}, \bibinfo
  {author} {\bibfnamefont {A.}~\bibnamefont {Pai}}, \bibinfo {author}
  {\bibfnamefont {I.~W.}\ \bibnamefont {Harry}}, \bibinfo {author}
  {\bibfnamefont {G.~S.~C.}\ \bibnamefont {Davies}}, \ and\ \bibinfo {author}
  {\bibfnamefont {K.}~\bibnamefont {Soni}},\ }\href {\doibase
  10.1103/PhysRevD.104.042004} {\bibfield  {journal} {\bibinfo  {journal}
  {Phys. Rev. D}\ }\textbf {\bibinfo {volume} {104}},\ \bibinfo {pages}
  {042004} (\bibinfo {year} {2021})},\ \Eprint
  {http://arxiv.org/abs/2106.00193} {arXiv:2106.00193 [gr-qc]} \BibitemShut
  {NoStop}%
\bibitem [{\citenamefont {Abbott}\ \emph {et~al.}(2018)\citenamefont {Abbott}
  \emph {et~al.}}]{LIGOScientific:2017tza}%
  \BibitemOpen
  \bibfield  {author} {\bibinfo {author} {\bibfnamefont {B.~P.}\ \bibnamefont
  {Abbott}} \emph {et~al.} (\bibinfo {collaboration} {LIGO Scientific,
  Virgo}),\ }\href {\doibase 10.1088/1361-6382/aaaafa} {\bibfield  {journal}
  {\bibinfo  {journal} {Class. Quant. Grav.}\ }\textbf {\bibinfo {volume}
  {35}},\ \bibinfo {pages} {065010} (\bibinfo {year} {2018})},\ \Eprint
  {http://arxiv.org/abs/1710.02185} {arXiv:1710.02185 [gr-qc]} \BibitemShut
  {NoStop}%
\bibitem [{\citenamefont {Allen}(2005)}]{Allen:2004gu}%
  \BibitemOpen
  \bibfield  {author} {\bibinfo {author} {\bibfnamefont {B.}~\bibnamefont
  {Allen}},\ }\href {\doibase 10.1103/PhysRevD.71.062001} {\bibfield  {journal}
  {\bibinfo  {journal} {Phys. Rev. D}\ }\textbf {\bibinfo {volume} {71}},\
  \bibinfo {pages} {062001} (\bibinfo {year} {2005})},\ \Eprint
  {http://arxiv.org/abs/gr-qc/0405045} {arXiv:gr-qc/0405045} \BibitemShut
  {NoStop}%
\bibitem [{\citenamefont {Usman}\ \emph {et~al.}(2016)\citenamefont {Usman}
  \emph {et~al.}}]{Usman:2015kfa}%
  \BibitemOpen
  \bibfield  {author} {\bibinfo {author} {\bibfnamefont {S.~A.}\ \bibnamefont
  {Usman}} \emph {et~al.},\ }\href {\doibase 10.1088/0264-9381/33/21/215004}
  {\bibfield  {journal} {\bibinfo  {journal} {Class. Quant. Grav.}\ }\textbf
  {\bibinfo {volume} {33}},\ \bibinfo {pages} {215004} (\bibinfo {year}
  {2016})},\ \Eprint {http://arxiv.org/abs/1508.02357} {arXiv:1508.02357
  [gr-qc]} \BibitemShut {NoStop}%
\bibitem [{\citenamefont {Cuoco}\ \emph {et~al.}(2021)\citenamefont {Cuoco}
  \emph {et~al.}}]{Cuoco:2020ogp}%
  \BibitemOpen
  \bibfield  {author} {\bibinfo {author} {\bibfnamefont {E.}~\bibnamefont
  {Cuoco}} \emph {et~al.},\ }\href {\doibase 10.1088/2632-2153/abb93a}
  {\bibfield  {journal} {\bibinfo  {journal} {Mach. Learn. Sci. Tech.}\
  }\textbf {\bibinfo {volume} {2}},\ \bibinfo {pages} {011002} (\bibinfo {year}
  {2021})},\ \Eprint {http://arxiv.org/abs/2005.03745} {arXiv:2005.03745
  [astro-ph.HE]} \BibitemShut {NoStop}%
\bibitem [{\citenamefont {Aasi}\ \emph {et~al.}(2015)\citenamefont {Aasi} \emph
  {et~al.}}]{LIGOScientific:2014pky}%
  \BibitemOpen
  \bibfield  {author} {\bibinfo {author} {\bibfnamefont {J.}~\bibnamefont
  {Aasi}} \emph {et~al.} (\bibinfo {collaboration} {LIGO Scientific}),\ }\href
  {\doibase 10.1088/0264-9381/32/7/074001} {\bibfield  {journal} {\bibinfo
  {journal} {Class. Quant. Grav.}\ }\textbf {\bibinfo {volume} {32}},\ \bibinfo
  {pages} {074001} (\bibinfo {year} {2015})},\ \Eprint
  {http://arxiv.org/abs/1411.4547} {arXiv:1411.4547 [gr-qc]} \BibitemShut
  {NoStop}%
\bibitem [{\citenamefont {Flandrin}(2018)}]{flandrin_2018}%
  \BibitemOpen
  \bibfield  {author} {\bibinfo {author} {\bibfnamefont {P.}~\bibnamefont
  {Flandrin}},\ }\href {\doibase 10.1017/9781108363181} {\emph {\bibinfo
  {title} {Explorations in Time-Frequency Analysis}}}\ (\bibinfo  {publisher}
  {Cambridge University Press},\ \bibinfo {year} {2018})\BibitemShut {NoStop}%
\bibitem [{\citenamefont {Szegedy}\ \emph {et~al.}(2014)\citenamefont {Szegedy}
  \emph {et~al.}}]{szegedy2014going}%
  \BibitemOpen
  \bibfield  {author} {\bibinfo {author} {\bibfnamefont {C.}~\bibnamefont
  {Szegedy}} \emph {et~al.},\ }\href@noop {} {\enquote {\bibinfo {title} {Going
  deeper with convolutions},}\ } (\bibinfo {year} {2014}),\ \Eprint
  {http://arxiv.org/abs/1409.4842} {arXiv:1409.4842 [cs.CV]} \BibitemShut
  {NoStop}%
\bibitem [{\citenamefont {Szegedy}\ \emph
  {et~al.}(2015{\natexlab{a}})\citenamefont {Szegedy} \emph
  {et~al.}}]{szegedy2015rethinking}%
  \BibitemOpen
  \bibfield  {author} {\bibinfo {author} {\bibfnamefont {C.}~\bibnamefont
  {Szegedy}} \emph {et~al.},\ }\href@noop {} {\enquote {\bibinfo {title}
  {Rethinking the inception architecture for computer vision},}\ } (\bibinfo
  {year} {2015}{\natexlab{a}}),\ \Eprint {http://arxiv.org/abs/1512.00567}
  {arXiv:1512.00567 [cs.CV]} \BibitemShut {NoStop}%
\bibitem [{\citenamefont {Blanchet}(2014)}]{Blanchet:2013haa}%
  \BibitemOpen
  \bibfield  {author} {\bibinfo {author} {\bibfnamefont {L.}~\bibnamefont
  {Blanchet}},\ }\href {\doibase 10.12942/lrr-2014-2} {\bibfield  {journal}
  {\bibinfo  {journal} {Living Rev. Rel.}\ }\textbf {\bibinfo {volume} {17}},\
  \bibinfo {pages} {2} (\bibinfo {year} {2014})},\ \Eprint
  {http://arxiv.org/abs/1310.1528} {arXiv:1310.1528 [gr-qc]} \BibitemShut
  {NoStop}%
\bibitem [{\citenamefont {Cotesta}\ \emph {et~al.}(2018)\citenamefont
  {Cotesta}, \citenamefont {Buonanno}, \citenamefont {Boh\'e}, \citenamefont
  {Taracchini}, \citenamefont {Hinder},\ and\ \citenamefont
  {Ossokine}}]{Cotesta:2018fcv}%
  \BibitemOpen
  \bibfield  {author} {\bibinfo {author} {\bibfnamefont {R.}~\bibnamefont
  {Cotesta}}, \bibinfo {author} {\bibfnamefont {A.}~\bibnamefont {Buonanno}},
  \bibinfo {author} {\bibfnamefont {A.}~\bibnamefont {Boh\'e}}, \bibinfo
  {author} {\bibfnamefont {A.}~\bibnamefont {Taracchini}}, \bibinfo {author}
  {\bibfnamefont {I.}~\bibnamefont {Hinder}}, \ and\ \bibinfo {author}
  {\bibfnamefont {S.}~\bibnamefont {Ossokine}},\ }\href {\doibase
  10.1103/PhysRevD.98.084028} {\bibfield  {journal} {\bibinfo  {journal} {Phys.
  Rev. D}\ }\textbf {\bibinfo {volume} {98}},\ \bibinfo {pages} {084028}
  (\bibinfo {year} {2018})},\ \Eprint {http://arxiv.org/abs/1803.10701}
  {arXiv:1803.10701 [gr-qc]} \BibitemShut {NoStop}%
\bibitem [{\citenamefont {Sathyaprakash}\ and\ \citenamefont
  {Dhurandhar}(1991)}]{Sathyaprakash:1991mt}%
  \BibitemOpen
  \bibfield  {author} {\bibinfo {author} {\bibfnamefont {B.~S.}\ \bibnamefont
  {Sathyaprakash}}\ and\ \bibinfo {author} {\bibfnamefont {S.~V.}\ \bibnamefont
  {Dhurandhar}},\ }\href {\doibase 10.1103/PhysRevD.44.3819} {\bibfield
  {journal} {\bibinfo  {journal} {Phys. Rev. D}\ }\textbf {\bibinfo {volume}
  {44}},\ \bibinfo {pages} {3819} (\bibinfo {year} {1991})}\BibitemShut
  {NoStop}%
\bibitem [{\citenamefont {Klimenko}\ \emph {et~al.}(2016)\citenamefont
  {Klimenko} \emph {et~al.}}]{Klimenko:2015ypf}%
  \BibitemOpen
  \bibfield  {author} {\bibinfo {author} {\bibfnamefont {S.}~\bibnamefont
  {Klimenko}} \emph {et~al.},\ }\href {\doibase 10.1103/PhysRevD.93.042004}
  {\bibfield  {journal} {\bibinfo  {journal} {Phys. Rev. D}\ }\textbf {\bibinfo
  {volume} {93}},\ \bibinfo {pages} {042004} (\bibinfo {year} {2016})},\
  \Eprint {http://arxiv.org/abs/1511.05999} {arXiv:1511.05999 [gr-qc]}
  \BibitemShut {NoStop}%
\bibitem [{\citenamefont {Babak}\ \emph {et~al.}(2013)\citenamefont {Babak}
  \emph {et~al.}}]{Babak:2012zx}%
  \BibitemOpen
  \bibfield  {author} {\bibinfo {author} {\bibfnamefont {S.}~\bibnamefont
  {Babak}} \emph {et~al.},\ }\href {\doibase 10.1103/PhysRevD.87.024033}
  {\bibfield  {journal} {\bibinfo  {journal} {Phys. Rev. D}\ }\textbf {\bibinfo
  {volume} {87}},\ \bibinfo {pages} {024033} (\bibinfo {year} {2013})},\
  \Eprint {http://arxiv.org/abs/1208.3491} {arXiv:1208.3491 [gr-qc]}
  \BibitemShut {NoStop}%
\bibitem [{\citenamefont {Nitz}\ \emph {et~al.}(2017)\citenamefont {Nitz},
  \citenamefont {Dent}, \citenamefont {Dal~Canton}, \citenamefont {Fairhurst},\
  and\ \citenamefont {Brown}}]{Nitz:2017svb}%
  \BibitemOpen
  \bibfield  {author} {\bibinfo {author} {\bibfnamefont {A.~H.}\ \bibnamefont
  {Nitz}}, \bibinfo {author} {\bibfnamefont {T.}~\bibnamefont {Dent}}, \bibinfo
  {author} {\bibfnamefont {T.}~\bibnamefont {Dal~Canton}}, \bibinfo {author}
  {\bibfnamefont {S.}~\bibnamefont {Fairhurst}}, \ and\ \bibinfo {author}
  {\bibfnamefont {D.~A.}\ \bibnamefont {Brown}},\ }\href {\doibase
  10.3847/1538-4357/aa8f50} {\bibfield  {journal} {\bibinfo  {journal}
  {Astrophys. J.}\ }\textbf {\bibinfo {volume} {849}},\ \bibinfo {pages} {118}
  (\bibinfo {year} {2017})},\ \Eprint {http://arxiv.org/abs/1705.01513}
  {arXiv:1705.01513 [gr-qc]} \BibitemShut {NoStop}%
\bibitem [{\citenamefont {Nitz}(2018)}]{Nitz:2017lco}%
  \BibitemOpen
  \bibfield  {author} {\bibinfo {author} {\bibfnamefont {A.~H.}\ \bibnamefont
  {Nitz}},\ }\href {\doibase 10.1088/1361-6382/aaa13d} {\bibfield  {journal}
  {\bibinfo  {journal} {Class. Quant. Grav.}\ }\textbf {\bibinfo {volume}
  {35}},\ \bibinfo {pages} {035016} (\bibinfo {year} {2018})},\ \Eprint
  {http://arxiv.org/abs/1709.08974} {arXiv:1709.08974 [gr-qc]} \BibitemShut
  {NoStop}%
\bibitem [{\citenamefont {Cornish}\ \emph {et~al.}(2021)\citenamefont
  {Cornish}, \citenamefont {Littenberg}, \citenamefont {B\'ecsy}, \citenamefont
  {Chatziioannou}, \citenamefont {Clark}, \citenamefont {Ghonge},\ and\
  \citenamefont {Millhouse}}]{Cornish:2020dwh}%
  \BibitemOpen
  \bibfield  {author} {\bibinfo {author} {\bibfnamefont {N.~J.}\ \bibnamefont
  {Cornish}}, \bibinfo {author} {\bibfnamefont {T.~B.}\ \bibnamefont
  {Littenberg}}, \bibinfo {author} {\bibfnamefont {B.}~\bibnamefont {B\'ecsy}},
  \bibinfo {author} {\bibfnamefont {K.}~\bibnamefont {Chatziioannou}}, \bibinfo
  {author} {\bibfnamefont {J.~A.}\ \bibnamefont {Clark}}, \bibinfo {author}
  {\bibfnamefont {S.}~\bibnamefont {Ghonge}}, \ and\ \bibinfo {author}
  {\bibfnamefont {M.}~\bibnamefont {Millhouse}},\ }\href {\doibase
  10.1103/PhysRevD.103.044006} {\bibfield  {journal} {\bibinfo  {journal}
  {Phys. Rev. D}\ }\textbf {\bibinfo {volume} {103}},\ \bibinfo {pages}
  {044006} (\bibinfo {year} {2021})},\ \Eprint
  {http://arxiv.org/abs/2011.09494} {arXiv:2011.09494 [gr-qc]} \BibitemShut
  {NoStop}%
\bibitem [{\citenamefont {Arai}\ and\ \citenamefont
  {Kapoor}(2020)}]{dl_vs_st_in_cv}%
  \BibitemOpen
  \bibinfo {editor} {\bibfnamefont {K.}~\bibnamefont {Arai}}\ and\ \bibinfo
  {editor} {\bibfnamefont {S.}~\bibnamefont {Kapoor}},\ eds.,\ \href {\doibase
  10.1007/978-3-030-17795-9} {\emph {\bibinfo {title} {Advances in Computer
  Vision}}}\ (\bibinfo  {publisher} {Springer International Publishing},\
  \bibinfo {year} {2020})\BibitemShut {NoStop}%
\bibitem [{\citenamefont {Mahabal}\ \emph {et~al.}(2019)\citenamefont {Mahabal}
  \emph {et~al.}}]{Mahabal_2019}%
  \BibitemOpen
  \bibfield  {author} {\bibinfo {author} {\bibfnamefont {A.}~\bibnamefont
  {Mahabal}} \emph {et~al.},\ }\href {\doibase 10.1088/1538-3873/aaf3fa}
  {\bibfield  {journal} {\bibinfo  {journal} {Publications of the Astronomical
  Society of the Pacific}\ }\textbf {\bibinfo {volume} {131}},\ \bibinfo
  {pages} {038002} (\bibinfo {year} {2019})}\BibitemShut {NoStop}%
\bibitem [{\citenamefont {{Sanchez Saez}}(2022)}]{2022scio.confE..17S}%
  \BibitemOpen
  \bibfield  {author} {\bibinfo {author} {\bibfnamefont {P.}~\bibnamefont
  {{Sanchez Saez}}},\ }in\ \href {\doibase 10.5281/zenodo.6563034} {\emph
  {\bibinfo {booktitle} {SciOps 2022: Artificial Intelligence for Science and
  Operations in Astronomy (SCIOPS). Proceedings of the ESA/ESO SCOPS Workshop
  held 16-20 May}}}\ (\bibinfo {year} {2022})\ p.~\bibinfo {pages}
  {17}\BibitemShut {NoStop}%
\bibitem [{\citenamefont {Vibho}\ and\ \citenamefont
  {Assaf}(2022)}]{Vibho:2022jkn}%
  \BibitemOpen
  \bibfield  {author} {\bibinfo {author} {\bibfnamefont {A.}~\bibnamefont
  {Vibho}}\ and\ \bibinfo {author} {\bibfnamefont {R.}~\bibnamefont {Assaf}},\
  }\href {\doibase 10.1016/j.ascom.2022.100566} {\bibfield  {journal} {\bibinfo
   {journal} {Astron. Comput.}\ }\textbf {\bibinfo {volume} {39}},\ \bibinfo
  {pages} {100566} (\bibinfo {year} {2022})}\BibitemShut {NoStop}%
\bibitem [{\citenamefont {Barchi}\ \emph {et~al.}(2020)\citenamefont {Barchi},
  \citenamefont {de~Carvalho}, \citenamefont {Rosa}, \citenamefont {Sautter},
  \citenamefont {Soares-Santos}, \citenamefont {Marques}, \citenamefont {Clua},
  \citenamefont {Gon\c{c}alves}, \citenamefont {de~S\'a-Freitas},\ and\
  \citenamefont {Moura}}]{Barchi:2019yda}%
  \BibitemOpen
  \bibfield  {author} {\bibinfo {author} {\bibfnamefont {P.~H.}\ \bibnamefont
  {Barchi}}, \bibinfo {author} {\bibfnamefont {R.~R.}\ \bibnamefont
  {de~Carvalho}}, \bibinfo {author} {\bibfnamefont {R.~R.}\ \bibnamefont
  {Rosa}}, \bibinfo {author} {\bibfnamefont {R.~A.}\ \bibnamefont {Sautter}},
  \bibinfo {author} {\bibfnamefont {M.}~\bibnamefont {Soares-Santos}}, \bibinfo
  {author} {\bibfnamefont {B.~A.~D.}\ \bibnamefont {Marques}}, \bibinfo
  {author} {\bibfnamefont {E.}~\bibnamefont {Clua}}, \bibinfo {author}
  {\bibfnamefont {T.~S.}\ \bibnamefont {Gon\c{c}alves}}, \bibinfo {author}
  {\bibfnamefont {C.}~\bibnamefont {de~S\'a-Freitas}}, \ and\ \bibinfo {author}
  {\bibfnamefont {T.~C.}\ \bibnamefont {Moura}},\ }\href {\doibase
  10.1016/j.ascom.2019.100334} {\bibfield  {journal} {\bibinfo  {journal}
  {Astron. Comput.}\ }\textbf {\bibinfo {volume} {30}},\ \bibinfo {pages}
  {100334} (\bibinfo {year} {2020})},\ \Eprint
  {http://arxiv.org/abs/1901.07047} {arXiv:1901.07047 [astro-ph.IM]}
  \BibitemShut {NoStop}%
\bibitem [{\citenamefont {González}\ \emph {et~al.}(2018)\citenamefont
  {González}, \citenamefont {Muñoz},\ and\ \citenamefont
  {Hernández}}]{GONZALEZ2018103}%
  \BibitemOpen
  \bibfield  {author} {\bibinfo {author} {\bibfnamefont {R.}~\bibnamefont
  {González}}, \bibinfo {author} {\bibfnamefont {R.}~\bibnamefont {Muñoz}}, \
  and\ \bibinfo {author} {\bibfnamefont {C.}~\bibnamefont {Hernández}},\
  }\href {\doibase https://doi.org/10.1016/j.ascom.2018.09.004} {\bibfield
  {journal} {\bibinfo  {journal} {Astronomy and Computing}\ }\textbf {\bibinfo
  {volume} {25}},\ \bibinfo {pages} {103} (\bibinfo {year} {2018})}\BibitemShut
  {NoStop}%
\bibitem [{\citenamefont {Duev}\ \emph {et~al.}(2019)\citenamefont {Duev} \emph
  {et~al.}}]{Duev_2019}%
  \BibitemOpen
  \bibfield  {author} {\bibinfo {author} {\bibfnamefont {D.~A.}\ \bibnamefont
  {Duev}} \emph {et~al.},\ }\href {\doibase 10.1093/mnras/stz1096} {\bibfield
  {journal} {\bibinfo  {journal} {Monthly Notices of the Royal Astronomical
  Society}\ }\textbf {\bibinfo {volume} {486}},\ \bibinfo {pages} {4158}
  (\bibinfo {year} {2019})}\BibitemShut {NoStop}%
\bibitem [{\citenamefont {George}\ and\ \citenamefont
  {Huerta}(2017)}]{George:2017vlv}%
  \BibitemOpen
  \bibfield  {author} {\bibinfo {author} {\bibfnamefont {D.}~\bibnamefont
  {George}}\ and\ \bibinfo {author} {\bibfnamefont {E.~A.}\ \bibnamefont
  {Huerta}},\ }in\ \href@noop {} {\emph {\bibinfo {booktitle} {{NiPS Summer
  School 2017}}}}\ (\bibinfo {year} {2017})\ \Eprint
  {http://arxiv.org/abs/1711.07966} {arXiv:1711.07966 [gr-qc]} \BibitemShut
  {NoStop}%
\bibitem [{\citenamefont {Gabbard}\ \emph {et~al.}(2018)\citenamefont
  {Gabbard}, \citenamefont {Williams}, \citenamefont {Hayes},\ and\
  \citenamefont {Messenger}}]{Gabbard:2017lja}%
  \BibitemOpen
  \bibfield  {author} {\bibinfo {author} {\bibfnamefont {H.}~\bibnamefont
  {Gabbard}}, \bibinfo {author} {\bibfnamefont {M.}~\bibnamefont {Williams}},
  \bibinfo {author} {\bibfnamefont {F.}~\bibnamefont {Hayes}}, \ and\ \bibinfo
  {author} {\bibfnamefont {C.}~\bibnamefont {Messenger}},\ }\href {\doibase
  10.1103/PhysRevLett.120.141103} {\bibfield  {journal} {\bibinfo  {journal}
  {Phys. Rev. Lett.}\ }\textbf {\bibinfo {volume} {120}},\ \bibinfo {pages}
  {141103} (\bibinfo {year} {2018})},\ \Eprint
  {http://arxiv.org/abs/1712.06041} {arXiv:1712.06041 [astro-ph.IM]}
  \BibitemShut {NoStop}%
\bibitem [{\citenamefont {\'Alvares}\ \emph {et~al.}(2020)\citenamefont
  {\'Alvares} \emph {et~al.}}]{Alvares:2020bjg}%
  \BibitemOpen
  \bibfield  {author} {\bibinfo {author} {\bibfnamefont {J.~a.~D.}\
  \bibnamefont {\'Alvares}} \emph {et~al.}\ }(\bibinfo {year} {2020})\ \Eprint
  {http://arxiv.org/abs/2011.10425} {arXiv:2011.10425 [gr-qc]} \BibitemShut
  {NoStop}%
\bibitem [{\citenamefont {Chatterjee}\ \emph {et~al.}(2021)\citenamefont
  {Chatterjee}, \citenamefont {Wen}, \citenamefont {Diakogiannis},\ and\
  \citenamefont {Vinsen}}]{Chatterjee:2021lit}%
  \BibitemOpen
  \bibfield  {author} {\bibinfo {author} {\bibfnamefont {C.}~\bibnamefont
  {Chatterjee}}, \bibinfo {author} {\bibfnamefont {L.}~\bibnamefont {Wen}},
  \bibinfo {author} {\bibfnamefont {F.}~\bibnamefont {Diakogiannis}}, \ and\
  \bibinfo {author} {\bibfnamefont {K.}~\bibnamefont {Vinsen}},\ }\href
  {\doibase 10.1103/PhysRevD.104.064046} {\bibfield  {journal} {\bibinfo
  {journal} {Phys. Rev. D}\ }\textbf {\bibinfo {volume} {104}},\ \bibinfo
  {pages} {064046} (\bibinfo {year} {2021})},\ \Eprint
  {http://arxiv.org/abs/2105.03073} {arXiv:2105.03073 [gr-qc]} \BibitemShut
  {NoStop}%
\bibitem [{\citenamefont {Gebhard}\ \emph {et~al.}(2019)\citenamefont
  {Gebhard}, \citenamefont {Kilbertus}, \citenamefont {Harry},\ and\
  \citenamefont {Sch\"olkopf}}]{PhysRevD.100.063015}%
  \BibitemOpen
  \bibfield  {author} {\bibinfo {author} {\bibfnamefont {T.~D.}\ \bibnamefont
  {Gebhard}}, \bibinfo {author} {\bibfnamefont {N.}~\bibnamefont {Kilbertus}},
  \bibinfo {author} {\bibfnamefont {I.}~\bibnamefont {Harry}}, \ and\ \bibinfo
  {author} {\bibfnamefont {B.}~\bibnamefont {Sch\"olkopf}},\ }\href {\doibase
  10.1103/PhysRevD.100.063015} {\bibfield  {journal} {\bibinfo  {journal}
  {Phys. Rev. D}\ }\textbf {\bibinfo {volume} {100}},\ \bibinfo {pages}
  {063015} (\bibinfo {year} {2019})}\BibitemShut {NoStop}%
\bibitem [{\citenamefont {Sch\"afer}\ \emph {et~al.}(2022)\citenamefont
  {Sch\"afer}, \citenamefont {Zelenka}, \citenamefont {Nitz}, \citenamefont
  {Ohme},\ and\ \citenamefont {Br\"ugmann}}]{Schafer:2021fea}%
  \BibitemOpen
  \bibfield  {author} {\bibinfo {author} {\bibfnamefont {M.~B.}\ \bibnamefont
  {Sch\"afer}}, \bibinfo {author} {\bibfnamefont {O.}~\bibnamefont {Zelenka}},
  \bibinfo {author} {\bibfnamefont {A.~H.}\ \bibnamefont {Nitz}}, \bibinfo
  {author} {\bibfnamefont {F.}~\bibnamefont {Ohme}}, \ and\ \bibinfo {author}
  {\bibfnamefont {B.}~\bibnamefont {Br\"ugmann}},\ }\href {\doibase
  10.1103/PhysRevD.105.043002} {\bibfield  {journal} {\bibinfo  {journal}
  {Phys. Rev. D}\ }\textbf {\bibinfo {volume} {105}},\ \bibinfo {pages}
  {043002} (\bibinfo {year} {2022})},\ \Eprint
  {http://arxiv.org/abs/2106.03741} {arXiv:2106.03741 [astro-ph.IM]}
  \BibitemShut {NoStop}%
\bibitem [{\citenamefont {Sch\"afer}\ and\ \citenamefont
  {Nitz}(2022)}]{PhysRevD.105.043003}%
  \BibitemOpen
  \bibfield  {author} {\bibinfo {author} {\bibfnamefont {M.~B.}\ \bibnamefont
  {Sch\"afer}}\ and\ \bibinfo {author} {\bibfnamefont {A.~H.}\ \bibnamefont
  {Nitz}},\ }\href {\doibase 10.1103/PhysRevD.105.043003} {\bibfield  {journal}
  {\bibinfo  {journal} {Phys. Rev. D}\ }\textbf {\bibinfo {volume} {105}},\
  \bibinfo {pages} {043003} (\bibinfo {year} {2022})}\BibitemShut {NoStop}%
\bibitem [{\citenamefont {L\'opez}\ \emph {et~al.}(2021)\citenamefont
  {L\'opez}, \citenamefont {Drago}, \citenamefont {Palma}, \citenamefont
  {Ricci},\ and\ \citenamefont {Cerd\'a-Dur\'an}}]{Lopez:2021rci}%
  \BibitemOpen
  \bibfield  {author} {\bibinfo {author} {\bibfnamefont {M.}~\bibnamefont
  {L\'opez}}, \bibinfo {author} {\bibfnamefont {M.}~\bibnamefont {Drago}},
  \bibinfo {author} {\bibfnamefont {I.~D.}\ \bibnamefont {Palma}}, \bibinfo
  {author} {\bibfnamefont {F.}~\bibnamefont {Ricci}}, \ and\ \bibinfo {author}
  {\bibfnamefont {P.}~\bibnamefont {Cerd\'a-Dur\'an}},\ }in\ \href {\doibase
  10.1109/CBMI50038.2021.9461885} {\emph {\bibinfo {booktitle} {{International
  Conference on Content-Based Multimedia Indexing}}}}\ (\bibinfo {year}
  {2021})\BibitemShut {NoStop}%
\bibitem [{\citenamefont {George}\ \emph {et~al.}(2018)\citenamefont {George},
  \citenamefont {Shen},\ and\ \citenamefont {Huerta}}]{George:2018awu}%
  \BibitemOpen
  \bibfield  {author} {\bibinfo {author} {\bibfnamefont {D.}~\bibnamefont
  {George}}, \bibinfo {author} {\bibfnamefont {H.}~\bibnamefont {Shen}}, \ and\
  \bibinfo {author} {\bibfnamefont {E.~A.}\ \bibnamefont {Huerta}},\ }\href
  {\doibase 10.1103/PhysRevD.97.101501} {\bibfield  {journal} {\bibinfo
  {journal} {Phys. Rev. D}\ }\textbf {\bibinfo {volume} {97}},\ \bibinfo
  {pages} {101501} (\bibinfo {year} {2018})}\BibitemShut {NoStop}%
\bibitem [{\citenamefont {Lopez}\ \emph {et~al.}(2022)\citenamefont {Lopez},
  \citenamefont {Gayathri}, \citenamefont {Pai}, \citenamefont {Heng},
  \citenamefont {Messenger},\ and\ \citenamefont {Gupta}}]{Lopez:2021ikt}%
  \BibitemOpen
  \bibfield  {author} {\bibinfo {author} {\bibfnamefont {D.}~\bibnamefont
  {Lopez}}, \bibinfo {author} {\bibfnamefont {V.}~\bibnamefont {Gayathri}},
  \bibinfo {author} {\bibfnamefont {A.}~\bibnamefont {Pai}}, \bibinfo {author}
  {\bibfnamefont {I.~S.}\ \bibnamefont {Heng}}, \bibinfo {author}
  {\bibfnamefont {C.}~\bibnamefont {Messenger}}, \ and\ \bibinfo {author}
  {\bibfnamefont {S.~K.}\ \bibnamefont {Gupta}},\ }\href {\doibase
  10.1103/PhysRevD.105.063024} {\bibfield  {journal} {\bibinfo  {journal}
  {Phys. Rev. D}\ }\textbf {\bibinfo {volume} {105}},\ \bibinfo {pages}
  {063024} (\bibinfo {year} {2022})},\ \Eprint
  {http://arxiv.org/abs/2112.06608} {arXiv:2112.06608 [gr-qc]} \BibitemShut
  {NoStop}%
\bibitem [{\citenamefont {Zevin}\ \emph {et~al.}(2017)\citenamefont {Zevin}
  \emph {et~al.}}]{Zevin:2016qwy}%
  \BibitemOpen
  \bibfield  {author} {\bibinfo {author} {\bibfnamefont {M.}~\bibnamefont
  {Zevin}} \emph {et~al.},\ }\href {\doibase 10.1088/1361-6382/aa5cea}
  {\bibfield  {journal} {\bibinfo  {journal} {Class. Quant. Grav.}\ }\textbf
  {\bibinfo {volume} {34}},\ \bibinfo {pages} {064003} (\bibinfo {year}
  {2017})},\ \Eprint {http://arxiv.org/abs/1611.04596} {arXiv:1611.04596
  [gr-qc]} \BibitemShut {NoStop}%
\bibitem [{\citenamefont {Cabero}\ \emph {et~al.}(2019)\citenamefont {Cabero}
  \emph {et~al.}}]{Cabero:2019orq}%
  \BibitemOpen
  \bibfield  {author} {\bibinfo {author} {\bibfnamefont {M.}~\bibnamefont
  {Cabero}} \emph {et~al.},\ }\href {\doibase 10.1088/1361-6382/ab2e14}
  {\bibfield  {journal} {\bibinfo  {journal} {Class. Quant. Grav.}\ }\textbf
  {\bibinfo {volume} {36}},\ \bibinfo {pages} {15} (\bibinfo {year} {2019})},\
  \Eprint {http://arxiv.org/abs/1901.05093} {arXiv:1901.05093
  [physics.ins-det]} \BibitemShut {NoStop}%
\bibitem [{\citenamefont {Davis}(2019)}]{Davis:2019mgu}%
  \BibitemOpen
  \bibfield  {author} {\bibinfo {author} {\bibfnamefont {D.}~\bibnamefont
  {Davis}},\ }\emph {\bibinfo {title} {{Improving the Sensitivity of Advanced
  Ligo Through Detector Characterization}}},\ \href@noop {} {Ph.D. thesis},\
  \bibinfo  {school} {Syracuse U., Syracuse U.} (\bibinfo {year}
  {2019})\BibitemShut {NoStop}%
\bibitem [{\citenamefont {Soni}\ \emph {et~al.}(2021)\citenamefont {Soni} \emph
  {et~al.}}]{Soni:2021cjy}%
  \BibitemOpen
  \bibfield  {author} {\bibinfo {author} {\bibfnamefont {S.}~\bibnamefont
  {Soni}} \emph {et~al.},\ }\href {\doibase 10.1088/1361-6382/ac1ccb}
  {\bibfield  {journal} {\bibinfo  {journal} {Class. Quant. Grav.}\ }\textbf
  {\bibinfo {volume} {38}},\ \bibinfo {pages} {195016} (\bibinfo {year}
  {2021})},\ \Eprint {http://arxiv.org/abs/2103.12104} {arXiv:2103.12104
  [gr-qc]} \BibitemShut {NoStop}%
\bibitem [{\citenamefont {Davis}\ \emph {et~al.}(2021)\citenamefont {Davis}
  \emph {et~al.}}]{LIGO:2021ppb}%
  \BibitemOpen
  \bibfield  {author} {\bibinfo {author} {\bibfnamefont {D.}~\bibnamefont
  {Davis}} \emph {et~al.} (\bibinfo {collaboration} {LIGO}),\ }\href {\doibase
  10.1088/1361-6382/abfd85} {\bibfield  {journal} {\bibinfo  {journal} {Class.
  Quant. Grav.}\ }\textbf {\bibinfo {volume} {38}},\ \bibinfo {pages} {135014}
  (\bibinfo {year} {2021})},\ \Eprint {http://arxiv.org/abs/2101.11673}
  {arXiv:2101.11673 [astro-ph.IM]} \BibitemShut {NoStop}%
\bibitem [{\citenamefont {Abbott}\ \emph {et~al.}(2016)\citenamefont {Abbott}
  \emph {et~al.}}]{LIGOScientific:2016gtq}%
  \BibitemOpen
  \bibfield  {author} {\bibinfo {author} {\bibfnamefont {B.~P.}\ \bibnamefont
  {Abbott}} \emph {et~al.} (\bibinfo {collaboration} {LIGO Scientific,
  Virgo}),\ }\href {\doibase 10.1088/0264-9381/33/13/134001} {\bibfield
  {journal} {\bibinfo  {journal} {Class. Quant. Grav.}\ }\textbf {\bibinfo
  {volume} {33}},\ \bibinfo {pages} {134001} (\bibinfo {year} {2016})},\
  \Eprint {http://arxiv.org/abs/1602.03844} {arXiv:1602.03844 [gr-qc]}
  \BibitemShut {NoStop}%
\bibitem [{\citenamefont {Nuttall}\ \emph {et~al.}(2015)\citenamefont {Nuttall}
  \emph {et~al.}}]{Nuttall:2015dqa}%
  \BibitemOpen
  \bibfield  {author} {\bibinfo {author} {\bibfnamefont {L.}~\bibnamefont
  {Nuttall}} \emph {et~al.},\ }\href {\doibase 10.1088/0264-9381/32/24/245005}
  {\bibfield  {journal} {\bibinfo  {journal} {Class. Quant. Grav.}\ }\textbf
  {\bibinfo {volume} {32}},\ \bibinfo {pages} {245005} (\bibinfo {year}
  {2015})},\ \Eprint {http://arxiv.org/abs/1508.07316} {arXiv:1508.07316
  [gr-qc]} \BibitemShut {NoStop}%
\bibitem [{\citenamefont {Santamaria}\ \emph {et~al.}(2010)\citenamefont
  {Santamaria} \emph {et~al.}}]{Santamaria:2010yb}%
  \BibitemOpen
  \bibfield  {author} {\bibinfo {author} {\bibfnamefont {L.}~\bibnamefont
  {Santamaria}} \emph {et~al.},\ }\href {\doibase 10.1103/PhysRevD.82.064016}
  {\bibfield  {journal} {\bibinfo  {journal} {Phys. Rev. D}\ }\textbf {\bibinfo
  {volume} {82}},\ \bibinfo {pages} {064016} (\bibinfo {year} {2010})},\
  \Eprint {http://arxiv.org/abs/1005.3306} {arXiv:1005.3306 [gr-qc]}
  \BibitemShut {NoStop}%
\bibitem [{\citenamefont {Campanelli}\ \emph {et~al.}(2006)\citenamefont
  {Campanelli}, \citenamefont {Lousto},\ and\ \citenamefont
  {Zlochower}}]{Campanelli:2006uy}%
  \BibitemOpen
  \bibfield  {author} {\bibinfo {author} {\bibfnamefont {M.}~\bibnamefont
  {Campanelli}}, \bibinfo {author} {\bibfnamefont {C.~O.}\ \bibnamefont
  {Lousto}}, \ and\ \bibinfo {author} {\bibfnamefont {Y.}~\bibnamefont
  {Zlochower}},\ }\href {\doibase 10.1103/PhysRevD.74.041501} {\bibfield
  {journal} {\bibinfo  {journal} {Phys. Rev. D}\ }\textbf {\bibinfo {volume}
  {74}},\ \bibinfo {pages} {041501} (\bibinfo {year} {2006})},\ \Eprint
  {http://arxiv.org/abs/gr-qc/0604012} {arXiv:gr-qc/0604012} \BibitemShut
  {NoStop}%
\bibitem [{\citenamefont {Abbott}\ \emph {et~al.}(2022)\citenamefont {Abbott}
  \emph {et~al.}}]{LIGOScientific:2021tfm}%
  \BibitemOpen
  \bibfield  {author} {\bibinfo {author} {\bibfnamefont {R.}~\bibnamefont
  {Abbott}} \emph {et~al.} (\bibinfo {collaboration} {LIGO Scientific, VIRGO,
  KAGRA}),\ }\href {\doibase 10.1051/0004-6361/202141452} {\bibfield  {journal}
  {\bibinfo  {journal} {Astron. Astrophys.}\ }\textbf {\bibinfo {volume}
  {659}},\ \bibinfo {pages} {A84} (\bibinfo {year} {2022})},\ \Eprint
  {http://arxiv.org/abs/2105.15120} {arXiv:2105.15120 [astro-ph.HE]}
  \BibitemShut {NoStop}%
\bibitem [{\citenamefont {Pratt}(1992)}]{NIPS1992_67e103b0}%
  \BibitemOpen
  \bibfield  {author} {\bibinfo {author} {\bibfnamefont {L.~Y.}\ \bibnamefont
  {Pratt}},\ }in\ \href
  {https://proceedings.neurips.cc/paper/1992/file/67e103b0761e60683e83c559be18d40c-Paper.pdf}
  {\emph {\bibinfo {booktitle} {Advances in Neural Information Processing
  Systems}}},\ Vol.~\bibinfo {volume} {5},\ \bibinfo {editor} {edited by\
  \bibinfo {editor} {\bibfnamefont {S.}~\bibnamefont {Hanson}}, \bibinfo
  {editor} {\bibfnamefont {J.}~\bibnamefont {Cowan}}, \ and\ \bibinfo {editor}
  {\bibfnamefont {C.}~\bibnamefont {Giles}}}\ (\bibinfo  {publisher}
  {Morgan-Kaufmann},\ \bibinfo {year} {1992})\BibitemShut {NoStop}%
\bibitem [{\citenamefont {Zhuang}\ \emph {et~al.}(2019)\citenamefont {Zhuang},
  \citenamefont {Qi}, \citenamefont {Duan}, \citenamefont {Xi}, \citenamefont
  {Zhu}, \citenamefont {Zhu}, \citenamefont {Xiong},\ and\ \citenamefont
  {He}}]{https://doi.org/10.48550/arxiv.1911.02685}%
  \BibitemOpen
  \bibfield  {author} {\bibinfo {author} {\bibfnamefont {F.}~\bibnamefont
  {Zhuang}}, \bibinfo {author} {\bibfnamefont {Z.}~\bibnamefont {Qi}}, \bibinfo
  {author} {\bibfnamefont {K.}~\bibnamefont {Duan}}, \bibinfo {author}
  {\bibfnamefont {D.}~\bibnamefont {Xi}}, \bibinfo {author} {\bibfnamefont
  {Y.}~\bibnamefont {Zhu}}, \bibinfo {author} {\bibfnamefont {H.}~\bibnamefont
  {Zhu}}, \bibinfo {author} {\bibfnamefont {H.}~\bibnamefont {Xiong}}, \ and\
  \bibinfo {author} {\bibfnamefont {Q.}~\bibnamefont {He}},\ }\href {\doibase
  10.48550/ARXIV.1911.02685} {\enquote {\bibinfo {title} {A comprehensive
  survey on transfer learning},}\ } (\bibinfo {year} {2019})\BibitemShut
  {NoStop}%
\bibitem [{\citenamefont {Ribani}\ and\ \citenamefont
  {Marengoni}(2019)}]{8920338}%
  \BibitemOpen
  \bibfield  {author} {\bibinfo {author} {\bibfnamefont {R.}~\bibnamefont
  {Ribani}}\ and\ \bibinfo {author} {\bibfnamefont {M.}~\bibnamefont
  {Marengoni}},\ }in\ \href {\doibase 10.1109/SIBGRAPI-T.2019.00010} {\emph
  {\bibinfo {booktitle} {2019 32nd SIBGRAPI Conference on Graphics, Patterns
  and Images Tutorials (SIBGRAPI-T)}}}\ (\bibinfo {year} {2019})\ pp.\ \bibinfo
  {pages} {47--57}\BibitemShut {NoStop}%
\bibitem [{\citenamefont {Yang}\ \emph {et~al.}(2022)\citenamefont {Yang},
  \citenamefont {Xiao}, \citenamefont {Zhang}, \citenamefont {Guo},
  \citenamefont {Zhao},\ and\ \citenamefont
  {Shen}}]{https://doi.org/10.48550/arxiv.2204.08610}%
  \BibitemOpen
  \bibfield  {author} {\bibinfo {author} {\bibfnamefont {S.}~\bibnamefont
  {Yang}}, \bibinfo {author} {\bibfnamefont {W.}~\bibnamefont {Xiao}}, \bibinfo
  {author} {\bibfnamefont {M.}~\bibnamefont {Zhang}}, \bibinfo {author}
  {\bibfnamefont {S.}~\bibnamefont {Guo}}, \bibinfo {author} {\bibfnamefont
  {J.}~\bibnamefont {Zhao}}, \ and\ \bibinfo {author} {\bibfnamefont
  {F.}~\bibnamefont {Shen}},\ }\href {\doibase 10.48550/ARXIV.2204.08610}
  {\enquote {\bibinfo {title} {Image data augmentation for deep learning: A
  survey},}\ } (\bibinfo {year} {2022})\BibitemShut {NoStop}%
\bibitem [{\citenamefont {Vakili}\ \emph {et~al.}(2020)\citenamefont {Vakili},
  \citenamefont {Ghamsari},\ and\ \citenamefont
  {Rezaei}}]{https://doi.org/10.48550/arxiv.2001.09636}%
  \BibitemOpen
  \bibfield  {author} {\bibinfo {author} {\bibfnamefont {M.}~\bibnamefont
  {Vakili}}, \bibinfo {author} {\bibfnamefont {M.}~\bibnamefont {Ghamsari}}, \
  and\ \bibinfo {author} {\bibfnamefont {M.}~\bibnamefont {Rezaei}},\ }\href
  {\doibase 10.48550/ARXIV.2001.09636} {\enquote {\bibinfo {title} {Performance
  analysis and comparison of machine and deep learning algorithms for iot data
  classification},}\ } (\bibinfo {year} {2020})\BibitemShut {NoStop}%
\bibitem [{\citenamefont {Goodfellow}\ \emph {et~al.}(2016)\citenamefont
  {Goodfellow}, \citenamefont {Bengio},\ and\ \citenamefont
  {Courville}}]{Goodfellow-et-al-2016}%
  \BibitemOpen
  \bibfield  {author} {\bibinfo {author} {\bibfnamefont {I.}~\bibnamefont
  {Goodfellow}}, \bibinfo {author} {\bibfnamefont {Y.}~\bibnamefont {Bengio}},
  \ and\ \bibinfo {author} {\bibfnamefont {A.}~\bibnamefont {Courville}},\
  }\href@noop {} {\emph {\bibinfo {title} {Deep Learning}}}\ (\bibinfo
  {publisher} {MIT Press},\ \bibinfo {year} {2016})\ \bibinfo {note}
  {\url{http://www.deeplearningbook.org}}\BibitemShut {NoStop}%
\bibitem [{\citenamefont {Abadi}\ \emph {et~al.}(2016)\citenamefont {Abadi}
  \emph {et~al.}}]{https://doi.org/10.48550/arxiv.1603.04467}%
  \BibitemOpen
  \bibfield  {author} {\bibinfo {author} {\bibfnamefont {M.}~\bibnamefont
  {Abadi}} \emph {et~al.},\ }\href@noop {} {\  (\bibinfo {year} {2016})},\
  \Eprint {http://arxiv.org/abs/1603.04467} {arXiv:1603.04467 [cs.DC]}
  \BibitemShut {NoStop}%
\bibitem [{\citenamefont {Bisong}(2019)}]{Bisong2019}%
  \BibitemOpen
  \bibfield  {author} {\bibinfo {author} {\bibfnamefont {E.}~\bibnamefont
  {Bisong}},\ }\enquote {\bibinfo {title} {Google colaboratory},}\ in\ \href
  {\doibase 10.1007/978-1-4842-4470-8_7} {\emph {\bibinfo {booktitle} {Building
  Machine Learning and Deep Learning Models on Google Cloud Platform: A
  Comprehensive Guide for Beginners}}}\ (\bibinfo  {publisher} {Apress},\
  \bibinfo {address} {Berkeley, CA},\ \bibinfo {year} {2019})\ pp.\ \bibinfo
  {pages} {59--64}\BibitemShut {NoStop}%
\bibitem [{\citenamefont {Boldaji}\ and\ \citenamefont
  {Semnani}(2021)}]{color_img_seg}%
  \BibitemOpen
  \bibfield  {author} {\bibinfo {author} {\bibfnamefont {M.~N.}\ \bibnamefont
  {Boldaji}}\ and\ \bibinfo {author} {\bibfnamefont {S.~H.}\ \bibnamefont
  {Semnani}},\ }\href {\doibase 10.48550/ARXIV.2110.09217} {\enquote {\bibinfo
  {title} {Color image segmentation using multi-objective swarm optimizer and
  multi-level histogram thresholding},}\ } (\bibinfo {year} {2021})\BibitemShut
  {NoStop}%
\bibitem [{\citenamefont {Pai}\ \emph {et~al.}(2001)\citenamefont {Pai},
  \citenamefont {Dhurandhar},\ and\ \citenamefont {Bose}}]{Pai:2000zt}%
  \BibitemOpen
  \bibfield  {author} {\bibinfo {author} {\bibfnamefont {A.}~\bibnamefont
  {Pai}}, \bibinfo {author} {\bibfnamefont {S.}~\bibnamefont {Dhurandhar}}, \
  and\ \bibinfo {author} {\bibfnamefont {S.}~\bibnamefont {Bose}},\ }\href
  {\doibase 10.1103/PhysRevD.64.042004} {\bibfield  {journal} {\bibinfo
  {journal} {Phys. Rev. D}\ }\textbf {\bibinfo {volume} {64}},\ \bibinfo
  {pages} {042004} (\bibinfo {year} {2001})},\ \Eprint
  {http://arxiv.org/abs/gr-qc/0009078} {arXiv:gr-qc/0009078} \BibitemShut
  {NoStop}%
\bibitem [{\citenamefont {Ying}(2019)}]{Ying_2019}%
  \BibitemOpen
  \bibfield  {author} {\bibinfo {author} {\bibfnamefont {X.}~\bibnamefont
  {Ying}},\ }\href {\doibase 10.1088/1742-6596/1168/2/022022} {\bibfield
  {journal} {\bibinfo  {journal} {Journal of Physics: Conference Series}\
  }\textbf {\bibinfo {volume} {1168}},\ \bibinfo {pages} {022022} (\bibinfo
  {year} {2019})}\BibitemShut {NoStop}%
\bibitem [{\citenamefont {Salman}\ and\ \citenamefont
  {Liu}(2019)}]{https://doi.org/10.48550/arxiv.1901.06566}%
  \BibitemOpen
  \bibfield  {author} {\bibinfo {author} {\bibfnamefont {S.}~\bibnamefont
  {Salman}}\ and\ \bibinfo {author} {\bibfnamefont {X.}~\bibnamefont {Liu}},\
  }\href {\doibase 10.48550/ARXIV.1901.06566} {\enquote {\bibinfo {title}
  {Overfitting mechanism and avoidance in deep neural networks},}\ } (\bibinfo
  {year} {2019})\BibitemShut {NoStop}%
\bibitem [{\citenamefont {Gholamalinezhad}\ and\ \citenamefont
  {Khosravi}(2020)}]{2020arXiv200907485G}%
  \BibitemOpen
  \bibfield  {author} {\bibinfo {author} {\bibfnamefont {H.}~\bibnamefont
  {Gholamalinezhad}}\ and\ \bibinfo {author} {\bibfnamefont {H.}~\bibnamefont
  {Khosravi}},\ }\href {\doibase 10.48550/ARXIV.2009.07485} {\enquote {\bibinfo
  {title} {Pooling methods in deep neural networks, a review},}\ } (\bibinfo
  {year} {2020})\BibitemShut {NoStop}%
\bibitem [{\citenamefont {Szegedy}\ \emph
  {et~al.}(2015{\natexlab{b}})\citenamefont {Szegedy} \emph
  {et~al.}}]{7298594}%
  \BibitemOpen
  \bibfield  {author} {\bibinfo {author} {\bibfnamefont {C.}~\bibnamefont
  {Szegedy}} \emph {et~al.},\ }in\ \href {\doibase 10.1109/CVPR.2015.7298594}
  {\emph {\bibinfo {booktitle} {2015 IEEE Conference on Computer Vision and
  Pattern Recognition (CVPR)}}}\ (\bibinfo {year} {2015})\ pp.\ \bibinfo
  {pages} {1--9}\BibitemShut {NoStop}%
\bibitem [{\citenamefont {Lee}\ \emph {et~al.}(2015)\citenamefont {Lee},
  \citenamefont {Xie}, \citenamefont {Gallagher}, \citenamefont {Zhang},\ and\
  \citenamefont {Tu}}]{pmlr-v38-lee15a}%
  \BibitemOpen
  \bibfield  {author} {\bibinfo {author} {\bibfnamefont {C.-Y.}\ \bibnamefont
  {Lee}}, \bibinfo {author} {\bibfnamefont {S.}~\bibnamefont {Xie}}, \bibinfo
  {author} {\bibfnamefont {P.}~\bibnamefont {Gallagher}}, \bibinfo {author}
  {\bibfnamefont {Z.}~\bibnamefont {Zhang}}, \ and\ \bibinfo {author}
  {\bibfnamefont {Z.}~\bibnamefont {Tu}},\ }in\ \href
  {https://proceedings.mlr.press/v38/lee15a.html} {\emph {\bibinfo {booktitle}
  {Proceedings of the Eighteenth International Conference on Artificial
  Intelligence and Statistics}}},\ \bibinfo {series} {Proceedings of Machine
  Learning Research}, Vol.~\bibinfo {volume} {38},\ \bibinfo {editor} {edited
  by\ \bibinfo {editor} {\bibfnamefont {G.}~\bibnamefont {Lebanon}}\ and\
  \bibinfo {editor} {\bibfnamefont {S.~V.~N.}\ \bibnamefont {Vishwanathan}}}\
  (\bibinfo  {publisher} {PMLR},\ \bibinfo {address} {San Diego, California,
  USA},\ \bibinfo {year} {2015})\ pp.\ \bibinfo {pages} {562--570}\BibitemShut
  {NoStop}%
\bibitem [{\citenamefont {Lin}\ \emph {et~al.}(2013)\citenamefont {Lin},
  \citenamefont {Chen},\ and\ \citenamefont
  {Yan}}]{https://doi.org/10.48550/arxiv.1312.4400}%
  \BibitemOpen
  \bibfield  {author} {\bibinfo {author} {\bibfnamefont {M.}~\bibnamefont
  {Lin}}, \bibinfo {author} {\bibfnamefont {Q.}~\bibnamefont {Chen}}, \ and\
  \bibinfo {author} {\bibfnamefont {S.}~\bibnamefont {Yan}},\ }\href {\doibase
  10.48550/ARXIV.1312.4400} {\enquote {\bibinfo {title} {Network in network},}\
  } (\bibinfo {year} {2013})\BibitemShut {NoStop}%
\bibitem [{\citenamefont {Huang}\ \emph {et~al.}(2016)\citenamefont {Huang},
  \citenamefont {Liu}, \citenamefont {van~der Maaten},\ and\ \citenamefont
  {Weinberger}}]{https://doi.org/10.48550/arxiv.1608.06993}%
  \BibitemOpen
  \bibfield  {author} {\bibinfo {author} {\bibfnamefont {G.}~\bibnamefont
  {Huang}}, \bibinfo {author} {\bibfnamefont {Z.}~\bibnamefont {Liu}}, \bibinfo
  {author} {\bibfnamefont {L.}~\bibnamefont {van~der Maaten}}, \ and\ \bibinfo
  {author} {\bibfnamefont {K.~Q.}\ \bibnamefont {Weinberger}},\ }\href
  {\doibase 10.48550/ARXIV.1608.06993} {\enquote {\bibinfo {title} {Densely
  connected convolutional networks},}\ } (\bibinfo {year} {2016})\BibitemShut
  {NoStop}%
\bibitem [{\citenamefont {Srivastava}\ \emph {et~al.}(2014)\citenamefont
  {Srivastava}, \citenamefont {Hinton}, \citenamefont {Krizhevsky},
  \citenamefont {Sutskever},\ and\ \citenamefont
  {Salakhutdinov}}]{JMLR:v15:srivastava14a}%
  \BibitemOpen
  \bibfield  {author} {\bibinfo {author} {\bibfnamefont {N.}~\bibnamefont
  {Srivastava}}, \bibinfo {author} {\bibfnamefont {G.}~\bibnamefont {Hinton}},
  \bibinfo {author} {\bibfnamefont {A.}~\bibnamefont {Krizhevsky}}, \bibinfo
  {author} {\bibfnamefont {I.}~\bibnamefont {Sutskever}}, \ and\ \bibinfo
  {author} {\bibfnamefont {R.}~\bibnamefont {Salakhutdinov}},\ }\href
  {http://jmlr.org/papers/v15/srivastava14a.html} {\bibfield  {journal}
  {\bibinfo  {journal} {Journal of Machine Learning Research}\ }\textbf
  {\bibinfo {volume} {15}},\ \bibinfo {pages} {1929} (\bibinfo {year}
  {2014})}\BibitemShut {NoStop}%
\end{thebibliography}%

\end{document}